\documentclass[useAMS,usenatbib]{mn2e}
\usepackage{amsmath}
\usepackage{amssymb}
\usepackage{graphicx}
\usepackage{float}
\usepackage{hyperref}
\usepackage{morefloats}
\renewcommand{\vec}[1]{ {\bf #1} }
\title[Angular momentum of haloes and their baryons]{Angular momentum
  properties of haloes and their baryon content in the Illustris
  simulation}

\author[J. Zjupa and V. Springel]
{Jolanta Zjupa$^{1,2}$\thanks{E-mail: jolanta.zjupa@h-its.org} and
 Volker Springel$^{1,3}$\vspace*{0.2cm}\\
  $^{1}$Heidelberger Institut f\"ur Theoretische Studien, Schloss-Wolfsbrunnenweg 35, 69118 Heidelberg, Germany\\
  $^{2}$Institut f\"ur Theoretische Physik, Philosophenweg 16, 69120 Heidelberg, Germany\\
  $^{3}$Zentrum f\"ur Astronomie der Universit\"at Heidelberg,
  Astronomisches Recheninstitut, M\"onchhofstr. 12-14, 69120
  Heidelberg, Germany} 
 
\begin{document}

\date{Accepted, Received; in original form}

\pagerange{\pageref{firstpage}--\pageref{lastpage}} \pubyear{2016}

\maketitle

\label{firstpage}

\begin{abstract} 
  The angular momentum properties of virialised dark matter haloes
  have been measured with good statistics in collisionless N-body
  simulations, but an equally accurate analysis of the baryonic spin
  is still missing. We employ the \emph{Illustris} simulation suite,
  one of the first simulations of galaxy formation with full
  hydrodynamics that produces a realistic galaxy population in a
  sizeable volume, to quantify the baryonic spin properties for more
  than $\sim 320,000$ haloes. We first compare the systematic
  differences between different spin parameter and halo definitions,
  and the impact of sample selection criteria on the derived
  properties. We confirm that dark matter only haloes exhibit a close
  to self-similar spin distribution in mass and redshift of lognormal
  form. However, the physics of galaxy formation radically changes the
  baryonic spin distribution. While the dark matter component remains
  largely unaffected, strong trends with mass and redshift appear for
  the spin of diffuse gas and the formed stellar component. With time
  the baryons staying bound to the halo develop a misalignment of
  their spin vector with respect to dark matter, and increase their
  specific angular momentum by a factor of $\sim 1.3$ in the
  non-radiative case and $\sim 1.8$ in the full physics setup at
  $z=0$. We show that this enhancement in baryonic spin can be
  explained by the combined effect of specific angular momentum
  transfer from dark matter onto gas during mergers and from feedback
  expelling low specific angular momentum gas from the halo. Our
  results challenge certain models for spin evolution and underline
  the significant changes induced by baryonic physics in the structure
  of haloes.
\end{abstract}

\section{Introduction} \label{Sec_Intro}
  
The origin of the angular momentum of galaxies is an important
question in cosmic structure formation, as the spin directly
determines the size of rotationally supported objects such as disk
galaxies. In the now well established standard paradigm of the
$\Lambda$CDM concordance cosmology, primordial dark matter density
perturbations seeded in an inflationary epoch grow with time due to
gravitational instability. Eventually, they decouple from the
background expansion, turn around and collapse to form virialised
structures. This happens first for small mass systems, which then
hierarchically merge into bigger structures \citep{Blumenthal84, 
Davis1985}. The baryons collected within haloes cool out and form
galaxies at their centres, giving rise to a hierarchical galaxy
formation process \citep{WhiteRees78}. If the baryons have a
non-vanishing specific angular momentum, it should be preserved in the
radiative cooling process, such that the gas settles into a
rotationally supported disk that forms inside out \citep{Fall80, 
MoMaoWhite98}, with a size directly related to the magnitude of the
spin.

It is thus important to clarify the amount of angular momentum
imparted on haloes and on the baryons they contain. Generally,
gravitationally self-bound structures gain their initial angular
momentum from interactions with the surrounding gravitational tidal
field \citep{Hoyle49}. This in particular makes it possible that
haloes acquire substantial non-vanishing angular momentum even though
the gravitational potential is irrotational in character.  The amount
of angular momentum contained in a galaxy as well as the growth rate
in the linear regime was first calculated by \cite{Peebles69}.
\cite{Doroshkevich70} pointed out that this particular growth rate is
a consequence of the imposed spherical symmetry and carried out a
calculation predicting the angular momentum to grow linearly with time
before non-linear effects start to play a significant role. These
results were confirmed by early collisionless dark matter only N-body
simulations by \cite{White84} and \cite{BarnesEf87}, and form the
basis of the so-called tidal torque theory, which describes the
acquisition of angular momentum for dark matter haloes \citep[see
also][]{Schaefer2012}.

However, once the subsequent evolution of haloes enters the non-linear
regime, simple tidal torque theory breaks down, as haloes cannot be
regarded any more as isolated objects. Instead, they undergo multiple
minor and major mergers. Analytic and semi-analytic models for the
acquisition of angular momentum from the orbital angular momentum of
infalling mergers \citep[e.g.][]{Vitvitska02, Maller02, MallerDekel02}
can successfully extend tidal torque theory and reproduce the spin
parameter distribution of haloes as well as the distribution of
specific angular momentum inside haloes at $z=0$. However, such
analytic descriptions rely on simplifying and ultimately uncertain
assumptions. On the other hand, the acquisition of angular momentum
through non-linear processes such as mergers is followed faithfully in
numerical simulations, making them a particularly powerful approach to
study this problem.

This has motivated numerous analysis of the angular momentum
properties of simulated dark matter only haloes.  \cite{AvilaReese05}
studied the dependence of halo spin on environment, \cite{Maccio07}
investigated the correlation of halo spin with mass and concentration,
and \cite{Maccio08} extended this analysis to the dependence of halo
spin on mass and cosmology. \citet{Bett07} have derived spin
parameters for $\sim 1.5\times 10^6$ dark matter only haloes from the
Millennium simulation \citep{Springel05Nature} and accurately
quantified the spin parameter distribution with the highest
statistical power so far. \cite{Bullock01} extended the analysis to
the distribution of specific angular momentum within haloes and found
a universal angular momentum profile within the virial radius. The
most important result from these studies has been the finding of a
nearly universal spin parameter distribution of approximately
lognormal form. A closer look at some of the results reported in the
literature however also reveals some small quantitative differences,
as we will discuss in detail in this paper.

When it comes to baryonic processes, even more interesting differences
appear. Although the dynamics of the galaxy and the dark matter are to
a large extent determined by the common gravitational potential of the
dark matter halo \citep{Rubin70}, so-called feedback processes play an
important role in shaping galaxy formation and evolution, primarily
through changing the gas dynamics. However, even in absence 
of feedback mechanisms, it is not trivially possible to extrapolate 
from dark matter onto baryonic spin properties. This can be already 
seen in early results from non-radiative hydrodynamical simulations.
In particular, \cite{vdBosch02} showed in their
non-radiative hydrodynamical simulation that even though the `initial'
spin distributions of dark matter and the gas component of haloes are
indistinguishable, there is a substantial misalignment between dark
matter and the gas component at $z=3$, with a significant fraction of
the gas being counter-rotating. The median misalignment angle reported
is $\sim 30^\circ$. \cite{Chen03}, \cite{Sharma05}, and
\cite{Gottloeber07} extended this analysis to $z=0$ and found a
relative enhancement of the gas to dark matter spin parameter of
$1.19$, $1.44$, and $1.39$, respectively. However, until recently the
analysis of the spin was largely restricted to non-radiative
simulations, as full physics simulations of galaxy formation were
simply too costly and did not produce realistic galaxy populations.

The impact of baryons onto dark matter in models where star
formation and feedback is taken into account was first studied by
\cite{Bett10}, who employed a sample of 67 haloes and their dark
matter only counterparts and found an increase of the specific angular
momentum of dark matter in the inner regions of haloes in the presence
of baryons. \cite{Bryan13} looked at larger statistical samples of
$\gtrsim 3000$ haloes taken from the OWLS simulations and confirmed this
finding. 

First results on the angular momentum properties of the stellar
component of galaxies from a realistic galaxy population taken from
the \emph{Illustris} simulation were obtained by \cite{Genel15GAM} who
showed a correlation between galaxy type and specific angular
momentum. This correlation was further confirmed by \cite{Zavala16}
using the {\small EAGLE} simulation.
Furthermore, \cite{Genel15GAM} observed that galactic winds
enhance the spin of stars compared to the dark matter, and that AGN
feedback counteracts this effect by damping this enhancement, as also
indirectly observed by \cite{Bryan13}.
\cite{Baldi16} studied rotational support in clusters and found, based 
on their sample of 258 both relaxed and unrelaxed clusters from the 
{\small MUSIC} simulation, little dependence of the gas spin parameter 
on the implemented baryonic physics.

\cite{Teklu15} and \cite{RodriguezGomez16} further investigated 
the correlation between galaxies and their host haloes. Employing 
a sample of 622 haloes, with no restriction on their dynamical state, 64 of
which host spiral galaxies and 110 ellipticals, \cite{Teklu15} found
that haloes hosting spiral galaxies exhibit on average higher spins,
and haloes hosting elliptical on average lower spins. This trend was 
previously only weakly observed by \cite{Sales12}, who found little 
evidence for galaxy morphology of 100 Milky Way sized galaxies 
form the {\small GIMIC} Simulation to be connected to halo spin and 
merging history, and argued that galaxy morphology is rather determined 
by the misalignment of angular momentum inside the galaxy and its host 
halo at turnaround.
\cite{RodriguezGomez16} showed that the correlation between galaxy
morphology and host halo spin is a strong function of halo mass, as in
more massive haloes mergers play an increasingly important role in
perturbing the gas distribution of the central galaxy. 
\cite{Zavala16} followed the time evolution of the specific angular 
momentum of different halo components and presented different evolution 
scenarios for the baryonic component, and their connection to the 
morphology of the galaxy forming at the halo centre.
 
In this paper we investigate how the dynamics of the baryonic and the
dark matter component of haloes are influenced by feedback processes
for a large statistical sample of $\sim 320,000$ haloes from the
\emph{Illustris} simulation. Our analysis focuses on the systematic
properties of the angular momentum content of whole haloes, and how it
changes relative to dark matter only simulations when baryonic physics
is included.

This paper is structured as follows. We begin in
Section~\ref{Sec_Methods} with a brief description of our simulation
methodology and details on our halo identification and sample
selection. We furthermore discuss in detail the effects of sample
selection criteria and adopted spin parameter definition on the spin
statistics. We then present results for the angular momentum
statistics of dark matter only haloes in Section~\ref{Sec_DM},
followed by results from non-radiative baryonic simulations in
Section~\ref{Sec_NR} and from the full physics \emph{Illustris}
simulation in Section~\ref{Sec_FP}. In each of these sections we also
examine the robustness of our results with respect to resolution, the
redshift evolution of the spin statistics, and the dependence of the
spin properties on halo mass. Section~\ref{Sec_FP} contains a
discussion of the effect of feedback onto the baryonic and dark matter
spin properties and highlights the main mechanisms responsible for a
substantially enhanced spin of the gas component. We give a discussion
and our conclusions in Section~\ref{Sec_Dis}, and summarise extensions
to the {\em Illustris} group catalogue in an Appendix.

\section{Methodology} \label{Sec_Methods}

\subsection{The Illustris simulation suite} \label{Sec_SimCode}

The \emph{Illustris} simulation suite consists of a set of
cosmological hydrodynamical simulations of a $75\,h^{-1}{\rm Mpc}$
wide periodic cosmological box carried out with the moving mesh code
{\small AREPO} \citep{Springel2010Arepo}. Initial Conditions were
generated at $z = 127$ and evolved to $z = 0$ with $1820^3$ dark
matter particles and $1820^3$ initial gas cells in the highest
resolution run, achieving a mass resolution of
$6.26 \times 10^6 {\rm M}_{\odot}$ in dark matter and
$1.26 \times 10^6 {\rm M}_{\odot}$ in baryonic matter. To investigate
numerical convergence, runs with a reduced number of initial dark
matter particles and gas cells were performed as well. Furthermore,
for every resolution the simulations were carried out with three
different physical setups, a dark matter only, a non-radiative, and a
full galaxy formation physics setup. An overview of the different
\emph{Illustris} simulations and their principal parameters is given
in Tab.~\ref{TabSims}.

In the dark matter only simulations, all mass is treated as
collisionless dark matter, while the non-radiative setup follows in
addition the hydrodynamics of the gas but ignores radiative cooling
and star formation. The full physics simulation includes these and
further processes related to galaxy formation through a model
described in full in \citet{Vogelsberger13}. In brief, unresolved
physics of the interstellar medium is modelled in a subgrid fashion,
where star formation is regulated by a pressure model that accounts
for supernova feedback in the interstellar medium. This model also 
includes chemical enrichment through explicit tracking of 9 elements.
Furthermore, black hole growth through gas accretion and associated 
energy feedback processes are included. The \emph{Illustris} simulation 
is one of the first cosmological hydrodynamical simulations of galaxy formation 
that produces a realistic population of galaxies at $z=0$. Other recent
projects that show similar successes are the {\small EAGLE} simulation
\citep{Schaye15eagle} and the Horizon-{\small AGN} simulation \citep{Dubois16}.

The galaxy formation physics model of \emph{Illustris} simultaneously
reproduces with reasonable accuracy a number of observed small-scale
properties, such as galaxy stellar masses and morphologies, as well as
large-scale properties, such as the metal abundance in neutral
hydrogen absorption systems, or the radial distribution of galaxies in
galaxy clusters \citep[e.g.][]{Vogelsberger14Nature}. It is thus
interesting to examine the spin distribution of the baryons in such a
calculation, which can be viewed as representing one of the most
realistic predictions for the large-scale dynamics of the baryons
available thus far. This motivation is further strengthened by the use
of the {\small AREPO} code for the simulation, which follows the gas
mass in a quasi-Lagrangian form by means of a fully adaptive mesh that
moves with the flow. This approach avoids classical disadvantages of
Cartesian adaptive mesh refinement (AMR) codes, such as the occurrence
of preferred spin directions along the coordinate axes
\citep[e.g.][]{Hahn10}. Simultaneously, it eliminates traditional
problems of smoothed particle hydrodynamics (SPH), such as relatively
high numerical noise and the need for an artificial viscosity
\citep{Springel2010SPH}. We thus expect that {\small AREPO} follows
the hydrodynamics more accurately than competing numerical approaches,
making its predictions for the spin of the baryonic component of
haloes all the more interesting.

However, we note that like any other numerical code, {\small AREPO} is
constantly improved further. In particular, \cite{Pakmor16} has
recently proposed changes in the gradient estimation as well as time
integration of the code that improve its accuracy and convergence
order in certain situations. The \emph{Illustris} suite analysed here
was carried out with a version of the code that did not include these
improvements, but as \cite{Pakmor16} show, they do not affect the
results of cosmological simulations of galaxy formation.

\begin{table}\centering
\begin{tabular}{llrr}
\hline
\textbf{Simulation}& \phantom{1} \textbf{Simulation type} & \textbf{dm} & \textbf{gas}\\ 
\hline
Illustris-1& \phantom{1} full physics hydro.& $1820^3$ & $1820^3$\\
Illustris-2& \phantom{1} full physics hydro.& $910^3$ & $910^3$\\
Illustris-3& \phantom{1} full physics hydro.& $455^3$ & $455^3$\\
\hline
Illustris-2-NR& \phantom{1} non-radiative hydro.& $910^3$ & $910^3$\\
Illustris-3-NR& \phantom{1} non-radiative hydro.& $455^3$ & $455^3$\\
\hline
Illustris-1-Dark& \phantom{1} collisionless dm only& $1820^3$ & -\\
Illustris-2-Dark& \phantom{1} collisionless dm only& $910^3$ & -\\
Illustris-3-Dark& \phantom{1} collisionless dm only& $455^3$& -\\
\hline
\end{tabular}
\caption{\emph{Illustris} simulation suite: listed are the symbolic 
  name, the physics included, the number of initial dark matter
  particles, and the number of initial gas cells for every simulation. \label{TabSims}}
\end{table}

\subsection{Measurement of halo properties through an extension of SUBFIND} \label{Sec_SubExtension}

During the \emph{Illustris} simulation runs, the group finders {\small
  FOF} and {\small SUBFIND} \citep{Springel01} were applied on the
fly, determining a set of basic halo and subhalo properties, as
described in detail in the public data release of \emph{Illustris}
\citep{Nelson15}.  However, the spin properties we want to analyse
here, as well as information about the binding energy of haloes were
not part of these properties. Unfortunately, simply computing
additional halo properties post-hoc is technically complicated. While
the \emph{Illustris} data is stored such that the particle/cell data
comprising individual gravitationally bound subhaloes can be retrieved
relatively easily despite the large simulation size, this is not
readily possible for the particles/cells making up an object defined
by its spherical overdensity radius $R_{200}$. Also, computing the
binding energy of haloes that consist of a large number or resolution
elements (regularly in excess of $10^6$ elements) becomes
computationally costly unless sophisticated algorithms are employed.

In order to efficiently measure further halo and subhalo properties
for \emph{Illustris}, we have therefore developed an extension of
{\small AREPO}'s group finders that allows the parallel processing of
an already existing group catalogue. The membership of individual
resolution elements to groups, subhaloes and spherical overdensity
haloes is kept exactly as in the existing group catalogue, allowing
additional properties of haloes to be measured. The results are then
simply added as further fields to the catalogue. A full list of the
newly available halo and subhalo properties in the extended group
catalogue can be found in the Appendix\footnote{They will be added to
  the public data release of \emph{Illustris} described by
  \citet{Nelson15} upon publication of this paper.}. Thanks to the
parallel tree solver for gravity in {\small AREPO}, one of the
quantities we can calculate in this way efficiently is the exact
gravitational binding energy of haloes (including spherical
overdensity objects), something that has often been only determined in
an approximate way in previous analysis of halo spin. The code
extension of {\small FOF/SUBFIND} is written such that it can run both
as a postprocessing option to augment existing catalogues, or as part
of the regular group finding, either on-the-fly or in postprocessing.
We also note that since the group catalogue is stored in the
convenient HDF5 format, the I/O routines of existing analysis code
using the group catalogue does not have to be adjusted or changed
after the group catalogue has been extended.

For definiteness, we briefly summarise the group definitions adopted
by \emph{Illustris}, which we also employ in the
following. Friends-of-friends (FOF) groups are determined for dark
matter particles as a set of equivalence classes, where any pair of two
particles is in the same group if their distance is smaller than a
prescribed linking length. For the linking length we adopt the
standard value of $0.2$ times the mean particle spacing,
$l_{\rm mean} = (m_{\rm dm} / \rho_{\rm dm})^{1/3} $, where
$m_{\rm dm}$ is the dark matter particle mass, and $\rho_{\rm dm}$ is
the mean dark matter density. Baryonic particles (stars and black
holes) and gaseous cells, if present, are then assigned in a second
step to the same halo as their closest dark matter particle.  Any
group constructed in this way corresponds to what we from now on call
a FOF-halo. While all groups with at least 32 particles/cells are
stored for the \emph{Illustris} simulation, for our analysis we will
typically impose a considerably higher minimum particle number in
order to prevent numerical noise and possible biases from poorly
resolved haloes.

Each FOF-halo is then decomposed by {\small SUBFIND} into a set of
gravitationally self-bound subhaloes, based on the algorithm described
in \citet{Springel01}. To this end, the total mass density at each
point is estimated by an adaptive kernel estimation. The resulting
density field is then processed with an excursion set technique that
finds locally overdense candidate substructures. Each of these
overdensities is then subjected to a gravitational unbinding
procedure, keeping only the bound part as a genuine
substructure. Every resolution element can only be member of one
subhalo, and the remaining bound part of the halo, after all smaller
substructures have been removed, is called the background subhalo.
For each subhalo the particle/cell with the smallest gravitational
potential is adopted as its centre.

\begin{figure}
\centering
\includegraphics[width=0.49\textwidth,trim= 0 0 0 0,clip]{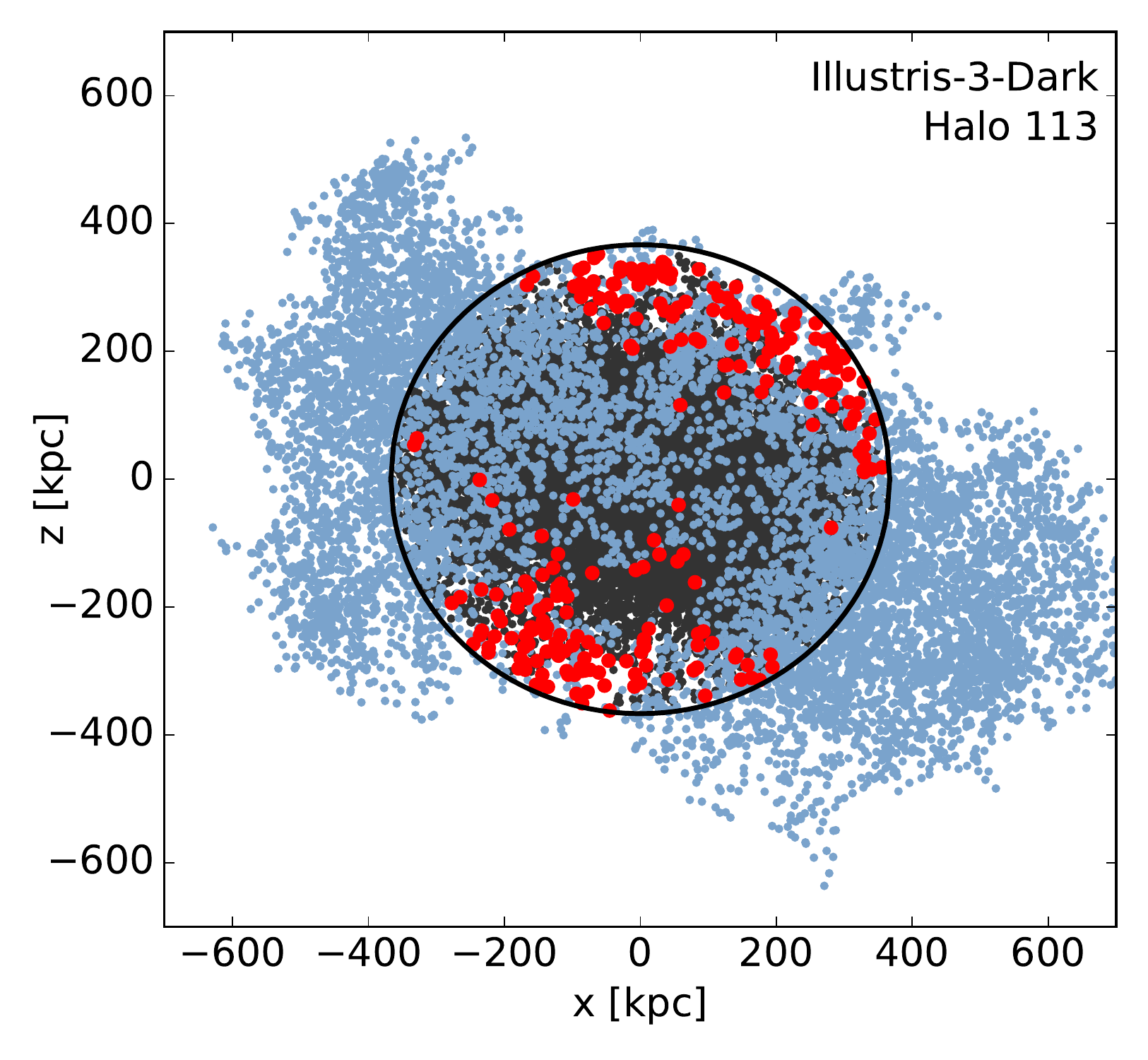}
\caption{Particles making up a randomly selected dark matter halo
  from Illustris-3-Dark. The black circle indicates
  $R_{200}$, FOF-halo dark matter particles inside $R_{200}$ are shown
  in $grey$, FOF-halo dark matter particles outside $R_{200}$ are
  shown in $blue$, dark matter particles part of the SO-halo
  definition but not the FOF-halo are shown as $big$ $red$ dots. This
  demonstrates the general fact that FOF-haloes are typically more
  extended and more massive than their SO-halo counterparts. 
  The number of particles that are part of the SO-halo
  but not the FOF-halo is usually
  relatively small, such that SO-haloes can be regarded as the 
  inner regions of FOF-haloes.
  \label{Figfofso}}
\end{figure}

Finally, around the centre of each background subhalo, which
corresponds to the point with the minimum gravitational potential of
the underlying FOF-halo, we determine spherical overdensity (SO)
groups. In this approach, one finds a spherical region around a given
point {\em in the full particle/cell set} that encloses a certain
overdensity with respect to the background density. We will generally
employ an overdensity of 200 with respect to the critical density, and
denote the corresponding radius and mass of the spherical region as
$R_{200}$ and $M_{200}$, respectively. In Fig.~\ref{Figfofso} we
depict a typical halo to illustrate the different halo definitions and
highlight some of the implications for halo geometry and the
calculation of halo properties.

\subsection{Halo sample selection} \label{Sec_SampSel}

For our analysis, we select all sufficiently well-resolved haloes that
appear to be reasonably relaxed systems. To avoid numerical biases due
to resolution effects, following the detailed resolution study of
\cite{Bett07}, we exclude all haloes resolved by less than 300 dark
matter particles from our FOF-halo and SO-halo samples.\footnote{ Note
  that this imposes a somewhat higher minimum halo mass in the
  \emph{Illustris} simulations including baryons compared to dark
  matter only.} We also exclude all haloes that have no
gravitationally self-bound component identified by {\small SUBFIND},
as these are not haloes in a physical sense.

Furthermore, we want to study only haloes close to virial equilibrium,
in order to avoid, for example, situations where the angular momentum
content is dominated by the contribution of the orbital angular
momentum of mergers. To this end we calculate the virial ratio,
\begin{equation} \label{q} 
q = \frac{2\,E_{\rm kin}}{E_{\rm pot}} + 1,
\end{equation}
both for our FOF- and SO-halo samples. The total kinetic energy in
Eq.~(\ref{q}) represents the sum of the bulk kinetic energy of the
halo particles/cells and the thermal energy of the gas, if present. In
calculating the virial ratio according to Eq.~(\ref{q}) we ignore the
surface pressure term, which appears in the virial theorem for
non-isolated systems \citep{Shapiro04} and represents the pressure
exerted on the halo by infalling matter. In our approach it is
sufficient to calculate the virial ratio in such an approximative way,
as we only employ it to filter out systems that have a huge surplus of
kinetic energy.

\begin{figure} \label{fig:1}
\centering
\includegraphics[width=0.5\textwidth,trim= 0 0 0 0,clip]{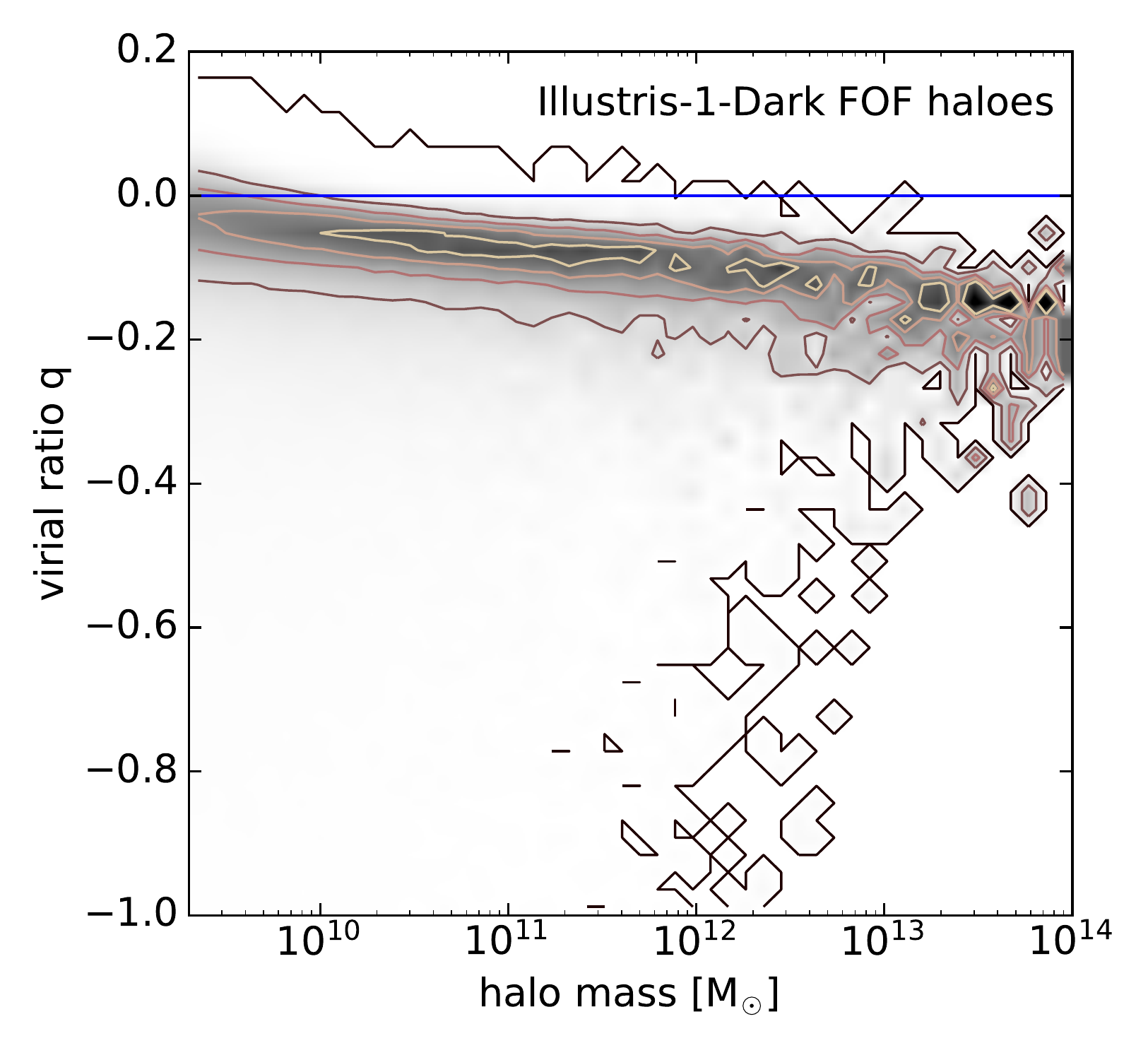}
\includegraphics[width=0.5\textwidth,trim= 0 0 0 0,clip]{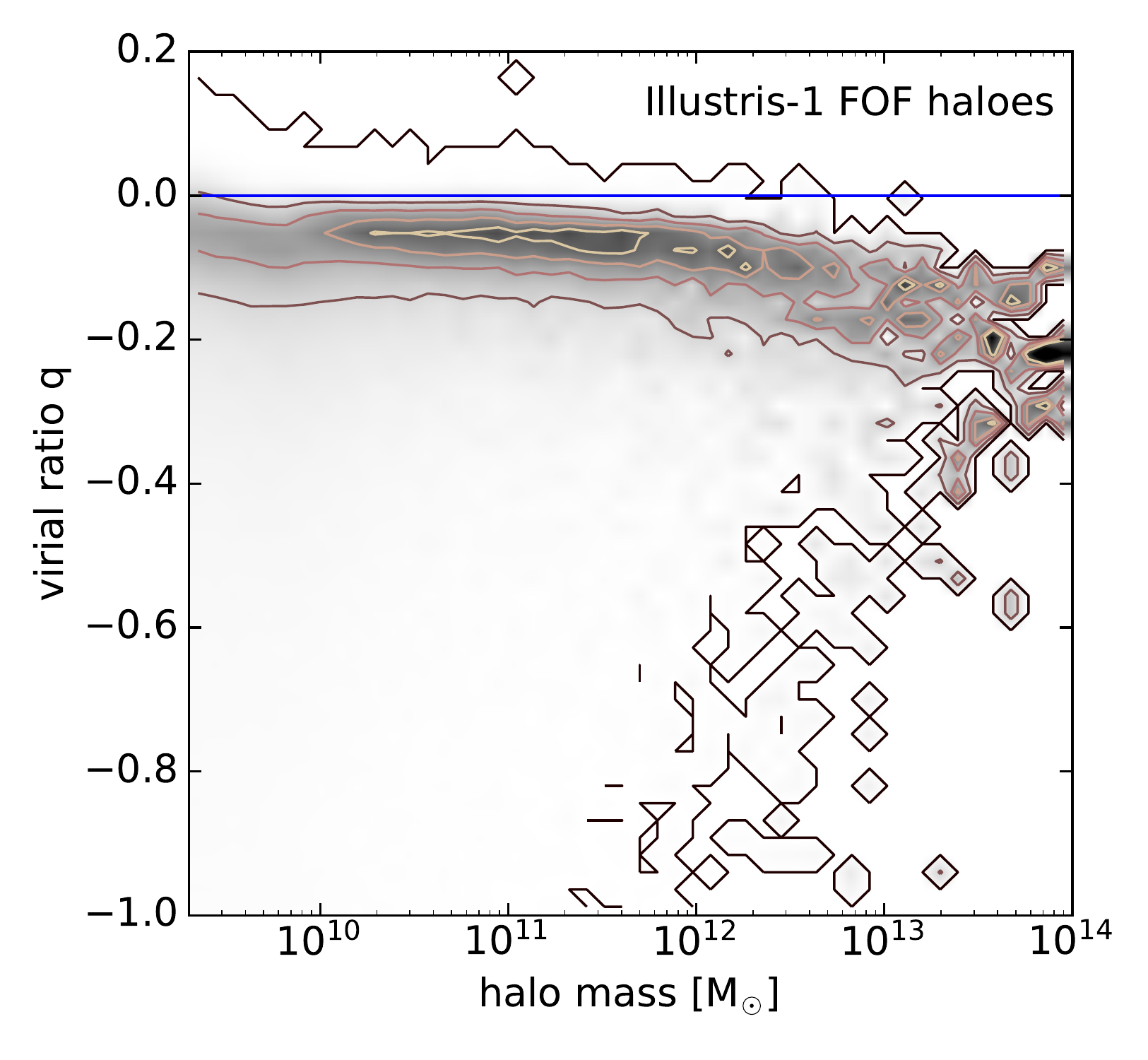}
\caption{Distribution of the virial ratio $q$ of the FOF-halo sample
  from Illustris-1-Dark and Illustris-1 normalised in every mass
  bin. The grey shading ranges from a fraction of $0$ to $0.2$ of all
  haloes in a mass bin having a given $q$-value. Contours are drawn at
  constant fractions of $0$, $0.05$, $0.1$, $0.15$, and $0.2$,
  respectively. The blue line denotes virial equilibrium. Negative
  virial ratios correspond to haloes being dominated by kinetic
  energy. In case of massive haloes this surplus of kinetic energy is
  due to the halo not being fully collapsed yet. The tail to extreme
  negative values for less massive haloes is due to
  mergers. \label{FigVirialRatioMass}}
\end{figure}

For an isolated structure in equilibrium the expected virial ratio
based on the virial theorem is $q = 0$. In
Fig.~\ref{FigVirialRatioMass} we show the distribution of virial
ratios of our FOF-halo samples from Illustris-1-Dark and from
Illustris-1 against halo mass. In this effectively two-dimensional
histogram, the individual mass bins have been independently normalised
to take out the variation of halo abundance with mass. The virial
equilibrium expectation is denoted by the horizontal blue line. Haloes
with positive $q$-values are dominated by potential energy, haloes
with negative $q$-values are dominated by kinetic energy. We dismiss
all haloes with $q < -1$ from our halo sample, corresponding to haloes
having more than twice as much kinetic energy as expected from virial
equilibrium. Those objects are often undergoing significant mergers,
which can produce highly negative $q$-values.

Both simulations show that the average virial ratio of haloes
decreases with halo mass. This reflects the fact that less massive
haloes form on average at higher redshifts and thus have more time to
virialise by $z = 0$, whereas more massive haloes are still in the
process of collapsing and virialising at $z = 0$, which is
additionally slowed down by further growth through accretion. More
massive haloes are also known to be more elongated on average,
consistent with their younger age and more active infall region.
 
\begin{figure}
\centering
\includegraphics[width=0.49\textwidth,trim= 10 0 0 0,clip]{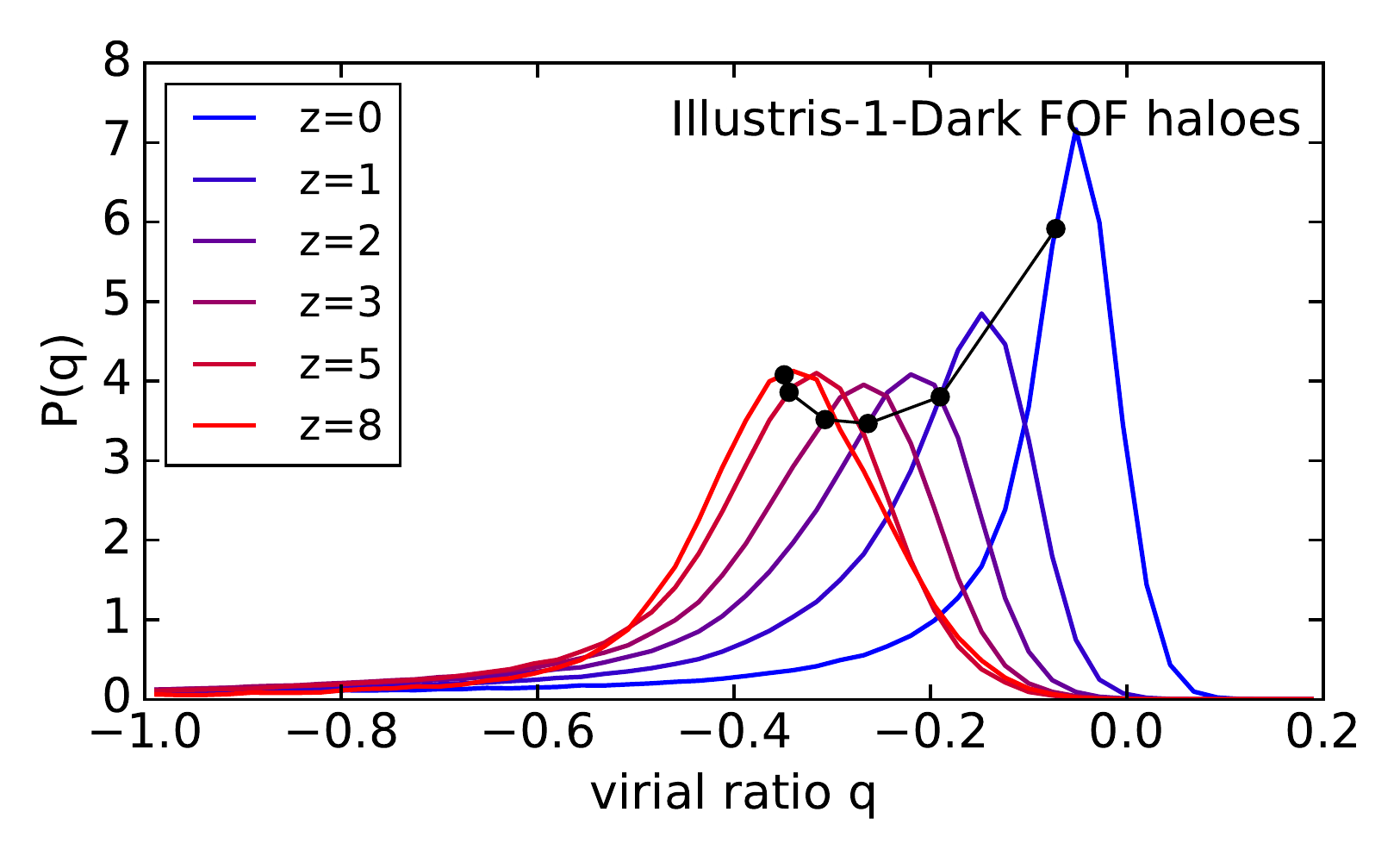}
\includegraphics[width=0.49\textwidth,trim= 10 0 0 0,clip]{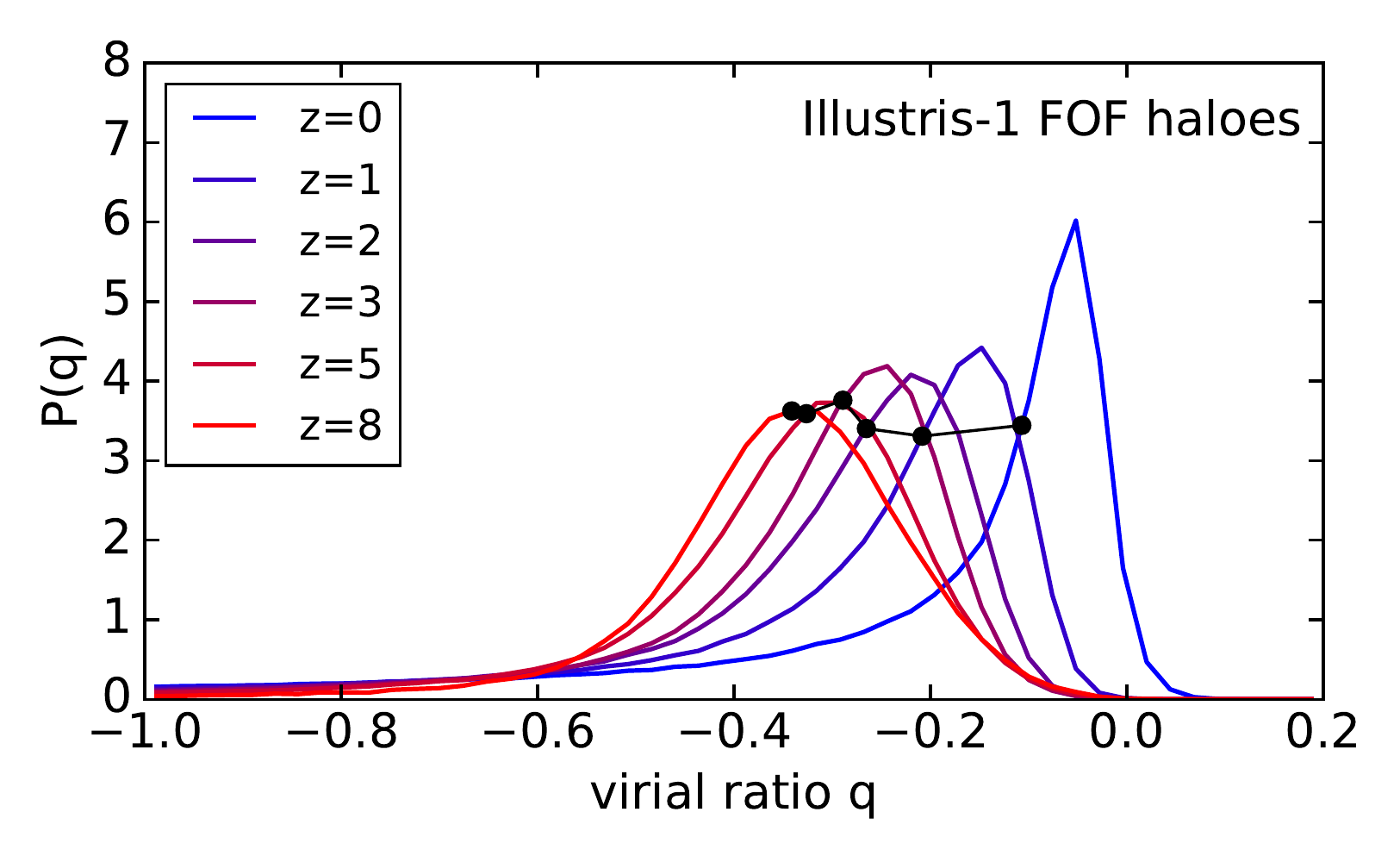}
\caption{Change of the virial ratio distribution of FOF-haloes as a
  function of redshift, both for Illustris-1-Dark (top panel) and
  Illustris-1 (bottom panel). Black dots indicate the median virial
  ratio at every redshift. With decreasing redshift, the FOF-halo
  sample becomes ever more relaxed, with the median virial ratio
  shifting continuously towards zero and hence to the expectation
  value for perfectly relaxed isolated
  systems. \label{FigVirialRatioZ}}
\end{figure}

In Fig.~\ref{FigVirialRatioZ} we show how the virial ratio
distributions of the FOF-halo samples of Illustris-1-Dark and
Illustris-1 evolve with time. As there are many more low mass haloes
than high mass haloes, the distributions are dominated by the mass
scale just above the enforced threshold of 300 dark matter particles.
Black dots indicate the median virial ratio at the displayed
redshifts. For both simulations the contribution of kinetic energy is
enhanced at high redshift, pushing the bulk of the haloes further away
from the equilibrium value, reflecting their ongoing rapid growth and
young age. The virial ratio distribution $P(q)$ normalised to the
total number of haloes as well as bin size thus peaks at low
$q$-values at high redshift and then progressively shifts towards
$q=0$ with decreasing redshift, corresponding to the halo sample
becoming more virialised with cosmic time. Our SO-halo sample shows
qualitatively the same behaviour. This reflects the fact that cosmic
structures are hardly ever in perfect virial equilibrium, but rather
in a slowly evolving quasi-equilibrium. 
However, we caution that the trends displayed here were derived neglecting the pressure surface term, which is smaller both for more massive and younger haloes, as both are undergoing more accretion. A higher absolute value of the negative pressure surface term results in the trends being weaker than shown here. Nevertheless, when comparing halo samples at different redshifts, one thus has to bear in mind that the samples will typically exhibit different degrees of relaxation.

Furthermore, we want to emphasise that there is no universally
accepted standard definition of what constitutes a well-resolved,
quasi-equilibrium structure in cosmological simulations. Besides a
different minimum number of particles and different virial ratio cuts,
also criteria such as the offset between the centre of mass and the
potential minimum, the abundance of dark matter substructures
\citep{Maccio07}, or the mass fraction in substructures not being
bound to the main potential \citep{Neto07} have been employed in the
literature to discriminate between relaxed and unrelaxed haloes. The
effect of the different sample selection criteria on the final values
of the derived properties, such as the spin parameter distribution and
median spin parameter value for dark matter only haloes, has not been
systematically investigated yet.

However, as we will show later, the inferred spin distribution of
haloes is equally sensitive to the halo and spin parameter definitions
as on the exact set of selection criteria for the halo sample. Which
halo definition is considered to be more physical or useful is largely
a matter of convention. In this paper, we will present results for
both the FOF-halo and SO-halo sample, which can be regarded as the two
most important limiting cases. The systematic differences arising from
the spin parameter definition are investigated in the next section.

\subsection{Spin parameter definitions} \label{Sec_SpinDefs}

The angular momentum content and degree of rotational support of
structures with different mass and spatial extent can be quantified
and compared by means of a suitably defined dimensionless spin
parameter $\lambda$. Its classic definition,
\begin{equation} \label{eq:1} 
\lambda_{\rm P} = j_{\rm sp} \frac{E_{\rm tot}^{1/2}}{G M_{\rm tot}^{3/2}},
\end{equation}
goes back to \cite{Peebles69}. Here $j_{\rm sp} = J / M$ gives the
magnitude of the specific angular momentum per unit mass of the
material in question. It is multiplied by a factor composed of the
total mass of the system $M_{\rm tot}$, the gravitational constant
$G$, and the absolute value of the total energy,
$E_{\rm tot} = |E_{\rm kin} + E_{\rm pot}|$, where the kinetic energy
is again the sum of the bulk kinetic energy of the halo
particles/cells and the thermal energy of the gas.  This
multiplicative factor expresses the specific angular momentum
$j_{\rm sp} = |\vec j_{\rm sp}|$ in dimensionless form.
 
Note that the specific angular momentum can be calculated for any
subset of the system. If the set is composed of $N$ mass elements,
with the $i$-th computational element having mass $m_i$, distance from
the halo centre $\vec{r}_i$, and velocity $\vec{v}_i$ with respect to
the centre of mass velocity of the halo, the specific angular momentum
is given as
\begin{equation}
  \vec j_{\rm sp} = \frac{\vec J}{M} = \frac{1}{M} { \sum_{i=1}^N m_i \vec r_i \times \vec v_i },
\end{equation}
with mass $M = \sum_i^N m_i$ of the subset. In a simulation including
baryons, the most interesting subsets include the dark matter, the gas,
and the stellar component of the halo. If the subset consists of the
whole halo, one arrives at $M = M_{\rm tot}$, and the definition of
the spin parameter corresponds to that of \cite{Peebles69}.

Early analytic studies of structure formation of dissipationless
haloes in the Einstein-de Sitter limit were able to estimate the
average value of the spin parameter imparted on density perturbations
by gravitational tidal torques.  Applying different approaches and
simplifying assumptions \citet{Heavens88} found a value of
$\lambda_{\rm P} \approx 0.05$, \citet{Ryden88} a value of
$\lambda_{\rm P} \approx 0.09$, and \citet{Steinmetz95} arrived at
$\lambda_{\rm P} \approx 0.07$.  Early dark matter only simulations of
the Einstein-de Sitter universe measured systematically lower Peebles
spin parameters than suggested by some of these analytic studies.
\citet{BarnesEf87} and \cite{Warren92} quote a value of
$\lambda_{\rm P} \approx 0.05$, and \citet{Cole96} found
$\lambda_{\rm P} \approx 0.04$ in their simulations.

The Peebles spin parameter definition has however an important
practical drawback. The potential binding energy $E_{\rm pot}$ of a
self-bound structure is needed to determine its total energy. This can
be computationally expensive to measure accurately for an N-body halo,
especially if its particle number is large. To avoid this
complication, \cite{Bullock01} proposed an alternative definition of
the spin parameter,
\begin{equation} \label{eq:2}
\lambda_{\rm B} = \frac{j_{\rm sp}}{\sqrt{2}R_{200} v_{200}}.
\end{equation}
In this definition, $R_{200}$ and $v_{200}$ are the virial radius and
the circular velocity at the virial radius of the halo, which can be
viewed as characteristic length and velocity scales of the object in
question, and are here used to express the specific angular momentum in
dimensionless form. The prefactor of $1/\sqrt{2}$ is introduced to
make this definition of the spin parameter yield the same value as the
definition of \cite{Peebles69} for the density distribution of a
singular isothermal sphere truncated at $R_{200}$, and where all
particles are put on circular orbits. \citet{MoMaoWhite98} furthermore
argue that the two spin parameter definitions are related by
$\lambda_{\rm P} = f(c)^{1/2} \lambda_{\rm B}$ for NFW-haloes, where
$f(c)$ is a function depending only on the concentration $c$ of the
halo.

\begin{figure}
\centering
\includegraphics[width=0.5\textwidth,trim= 10 0 0 0,clip]{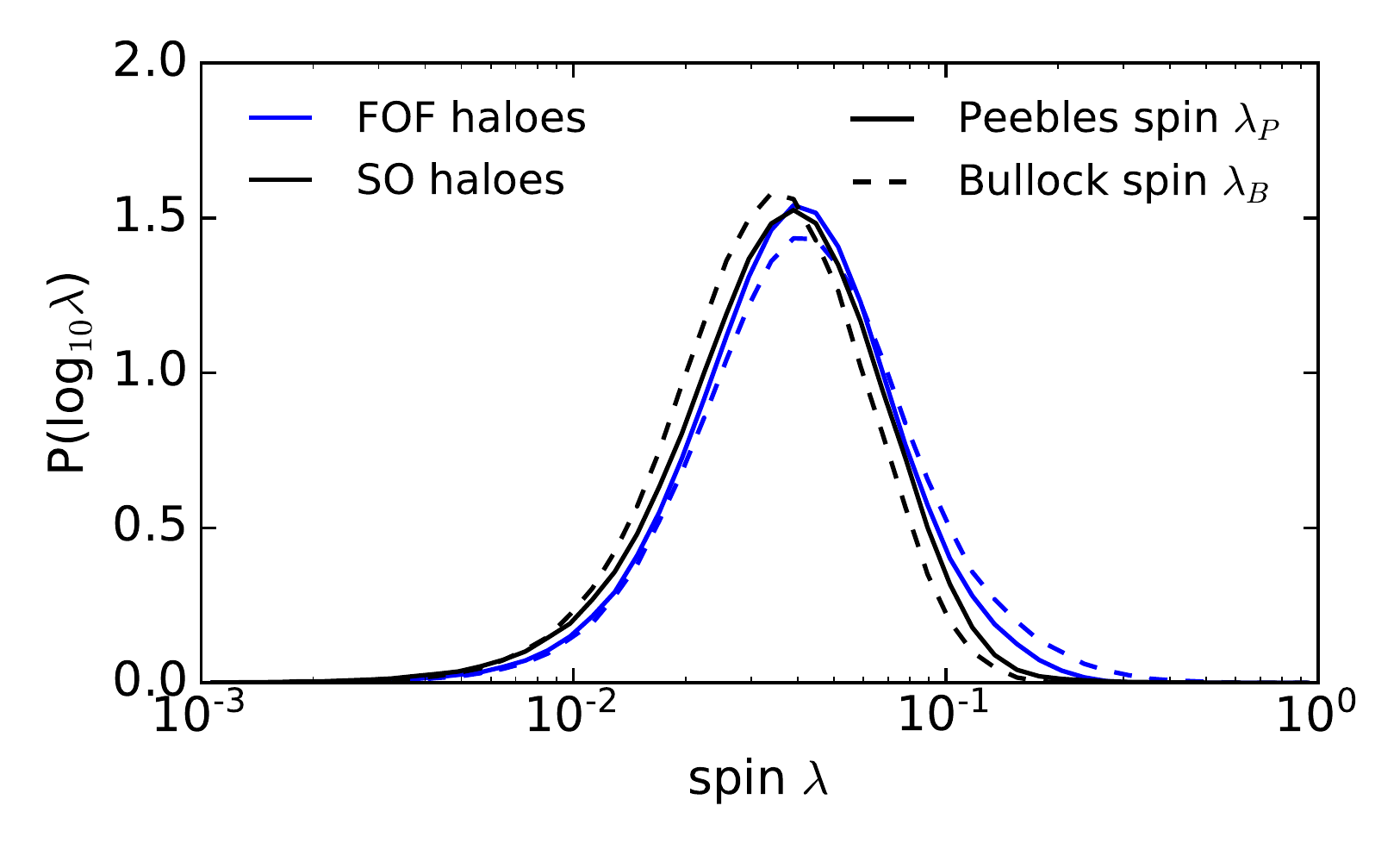}
\includegraphics[width=0.5\textwidth,trim= 10 0 0 0,clip]{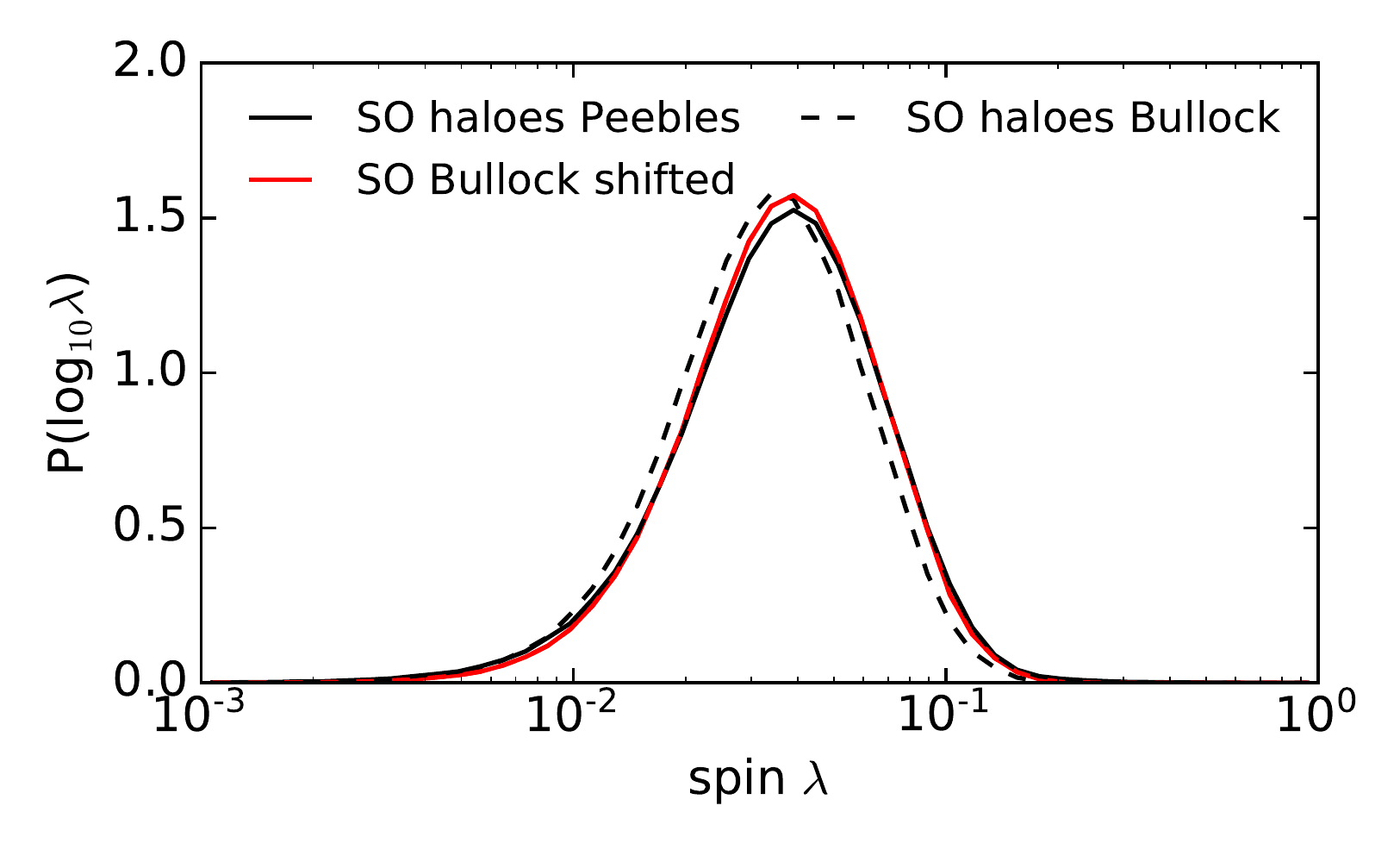}
\includegraphics[width=0.5\textwidth,trim= 10 0 0 0,clip]{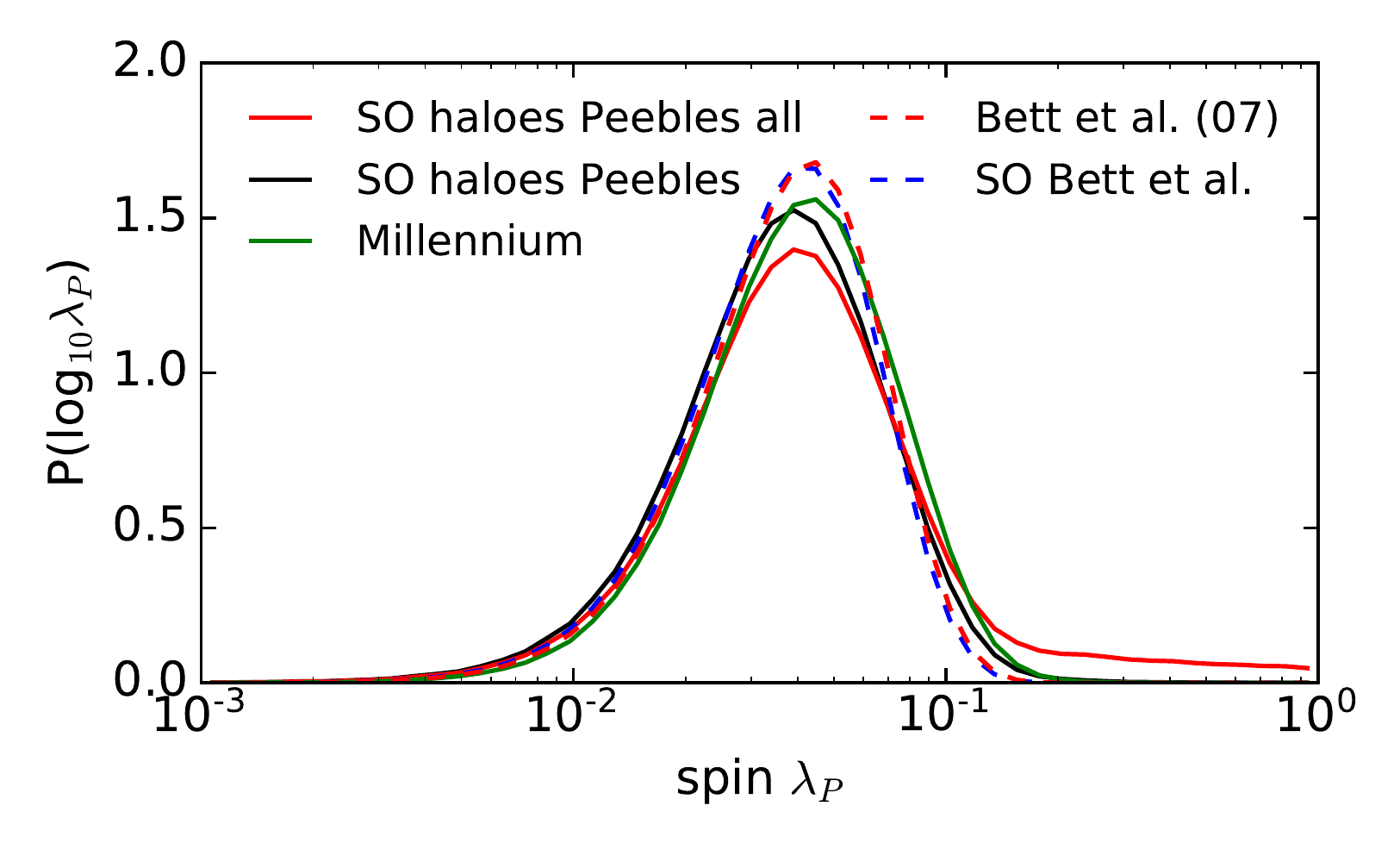}
\caption{{\em Upper panel}: Comparison of the Peebles and Bullock spin
  parameter distributions derived from Illustris-1-Dark for dark
  matter only FOF- and SO-haloes. The Peebles spin definition yields
  consistent results for the different halo definitions.  {\em Middle
    panel}: Comparing the spin values derived with the Bullock spin
  definition for SO-haloes to the Peebles values, ignoring the
  concentration dependence but taking into account a constant
  conversion factor of $1.1$.  {\em Lower panel}: Comparison of our
  SO-halo Peebles distribution from Illustris-1-Dark (black solid) and
  the Millennium Simulation (green solid) to the Peebles distribution
  from \citet{Bett07} derived for their cleaned halo sample (red
  dashed) and SO-halo sample (blue dashed). The red solid line shows
  the resulting spin distribution from Illustris-1-Dark when all
  restrictions on the dynamic state of haloes are abandoned. Different
  sample selection criteria thus explain the observed differences in
  the spin distributions and their effect has to be borne in mind when
  aiming for precision measurements.  \label{FigSpinDefs} }
\end{figure}
 
The Bullock definition has been popular and is widely employed in the
literature, as it is easy to calculate and does not require the
knowledge of the potential energy of a halo, which is often estimated
only approximately to reduce the computational cost. \cite{Maccio07}
quote a median Bullock spin of $\lambda_{\rm B} \approx 0.030$ for
their relaxed halo sample, and \cite{AvilaReese05} report a value of
$\lambda_{\rm B} \approx 0.033$. The values found for the Bullock spin
parameter are all systematically lower than spin parameters derived
with the Peebles spin parameter definition.  This is in agreement with
\cite{Bullock01} who calculated the median spin parameter according to
their new definition and the classic definition by \cite{Peebles69}
for the same SO-halo sample and found values of
$\lambda_{\rm B} \approx 0.035$ and $\lambda_{\rm P} \approx 0.042$,
respectively. Thus, although $f(c)$ is close to unity and the Peebles
and Bullock spin parameters are `approximately' equal, there are
clearly systematic differences that have to be taken into account when
performing precision measurements.

To highlight these systematic differences between the Peebles and
Bullock spin parameter definitions, we calculate the spin according to
both definitions for our FOF- and SO-halo samples. The total
energy needed for the Peebles spin definition is given by
$E_{\rm tot} = |E_{\rm kin} + E_{\rm pot}|$, where we calculate the
potential energy with high precision employing the parallel gravity
tree solver, as described in Section~\ref{Sec_SubExtension}, and the
kinetic energy\footnote{Note that in our extended group catalogue
  $E_{\rm kin}$ contains only the kinetic energy due to the particle
  and cell velocities. The thermal energy of the gas is stored in
  $E_{\rm thr}$ and has to be added to $E_{\rm kin}$ to obtain the
  total physical kinetic energy.} is again the sum of the bulk kinetic
energy of the halo particles/cells and the thermal energy of the gas.
To calculate the Bullock spin parameter we express the virial velocity
$v_{200} = 10 H(z) R_{200}$ via the virial radius $R_{200}$, which is
determined for SO-haloes by {\small SUBFIND} and contained in the
group catalogue. In the case of FOF-haloes, we use the FOF-halo mass
$M_{\rm FOF}$ to estimate an equivalent virial radius
$R_{\rm 200,FOF} = (GM_{\rm FOF}/100\,H^2)^{1/3}$, which corresponds
to the assumption that the FOF-halo has all of its mass contained
within a sphere of overdensity 200 relative to the critical density.
This is a very crude estimate and highlights the limitation of the
Bullock spin parameter definition being designed for spherically
symmetric haloes, because only then the values measured operationally
for the characteristic radius $R_{200}$ and velocity $v_{200}$ can be
expected to make sense. The Bullock spin parameter definition has no
built in mechanism to account for the complicated geometry of
FOF-haloes.

In the upper panel of Fig.~\ref{FigSpinDefs}, we show the spin
parameter distributions derived with the Peebles and Bullock spin
parameter definitions applied to our FOF- and SO-halo samples, which
comprise $\sim 400,000$ and $\sim 360,000$ objects, respectively. The
distributions are derived by binning the haloes in the given spin
parameter range in 50 equidistant logarithmic bins and normalising the
number of haloes per bin by the total number of haloes and the
logarithmic bin size, such that
$\int_{-\infty}^{\infty}P({\rm log}_{10}\lambda) \ {\rm
  dlog}_{10}\lambda = 1$.
With this normalisation the shown distribution is independent of the
chosen bin size and the total number of bins.

\begin{figure}
\centering
\includegraphics[width=0.5\textwidth,trim= 10 0 0 0,clip]{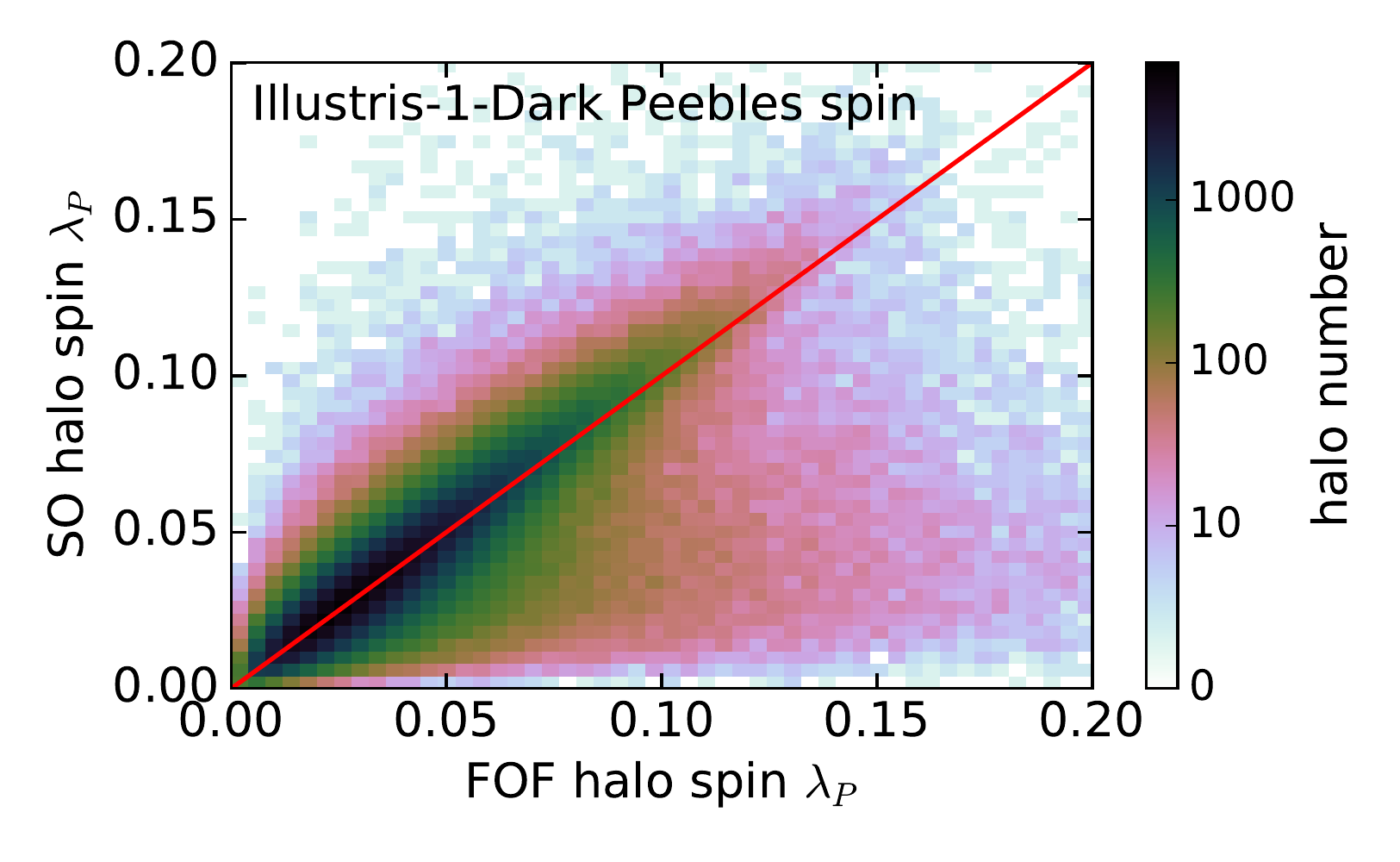}
\includegraphics[width=0.5\textwidth,trim= 10 0 0 0,clip]{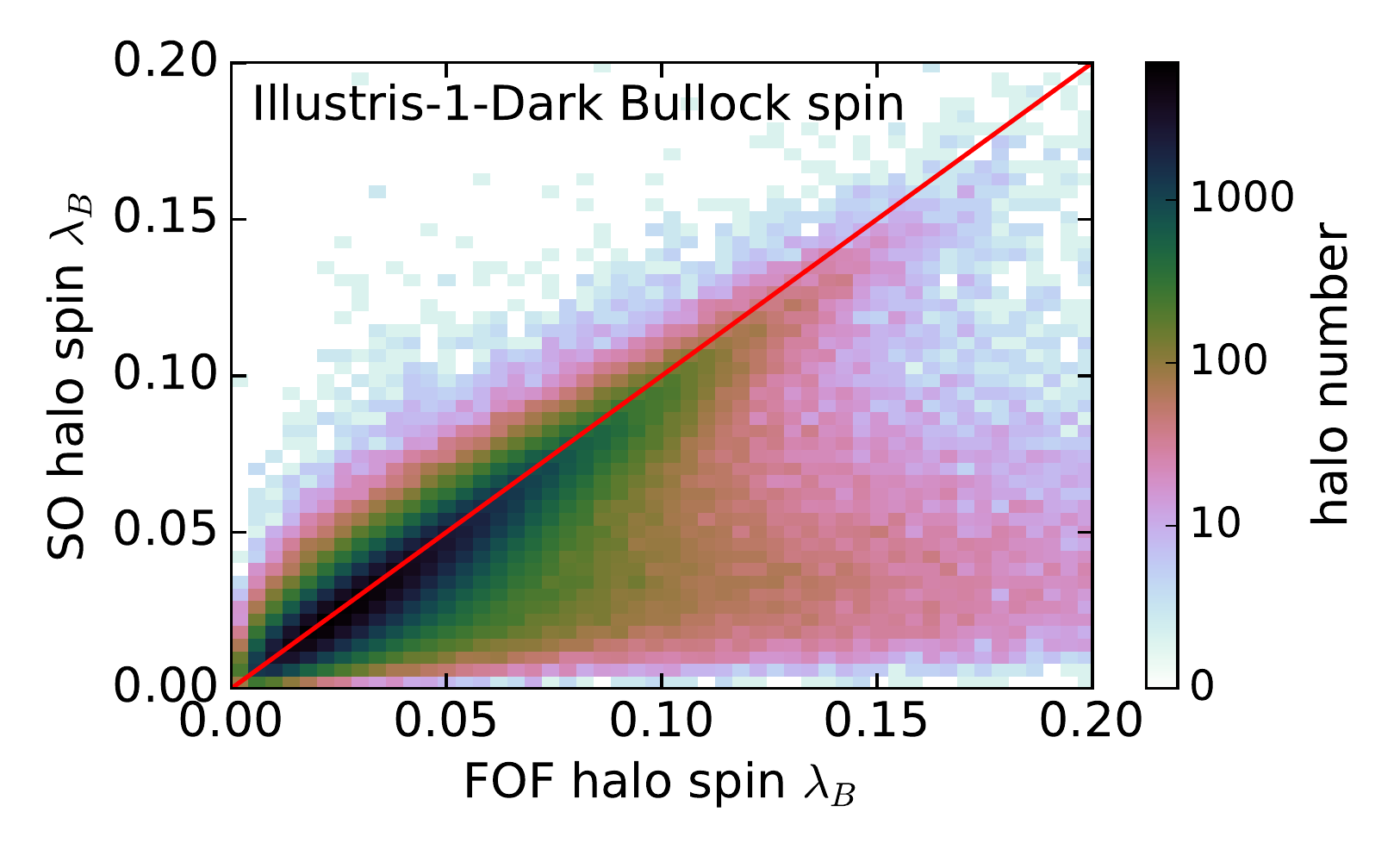}
\caption{Distribution of the Peebles (upper panel) and Bullock (lower
  panel) spin parameter obtained by applying the given spin definition
  to the FOF- and SO-counterparts of the same halo. Most haloes
  scatter around the red identity line. However, a fraction of the
  haloes, that is larger in case the Bullock spin is applied, exhibit
  a significant enhancement in spin for the FOF definition compared to
  SO. Those haloes are significantly extended beyond $R_{200}$ in
  their FOF-halo definition (compare to Fig.~\ref{Figfofso}), causing
  inaccuracies when the Bullock spin definition is
  used. \label{FigSpinDefsComp}}
\end{figure}

The Peebles spin parameter definition yields almost the same
distribution for FOF- and SO-haloes with median spin parameters of
$\lambda_{\rm P, FOF} = 0.0391$ and $\lambda_{\rm P, SO} = 0.0365$ for
the FOF- and SO-halo samples, respectively. This confirms that the
Peebles spin definition produces results that are quite robust with
respect to different halo geometries (see also
Fig.~\ref{FigSpinDefsComp}). The Bullock definition on the other hand
is more sensitive to the halo definition and the complicated geometry
of FOF-haloes. As a result, it yields systematically different spin
distributions for the two halo definitions. Also the median Bullock
spin parameter of the FOF-halo sample,
$\lambda_{\rm B, FOF} = 0.0414$, is substantially larger than the
median value for SO-haloes, $\lambda_{\rm B, SO} = 0.0333$.

The SO-halo result derived with the Bullock spin parameter however
resembles the distribution obtained with the Peebles spin parameter in
shape, except that the absolute values are systematically shifted to
somewhat lower values. In the case of statistically large samples
where it is not possible to estimate the concentrations $c$ of every
individual halo, the Bullock spin distribution can in fact be simply
rescaled by constant factor of $1.1$ to yield a spin distribution that
almost perfectly reproduces the Peebles spin parameter distribution
for a statistically large set of simulated haloes. We explicitly show
the shifted Bullock spin parameter distribution in the middle panel of
Fig.~\ref{FigSpinDefs} as red line, which comes to lie just on top of
the Peebles spin parameter distribution for SO-haloes. Thus, when
comparing mean spin values derived with the two different spin
definitions we stress that the constant offset factor of $1.1$ has to
be taken into account. We want to emphasise that this is true only for
SO-haloes, and that the Bullock spin should not be applied to
FOF-haloes, as this definition cannot properly take into account the
complex geometry of FOF-haloes and results in a spin distribution that
is different in shape from the others.

\begin{figure}
\centering
\includegraphics[width=0.5\textwidth,trim= 10 0 0 0,clip]{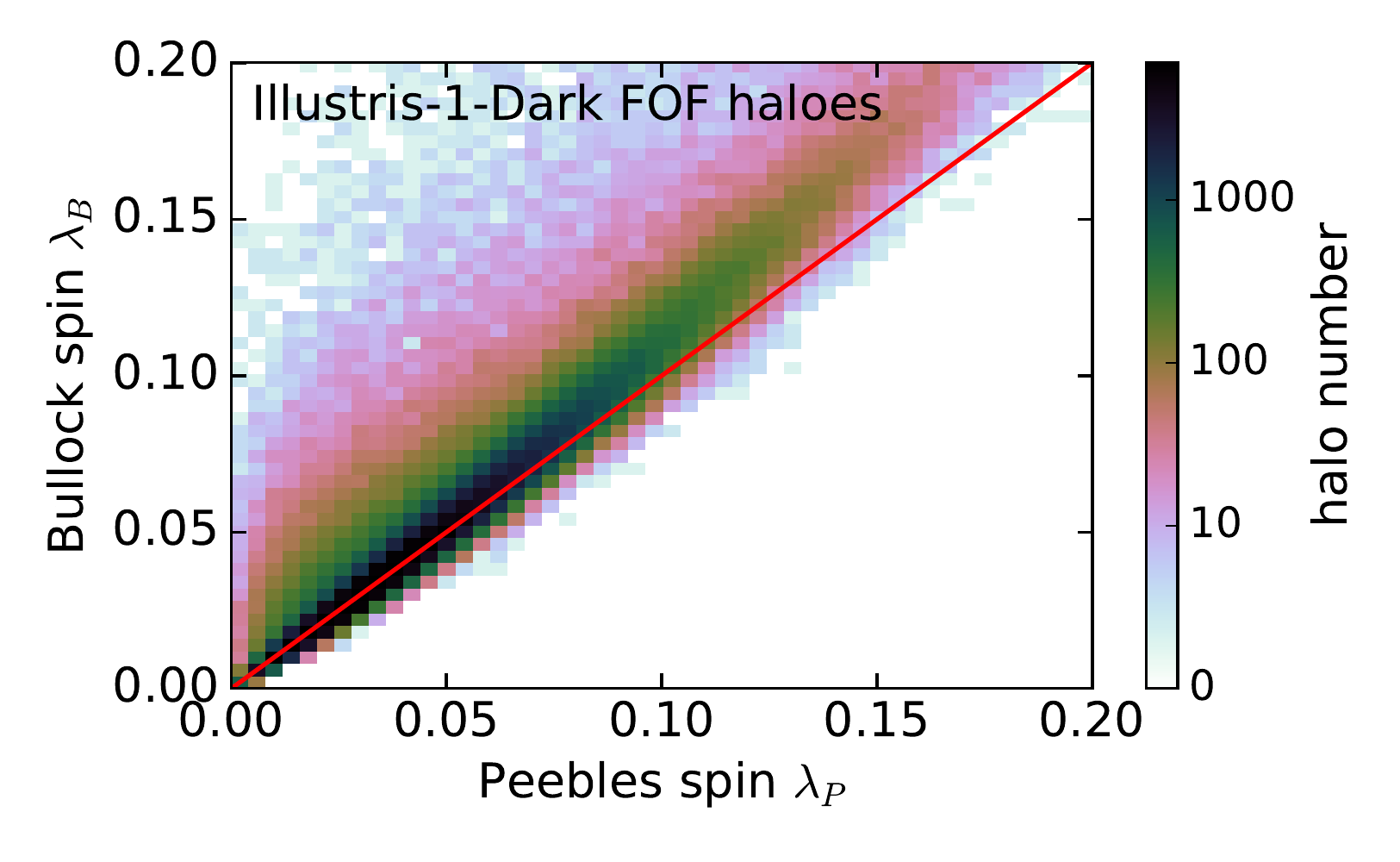}
\includegraphics[width=0.5\textwidth,trim= 10 0 0 0,clip]{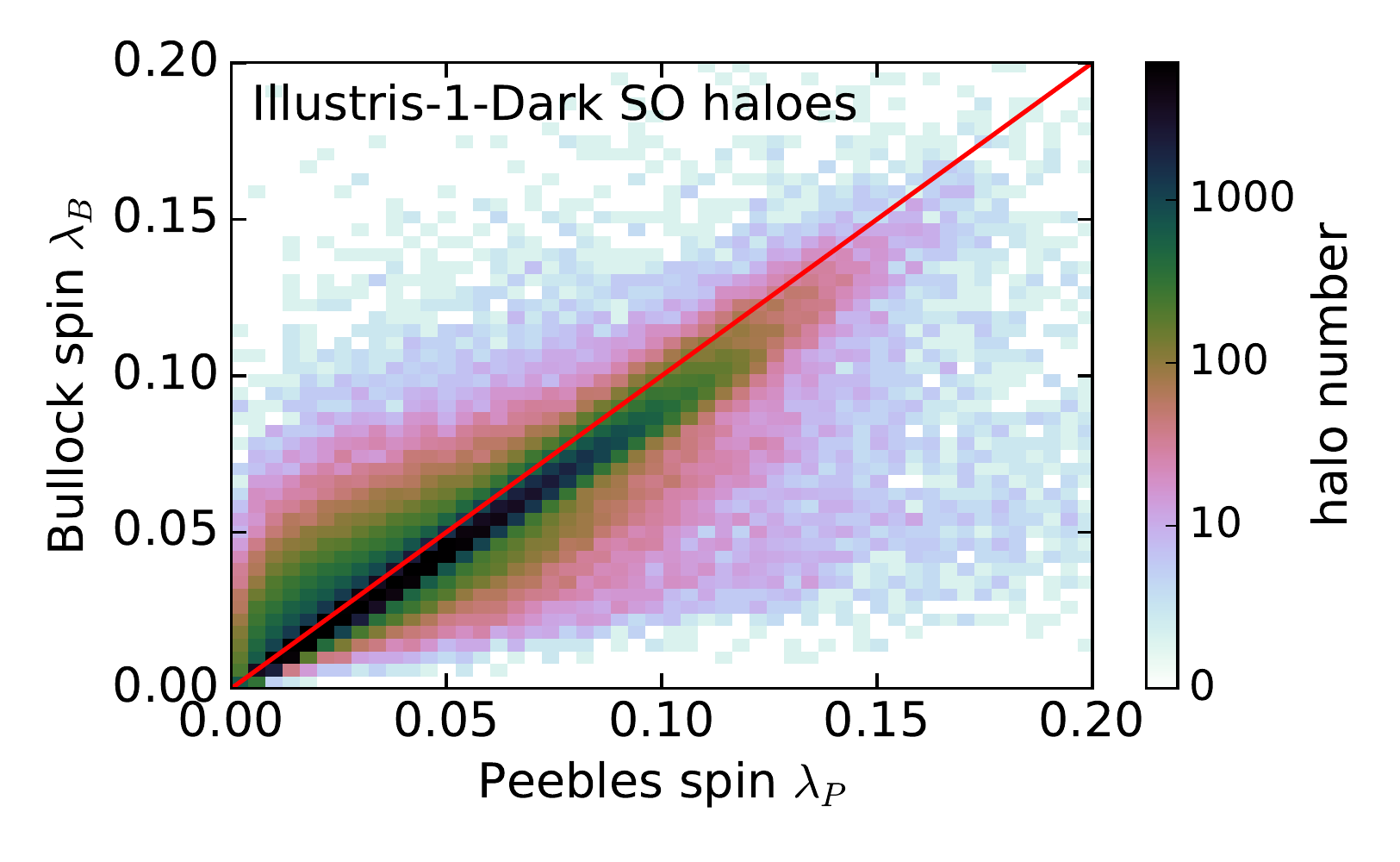}
\caption{Distribution of the spin parameter values derived with the
  Peebles and Bullock spin definitions when applied to the same
  FOF-halo (upper panel) or the same SO-halo (lower panel). The colour
  scale indicates the number of haloes with certain spin
  values. Haloes with identical Peebles and Bullock spin parameters
  would fall onto the red line. We find that for SO-haloes the Bullock
  spin has to be rescaled by a constant factor of 1.1 to reproduce the
  Peebles value. In case of FOF-haloes the Bullock definition
  significantly overestimates the spin for a fraction of haloes as it
  cannot properly account for the aspherically distributed material at
  large radii and thus should not be applied to
  FOF-haloes. \label{FigHaloDefsComp}}
\end{figure}

To further depict the differences between the two spin parameter
definitions, we compare the Peebles and Bullock spin parameter
definitions on a halo-by-halo basis. In Fig.~\ref{FigHaloDefsComp} we
show the distribution of spins resulting from the Peebles and Bullock
spin parameter definitions being applied to the same FOF-haloes (upper
panel) or the same SO-haloes (lower panel). Identity is indicated as
red line. In the case of FOF-haloes the Bullock spin severely
overestimates the spin value for a fraction of the haloes, forming the
extended tail in the left upper corner of the upper panel of
Fig.~\ref{FigHaloDefsComp}. Those are haloes significantly extended
beyond $R_{200}$ for which the Bullock spin fails to properly take
into account the angular momentum contained in the outer regions of
the halo. If SO-haloes are employed, no such systematic bias appears,
however, some scatter still remains. Also, the bulk of haloes tends to
somewhat higher Peebles than Bullock spin parameters which is another
illustration of the constant offset discussed in the middle panel of
Fig.~\ref{FigSpinDefs}.
 
Furthermore, in Fig.~\ref{FigSpinDefsComp} we compare the Peebles
(upper panel) and Bullock (lower panel) spin parameters derived for
the FOF- and SO-counterparts of the same halo. To this end we create a
combined halo sample that contains only haloes that match the
selection criteria in both the FOF- and the SO-halo definition. Most
haloes scatter around the identity line, which is again indicated in
red.  However, a few percent of the haloes exhibit a significant
enhancement in spin in the FOF definition compared to SO. This
fraction is somewhat larger for the Bullock spin definition compared
to the Peebles definition. The enhancement occurs in aspherical haloes
significantly extended beyond $R_{200}$ in their FOF definition
(compare to Fig.~\ref{Figfofso}).  In the case of the Peebles spin,
this enhancement is simply due to additional material, such as from
minor mergers at large radii, and affects a smaller number of
haloes. On the other hand, when dealing with the Bullock definition
this enhancement affects a larger number of haloes and leads to a bias
in the resulting mean spin, as this enhancement is not purely of
physical nature but partly caused by the fact that FOF-haloes are
normalised by virial properties not properly accounting for their true
shape.  This is also the reason why the Bullock spin parameter
distribution of FOF-haloes differs from the SO-halo distribution not
only in peak position but also in shape.  Thus we want to stress that
the Bullock spin parameter appears unsuitable for the FOF-halo
definition and should not be applied to intrinsically aspherical
FOF-haloes.

To verify our results, we have also applied our group finder and
extended halo property calculation to the Millennium Simulation
\citep{Springel05Nature}. From this run we construct an SO-halo sample
in the same way we did for {\em Illustris}, containing
$\sim 1.4\times 10^6$ objects. We plot the resulting Peebles spin
parameter distribution as a green line in the lower panel of
Fig.~\ref{FigSpinDefs}. It is similar to the distribution we obtain
from Illustris-1-Dark shown as black line, and exhibits a median value
of $\lambda_P = 0.0403$. Note that the imposed 300 dark matter
particle cut leads to a higher minimum halo mass in the Millennium
Simulation compared to \emph{Illustris} by approximately one order of
magnitude. Taking into account the subtle trend of spin with halo mass
(compare Fig. \ref{FigDMMass}) the Millennium spin distribution is
expected to have somewhat higher values. 
Furthermore, the Millennium Simulation was carried out with a WMAP1 cosmology, which has a slightly higher $\sigma_8$ value than \emph{Illustris}. This makes haloes of a given mass collapse earlier in Millennium than in \emph{Illustris}. Comparing the given distributions is thus the same as making a comparison between spin distributions obtained with the same cosmology but derived at somewhat different times. As shown in the bottom left panel of Fig.~\ref{FigSpinDMZ}, the Peebles spin parameter distribution of SO haloes exhibits a small trend to higher values with decreasing redshift, such that the higher $\sigma_8$ contributes to the spin distribution derived from the Millennium Simulation being slightly shifted to later times and thus to the right with respect to the one derived from \emph{Illustris}. To avoid such ambiguities and facilitate the comparison between different data sets, we argue in Section \ref{Sec_DMzdep} that it is best to use the Peebles spin parameter of FOF-haloes, as it is the only measure yielding a spin distribution self-similar in time, and as such is least affected by cosmology.

Our Millennium spin distribution can be directly compared to the most
precise dark matter only result from the literature, which is given by
the best fit from \cite{Bett07}, independently derived for a cleaned
halo sample from the Millennium Simulation using the Peebles spin
parameter definition and the new fitting function they proposed (red
dashed line, see also Section \ref{Sec_DMz0}). The fit to this halo
sample is almost identical to their SO-halo sample fit (blue dashed
line), and is very close, albeit not identical, to our result. The
small residual difference in the spin distribution from \cite{Bett07}
and our study should originate in the details of the selection
criteria used to define sets of haloes in quasi-equilibrium. We thus
investigate whether sample selection criteria can account for the
small differences observed between the spin distributions. We show as
red solid line in the lower panel of Fig.~\ref{FigSpinDefs} the spin
parameter distribution obtained for our SO-halo sample when requiring
only a minimum resolution of 300 dark matter particles
\citep[identical to][]{Bett07} and that haloes have at least one
gravitationally bound component. Imposing no limit on the virial ratio
allows ongoing mergers to be included in the sample. The angular
momentum of haloes in active merging phases has a large and often
dominating contribution of orbital angular momentum, leading to an
extended tail of the spin parameter distribution to high values. This
tail enters the overall normalisation of the spin distribution and
lowers its peak value, but the median of the spin distribution is not
affected significantly. The changes induced in the spin distribution
by different sample selection criteria are thus exactly of the order
of magnitude and of the type of the residual differences observed
between our results and the study of \cite{Bett07}, validating that
our independent analysis methods are consistent.

\section{Dark matter only results} \label{Sec_DM}

\subsection{Angular momentum statistics of dark matter haloes at $z=0$} \label{Sec_DMz0}

\begin{figure}
\centering
\includegraphics[width=0.49\textwidth,trim= 10 0 0 0,clip]{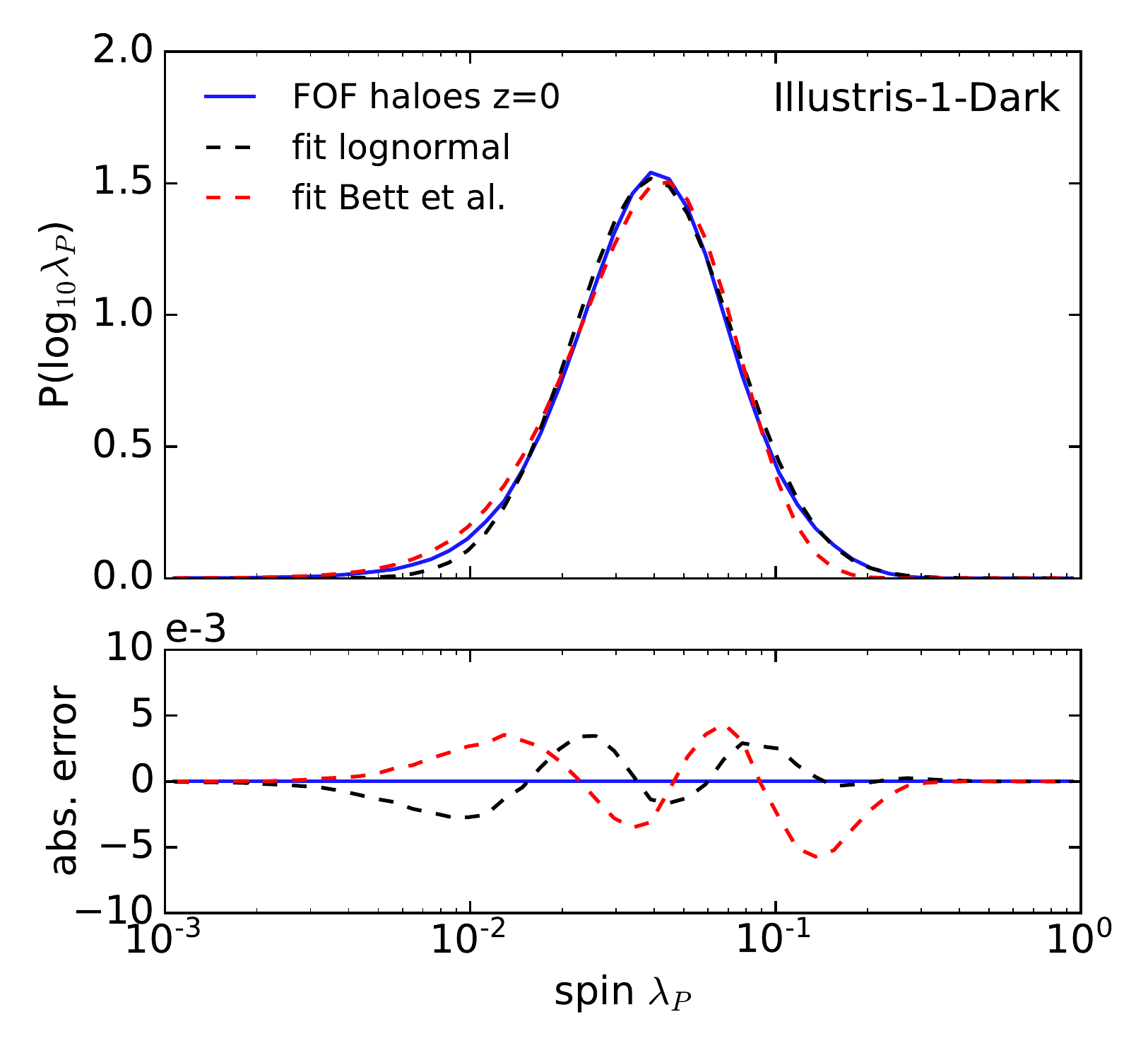}
\includegraphics[width=0.49\textwidth,trim= 10 0 0 0,clip]{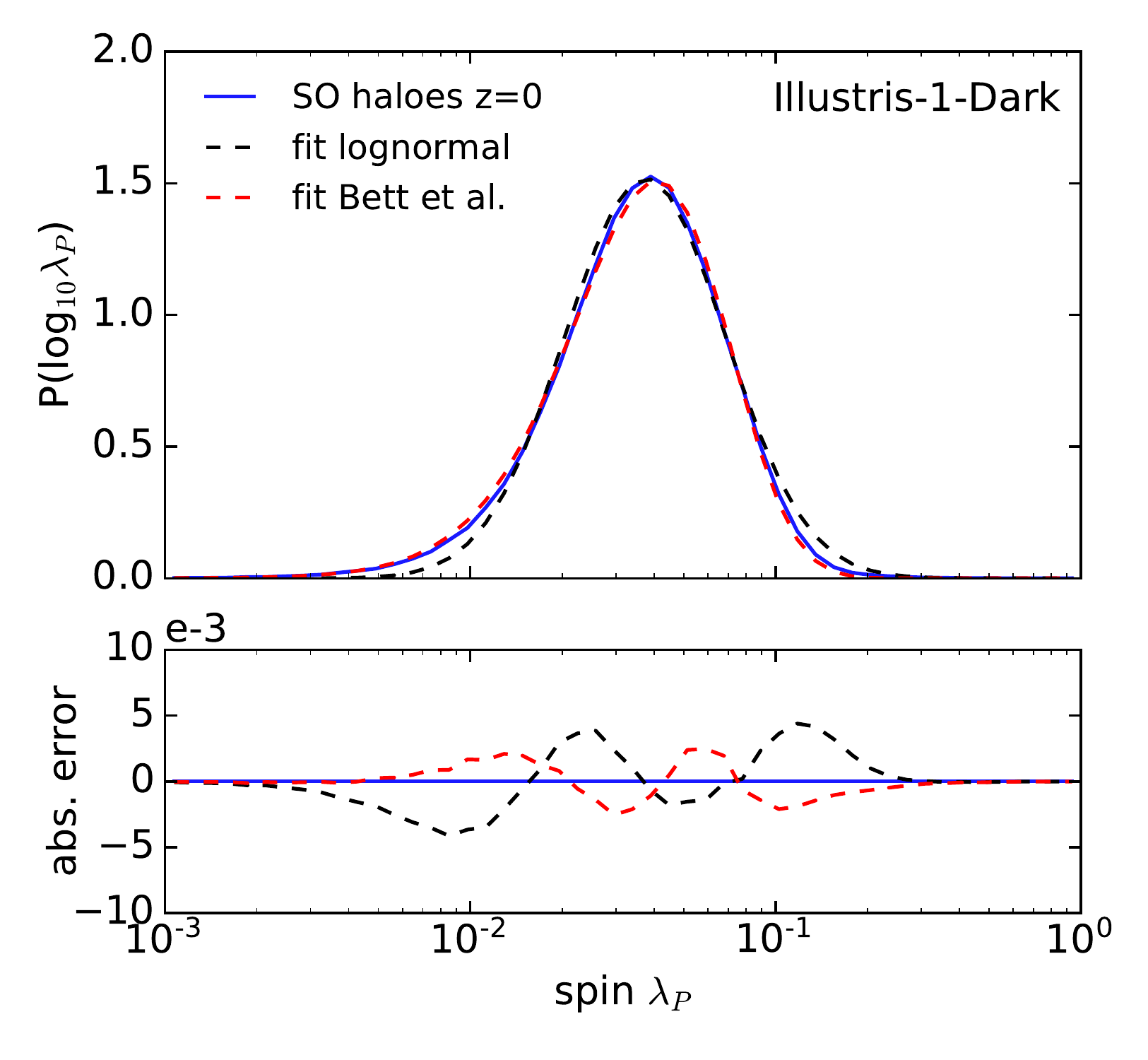}
\caption{Peebles spin parameter distribution (blue) of FOF-haloes
  (upper panel) and SO-haloes (lower panel) from
  Illustris-1-Dark. Both distributions are fitted with a lognormal
  function (dashed black) and a fitting function proposed by
  \citet{Bett07} (dashed red). For easier comparison we show the
  absolute error of the fitting functions with respect to the
  distribution obtained from Illustris-1-Dark. \label{FigSpinDMfit}}
\end{figure}

We begin by presenting the angular momentum properties of dark matter
only haloes from Illustris-1-Dark. In Fig.~\ref{FigSpinDMfit} we show
the spin parameter distributions for FOF- (upper panel) and SO-haloes
(lower panel) at $z = 0$. Our FOF- and SO-halo samples comprise
$\sim 400,000$ and $\sim 360,000$ objects, respectively. We provide
least-square error fits of the two most common fitting functions to
the derived Peebles spin parameter distributions. We chose to present
results derived with the Peebles spin parameter, as it yields nearly
identical spin distributions for both halo definitions when the
potential energy of the halo is estimated accurately. The classic
analytic approximation for the spin parameter distribution is the
lognormal function \citep[e.g.][]{vdBosch98},
\begin{equation} 
P({\rm log}_{10}\lambda) = \frac{1}{\sqrt{2\pi}\sigma} \exp \left[ -\frac 1 2
  \left(\frac{\log_{10}(\lambda / \lambda_0)}{
      \sigma}\right)^2\right],
\label{eqnlognormal}
\end{equation} 
where $\lambda_0$ is the peak position and $\sigma$ the width of the
distribution. The lognormal function is normalised such that
$\int_{-\infty}^{\infty}P({\rm log}_{10}\lambda) \ {\rm
  dlog}_{10}\lambda = 1$.
\cite{Bett07} performed an extended analysis of the Peebles spin
parameter distribution of dark matter only haloes from the Millennium
Simulation \citep{Springel05Nature} and found that their cleaned halo
sample is better described by the fitting function
\begin{equation} \label{Bett}
P_{\rm B}(\log_{10} \lambda) = A \left( \frac \lambda {\lambda_0} \right)^3 \exp\left[ - \alpha \left(\frac \lambda {\lambda_0} \right)^{3/ \alpha} \right],
\end{equation} 
where $\lambda_0$ is again the peak position, $\alpha$ a free fitting
parameter, and $A$ the normalisation, such that
$\int_{-\infty}^{\infty}P_B({\rm log}_{10}\lambda) \ {\rm
  dlog}_{10}\lambda = 1$.
This new fitting function is constructed to rise with the third power
for small values and falls off exponentially for large
values. However, this specific shape in combination with the free
fitting parameter $\alpha$ makes the given function highly flexible.

We fit both functions to the Peebles spin parameter distributions of
our FOF- and SO-halo samples. The small panels in
Fig.~\ref{FigSpinDMfit} show the absolute error of the fits relative
to the distribution derived from Illustris-1-Dark in every bin. From
the absolute errors in every bin we derive the root mean square error
$\epsilon_{\rm rms}$ of the fit and list it with the best fit
parameters in Tab.~\ref{TabDMfit}. Based on $\epsilon_{\rm rms}$ we
find that the new fitting function given by Eq.~(\ref{Bett}) describes
the SO-halo spin distribution slightly better, which is consistent
with \cite{Bett07}, who derived their fitting function based on a halo
sample very similar to their SO-halo sample. The FOF-halo spin
distribution on the other hand is better fit by the classic lognormal
function. We find the same behaviour when analysing the Bullock spin
parameter distributions. The differences between the two fitting
functions lie primarily in the different slopes in the wings of the
distribution and the detailed peak shape. However, these differences
are relatively small, and which analytic function describes the spin
parameter distribution best depends ultimately on the preferred halo
definition.

\begin{table}
\centering
\begin{tabular}{|l|l|l|l|}
\hline
FOF fit lognormal & $\lambda_0 = 0.040$  & $\sigma = 0.26$ & $\epsilon_{\rm rms}= 0.0015$ \\ 
FOF fit Bett et al. & $\lambda_0 = 0.043$ & $\alpha = 3.16$ & $\epsilon_{\rm rms}= 0.0022$ \\
SO fit lognormal & $\lambda_0 =  0.037$ & $\sigma = 0.26$ & $\epsilon_{\rm rms}=0.0020 $ \\ 
SO fit Bett et al. & $\lambda_0 = 0.041$ & $\alpha = 3.15$ & $\epsilon_{\rm rms}= 0.0011$ \\ \hline
\end{tabular}
\caption{Best fit parameters of the analytic fits 
  to the Peebles spin parameter distributions derived from
  Illustris-1-Dark. We give results both for FOF- and SO-haloes, and
  in each case for the classic log-normal function and the
  proposed modified fitting function by \citet{Bett07}. \label{TabDMfit} }
\end{table}

\subsection{Convergence} \label{Sec_DMconv}

\begin{figure}
\centering
\includegraphics[width=0.5\textwidth,trim= 10 0 0 0,clip]{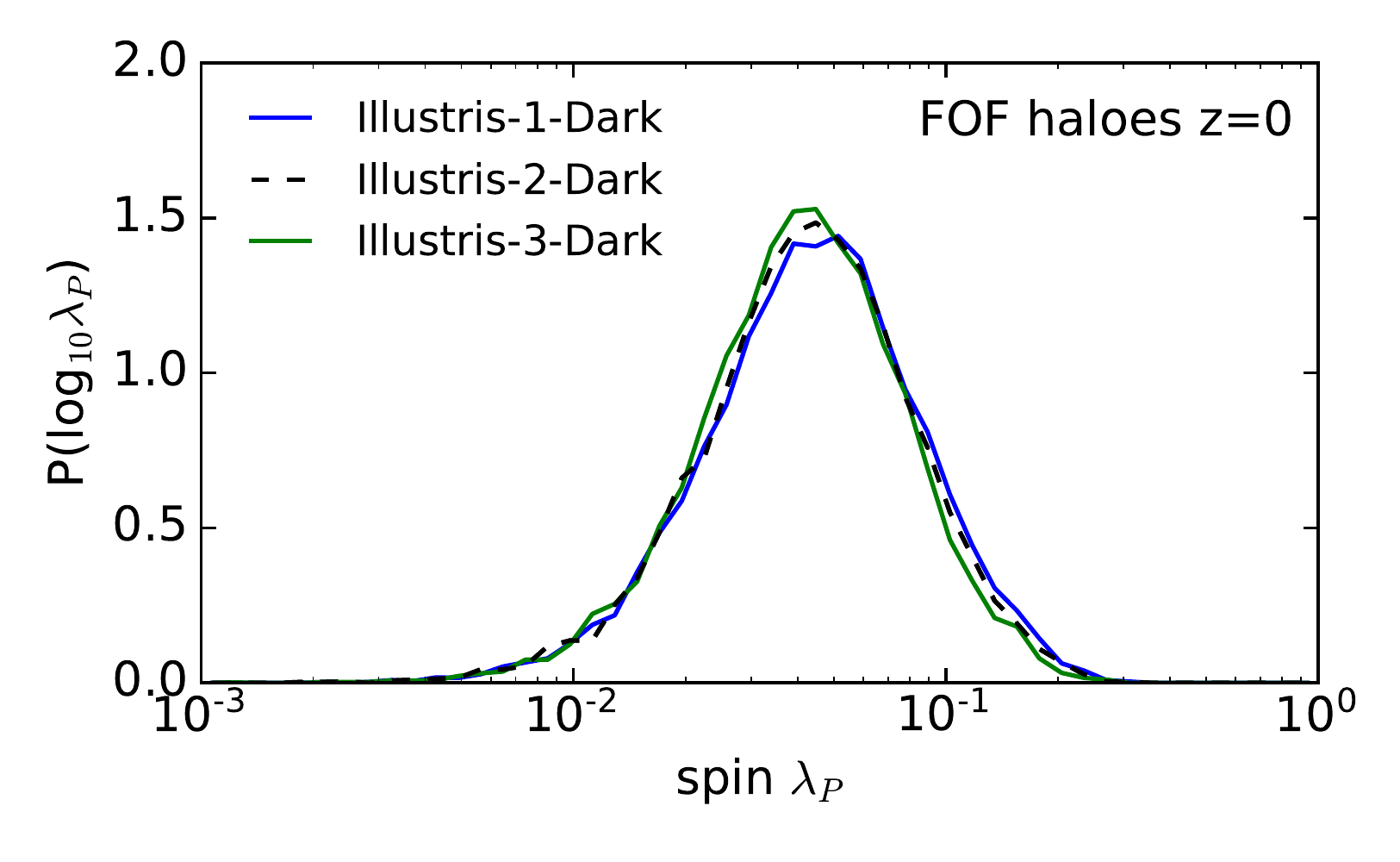}
\includegraphics[width=0.5\textwidth,trim= 10 0 0 0,clip]{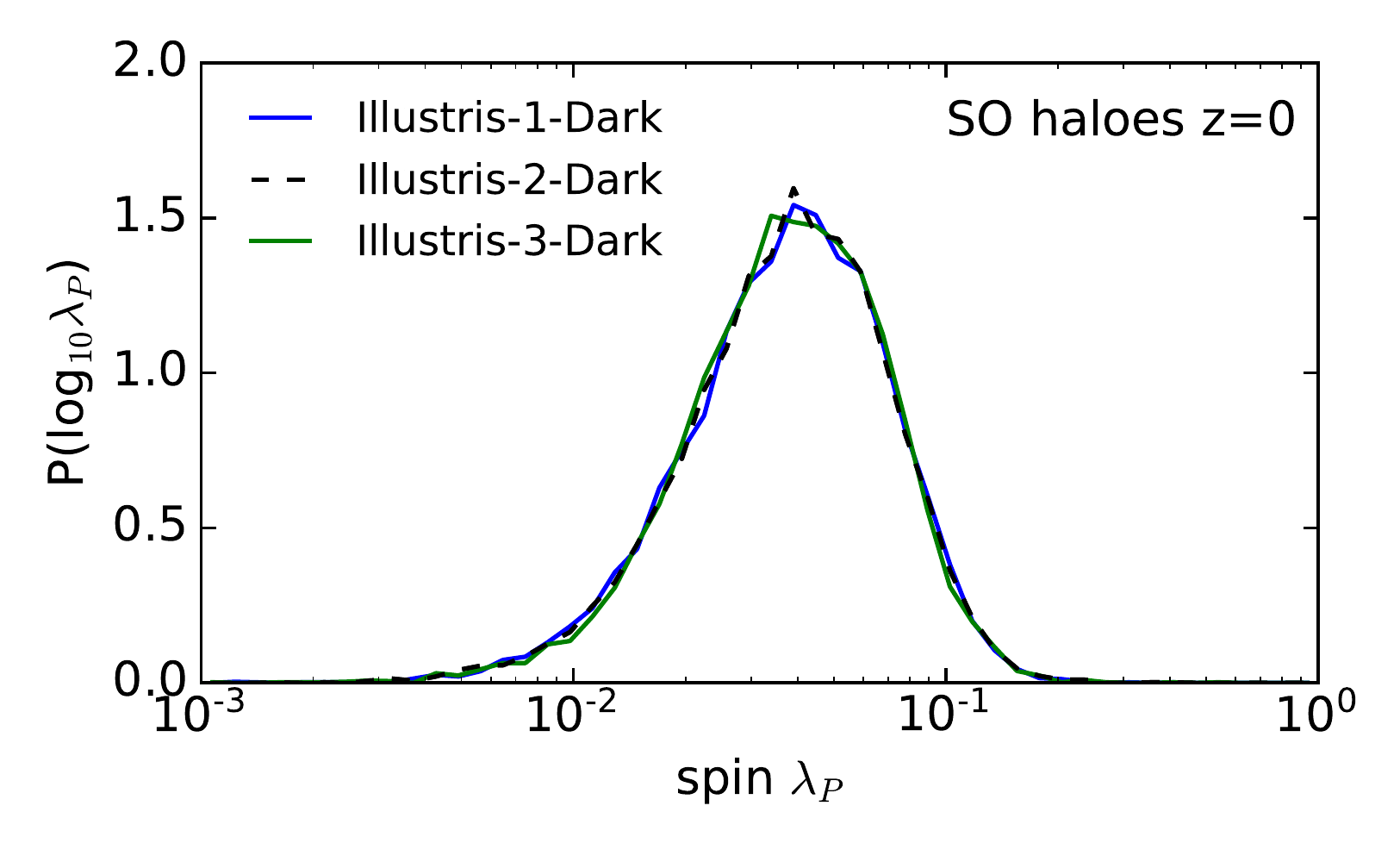}
\caption{Peebles spin parameter distributions of dark matter only
  FOF-haloes (upper panel) and SO-haloes (lower panel) derived for the
  three different resolutions of the {\em Illustris} simulation suite
  (see Tab.~\ref{TabSims}). We find very good convergence for the
  three resolutions; the small residual deviations are consistent with
  noise expected from the limited halo sample
  size. \label{FigSpinDMconv}}
\end{figure}

In Fig.~\ref{FigSpinDMconv} we show the Peebles spin parameter
distribution for FOF-haloes (upper panel) and SO-haloes (lower panel)
derived at three different resolutions of Illustris-Dark (see
Tab.~\ref{TabSims}) for halo masses above
$1.4\times 10^{11}\, {\rm M}_\odot$.  This mass limit is set by our
selection criteria, which require a halo to be resolved by at least
300 dark matter particles. When applied to Illustris-3-Dark with the
coarsest resolution this corresponds to the above value. As there are
subtle trends of spin with halo mass a common mass range must be
adopted when examining the numerical convergence of our measurements.

Reassuringly, we find very good convergence of the Peebles spin
parameter distribution for the three resolutions of
Illustris-Dark. The small residual deviations originate in the limited
halo sample size and are consistent with the associated counting
noise. The convergence is equally good for the Bullock spin parameter
distribution, which we refrain from showing explicitly.

\subsection{Redshift evolution of halo spin} \label{Sec_DMzdep}

\begin{figure*}
\centering
\includegraphics[width=0.49\textwidth,trim= 10 0 0 0,clip]{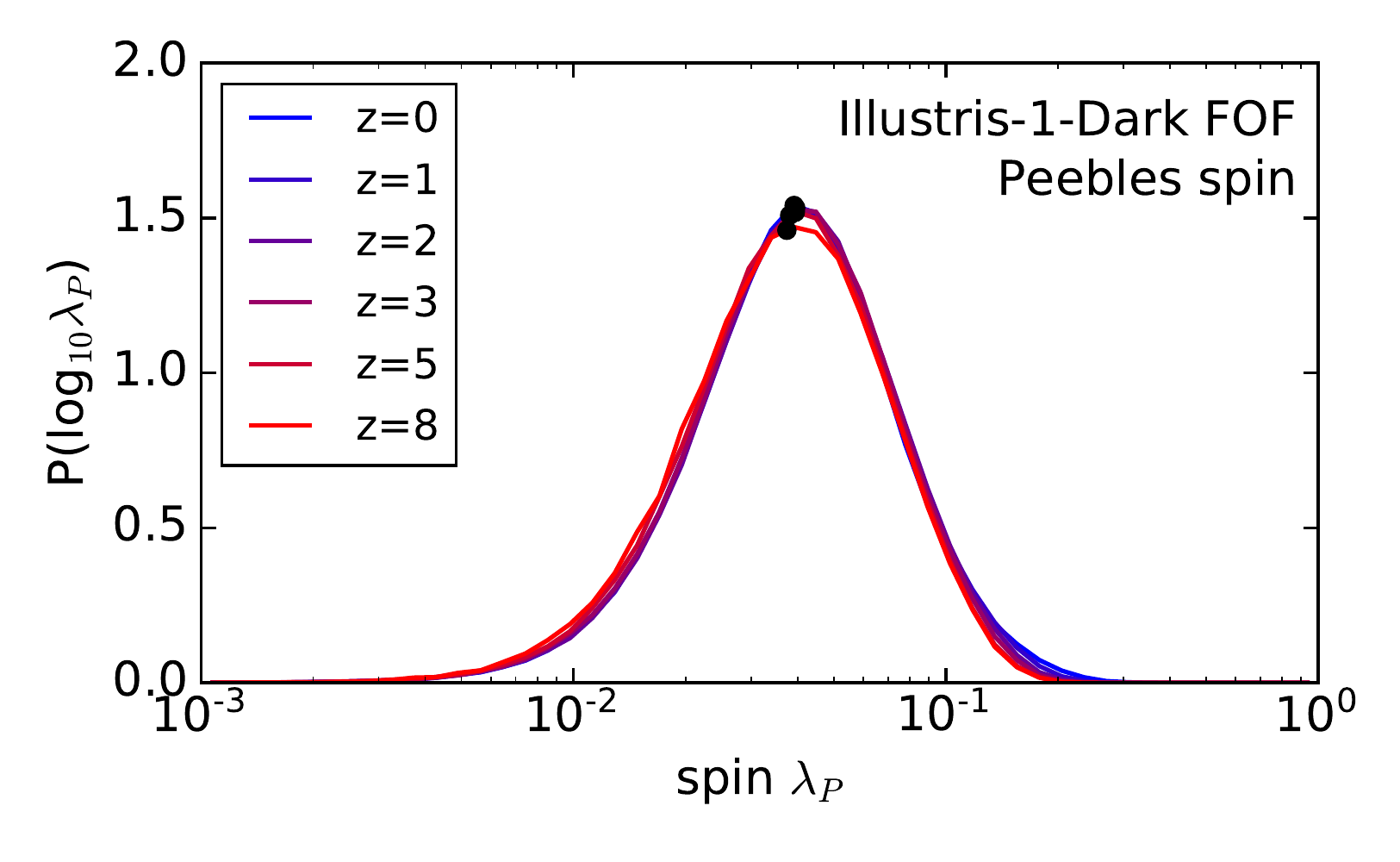}
\includegraphics[width=0.49\textwidth,trim= 10 0 0 0,clip]{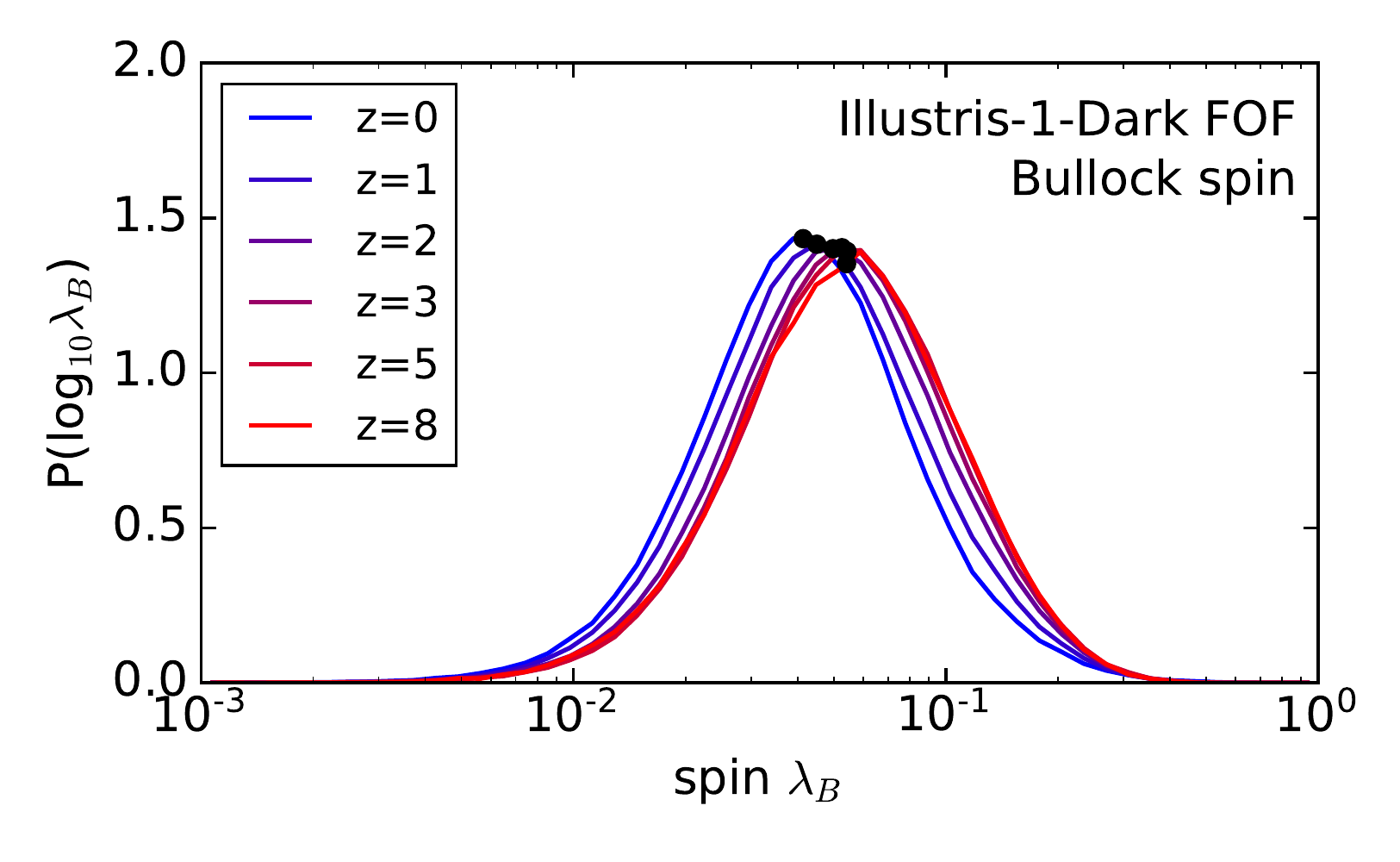}
\includegraphics[width=0.49\textwidth,trim= 10 0 0 0,clip]{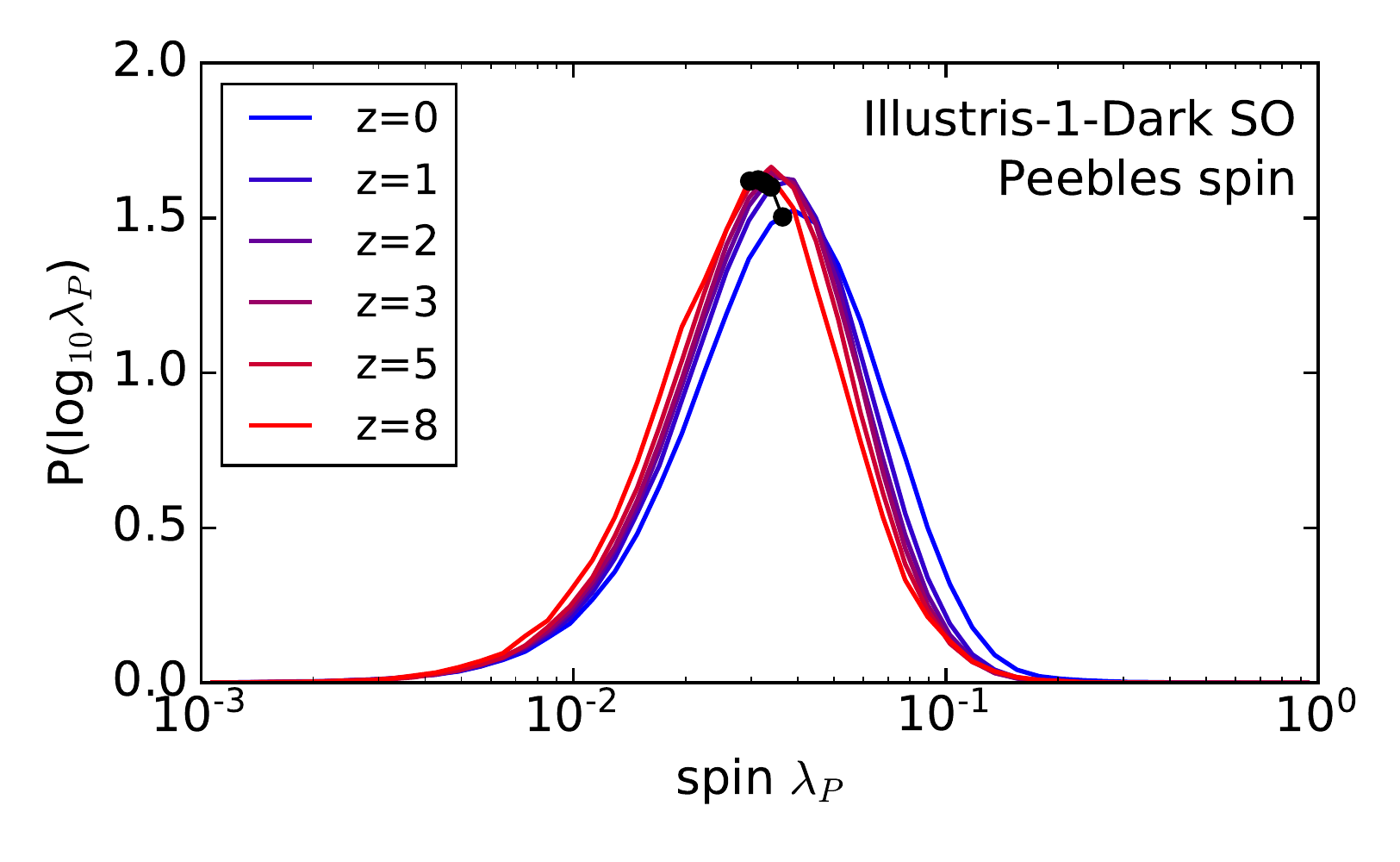}
\includegraphics[width=0.49\textwidth,trim= 10 0 0 0,clip]{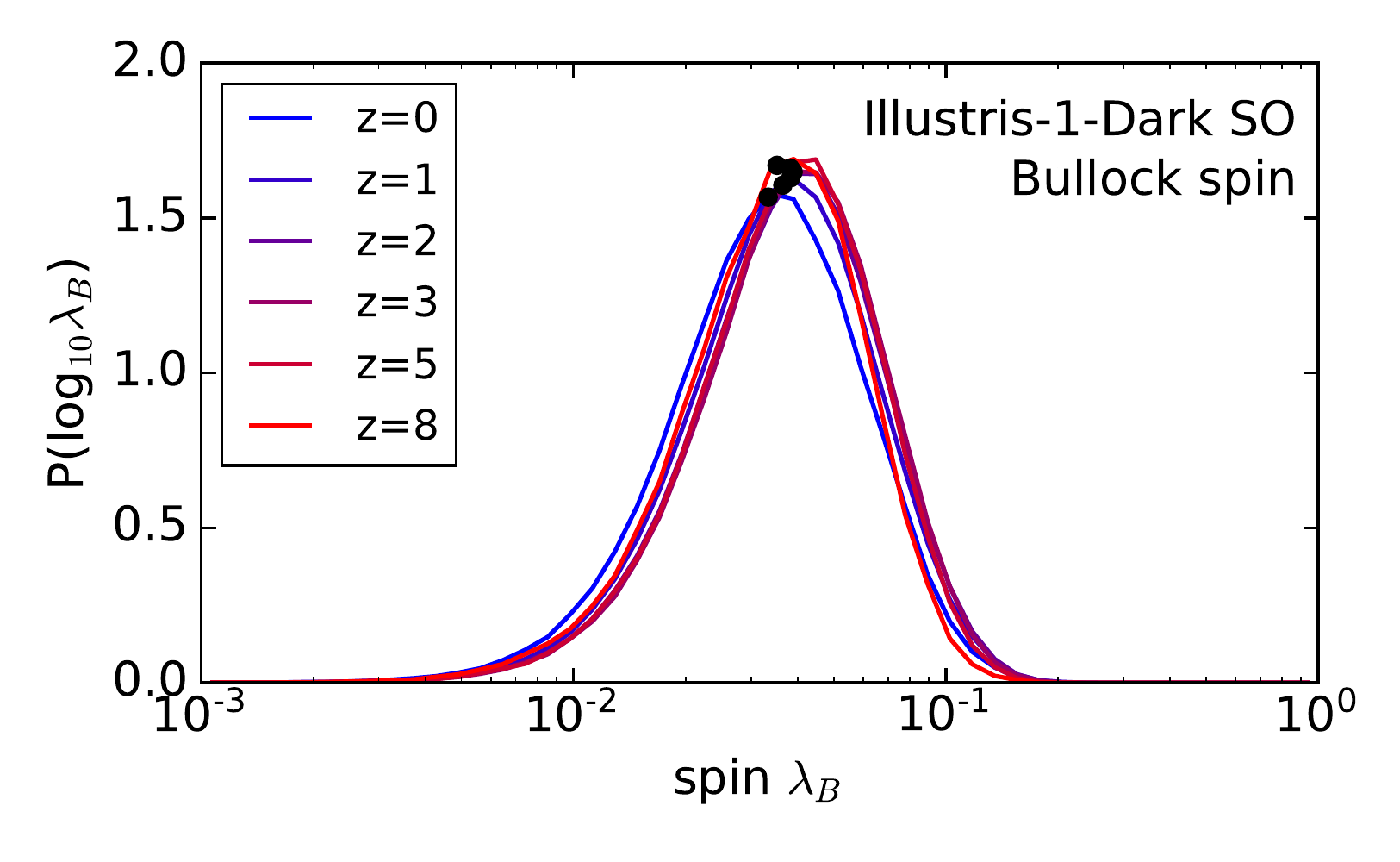}
\caption{Redshift evolution of the spin parameter distribution of
  FOF-haloes (upper panel) and SO-haloes (lower panel) derived with
  the Peebles (left column) and Bullock (right column) spin parameter
  from Illustris-1-Dark. Black dots mark the median spin parameter at
  every redshift. The Peebles spin parameter distribution of
  FOF-haloes is perfectly self-similar with respect to time. The other
  distributions exhibit residual trends with redshift, which are due
  to shortcomings in the Bullock spin and SO-halo definitions. We thus
  restrict ourselves in the following on the Peebles spin applied to
  FOF-haloes, which yields the most robust
  results.  \label{FigSpinDMZ}}
\end{figure*}

In Fig.~\ref{FigSpinDMZ} we show the spin parameter distributions
obtained with the Peebles (left column) and Bullock (right column)
spin parameter definition for FOF- (upper row) and SO-haloes (lower
row) from Illustris-1-Dark for different redshifts. Black dots mark
the median spin parameters at every redshift.

The Peebles spin parameter definition applied to FOF-haloes yields a
remarkably self-similar spin parameter distribution in time, as
expected from theory for self-similar dark matter structures. The
self-similarity of haloes is only broken by a varying mean
concentration $c$ of the NFW radial density profile with halo
mass. Less massive haloes are expected to be on average denser, which
corresponds to higher values of $c$, and to collapse earlier than more
massive haloes \citep{NFW97}. This introduces a time dependence of the
mean concentration for quasi-relaxed structures thus generally
breaking the self-similarity of haloes with respect to both halo mass
and time. However, as \cite{NFW97} found no correlation of the spin
with concentration \citep[see also][]{Bullock01}, we expect the spin
parameter distribution of haloes to be approximately self-similar with
mass and time. This feature is realised by the Peebles spin parameter
distribution of FOF-haloes which shows only a small evolution of the
median spin value from $\lambda_{z=8} = 0.0374$ at $z=8$ to
$\lambda_{z=0} = 0.0391$ at $z=0$.

The spin parameter distributions derived with other approaches however
exhibit non-vanishing trends with redshift. The strongest shift is
visible if the Bullock spin parameter definition is applied to
FOF-haloes. The residual trends in this distribution are a consequence
of imposing a spherical shape on the more complicated geometry of
FOF-haloes. The trends in the distributions obtained from the SO-halo
definition are probably caused by not taking into account the
gravitationally bound matter outside the virial radius $R_{200}$,
whose relative mass fraction with respect to the matter inside of
$R_{200}$ can change over time, inducing the observed trends.

We conclude that although the SO-halo definition is operationally very
clean, it is the Peebles spin parameter definition applied to
FOF-haloes that yields physically the most stable and reliable
results. Thus, in the following we will restrict ourselves to showing
mostly results obtained for FOF-haloes with the Peebles spin parameter
definition. We have checked that our analysis carried out for
SO-haloes yields qualitatively the same results.

\begin{figure} 
\centering
\includegraphics[width=0.5\textwidth,trim= 0 10 0 10,clip]{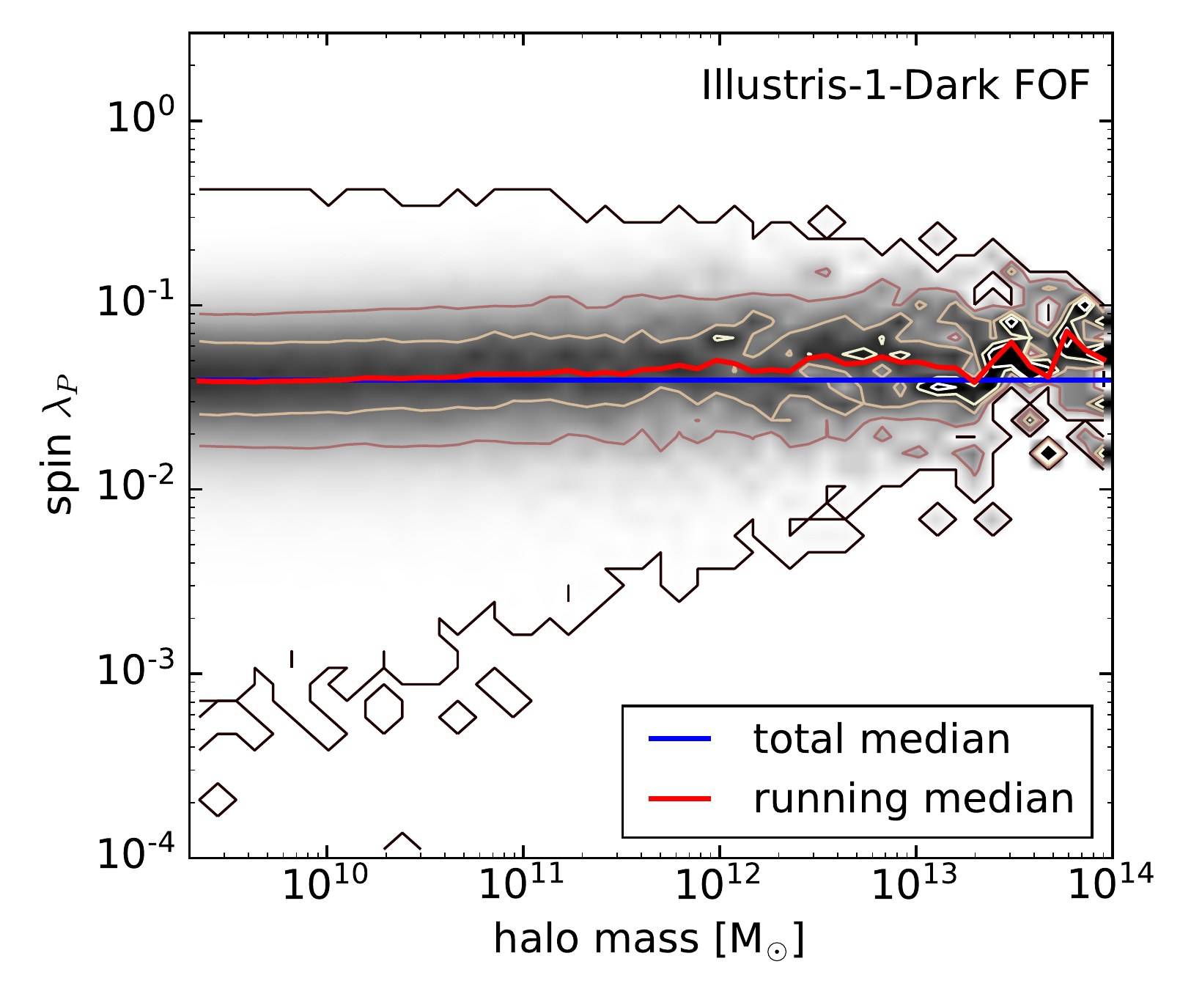}
\caption{Distribution of the Peebles spin parameter with respect to
  the mass $M_{\rm FOF}$ of dark matter only FOF-haloes from
  Illustris-1-Dark, separately normalised in every mass bin. The grey
  shading ranges from a fraction of $0$ to $0.2$ of all haloes in a
  mass bin having a given spin parameter. Contours are drawn at
  constant fractions of $0$, $0.05$, $0.1$, $0.15$, and $0.2$,
  respectively. The total median spin parameter is shown as a blue
  line, the median of every mass bin as a red line. The median spin
  parameter slightly increases with increasing halo
  mass. \label{FigDMMass}}
\end{figure}

\begin{figure}
\centering
\includegraphics[width=0.47\textwidth,trim= 10 0 0 0,clip]{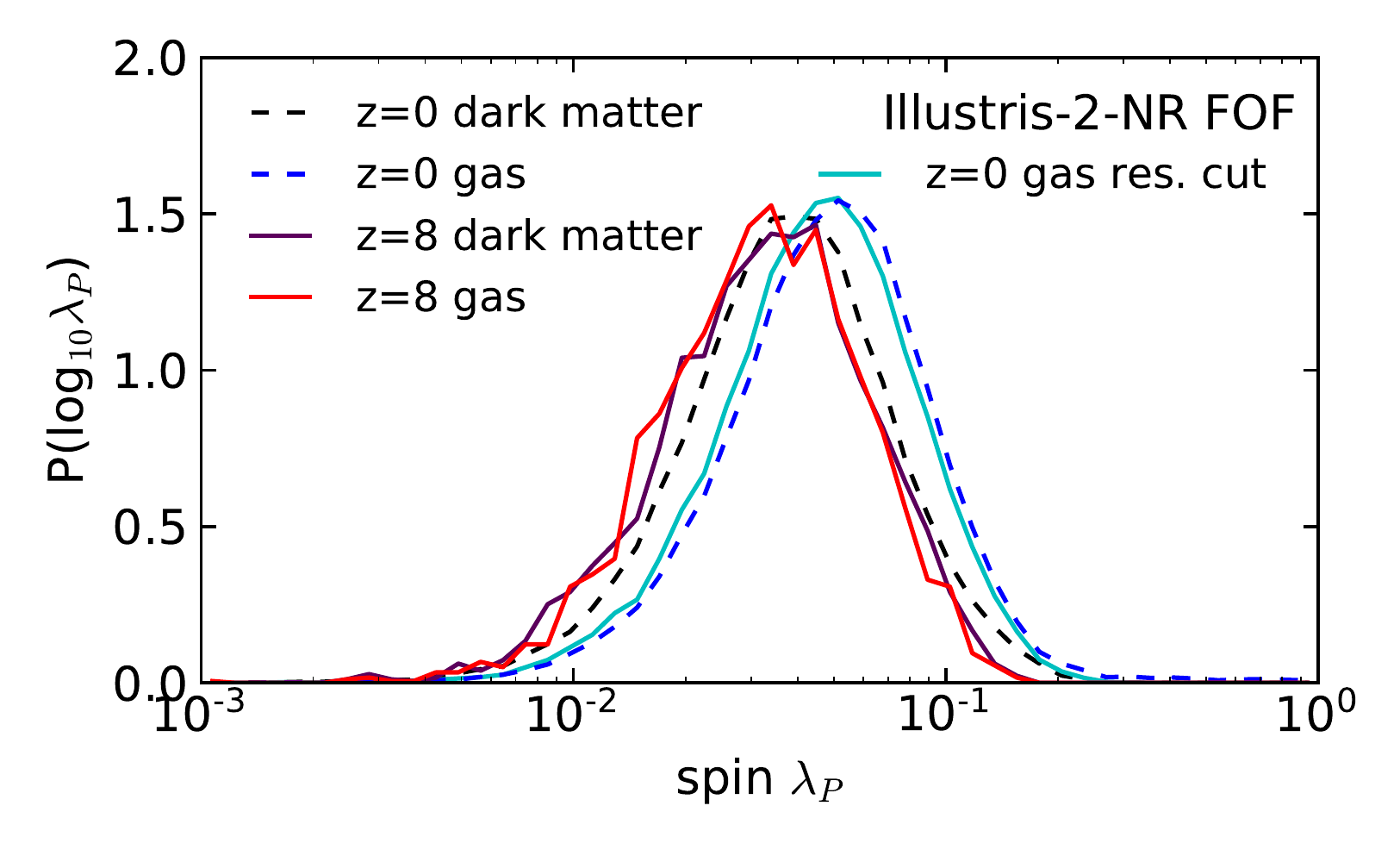}
\includegraphics[width=0.47\textwidth,trim= 10 0 0 0,clip]{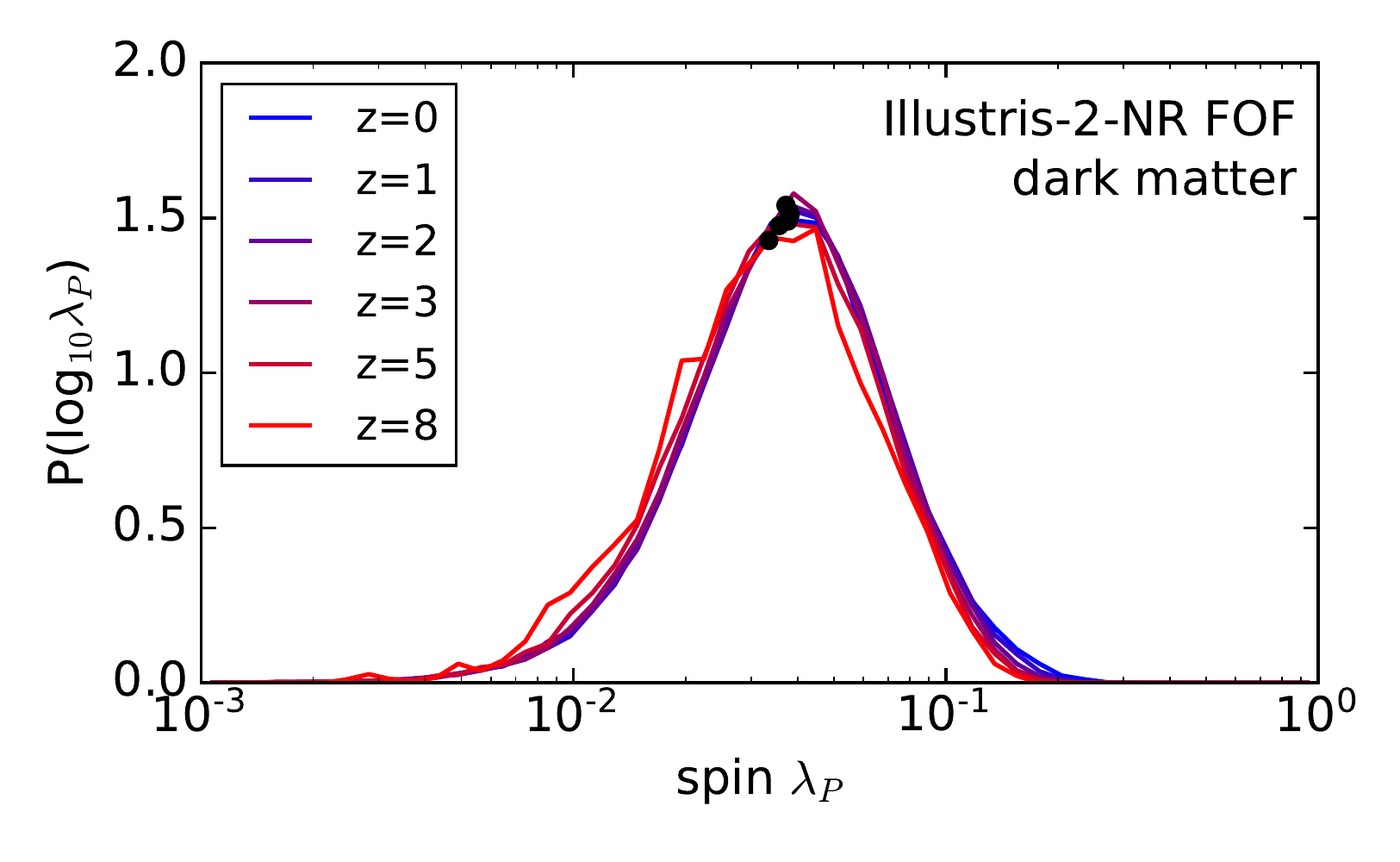}
\includegraphics[width=0.47\textwidth,trim= 10 0 0 0,clip]{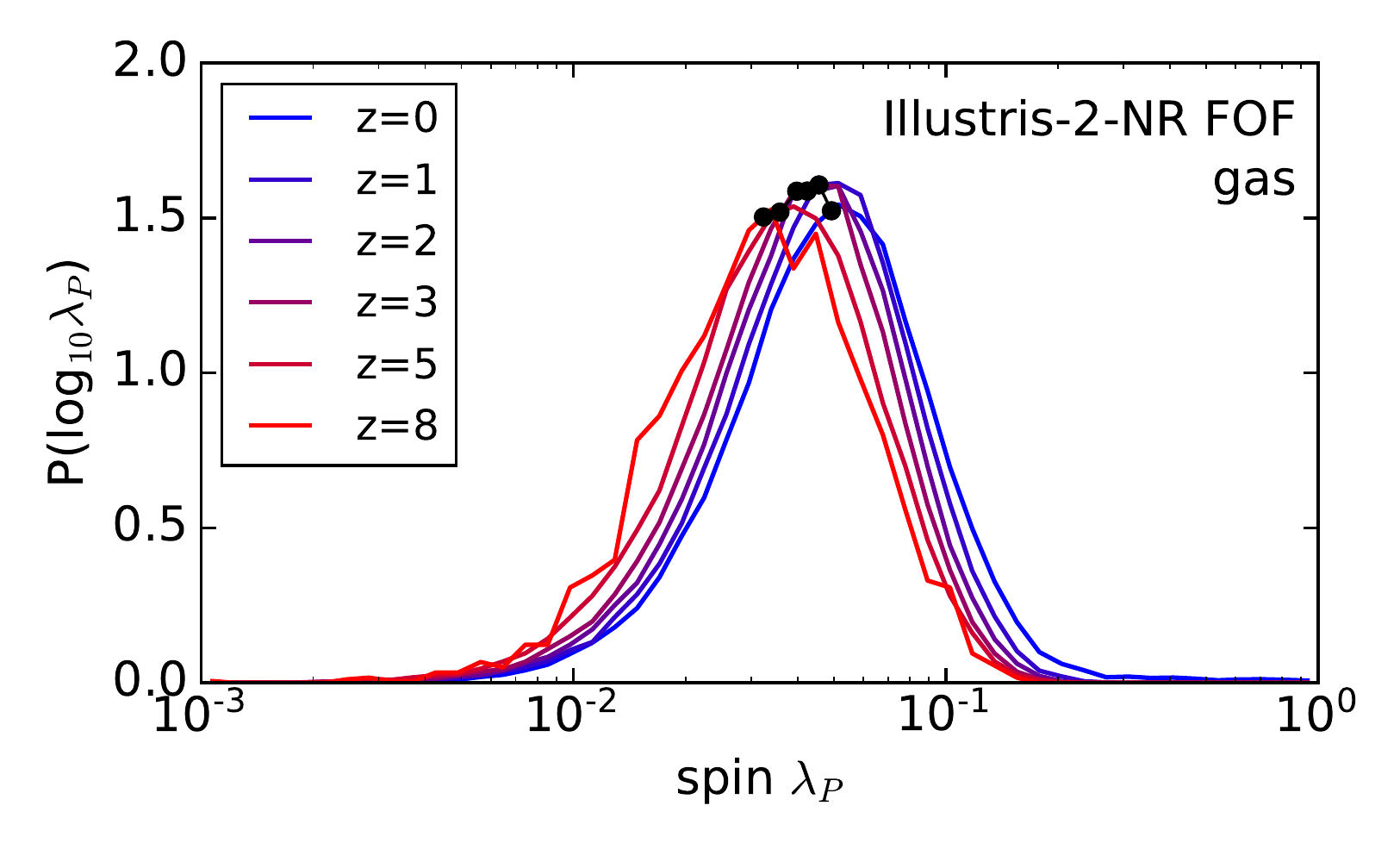}
\caption{Redshift evolution of the dark matter (middle panel) and gas
  (lower panel) Peebles spin parameter distributions of FOF-haloes
  from Illustris-2-NR. Black dots mark the median spin parameter at
  every redshift. The dark matter spin distribution exhibits the same
  self-similarity in time as the dark matter-only Illustris-1-Dark
  simulation, however, at somewhat lower spin values. The spin
  distribution of gas on the other hand systematically shifts to
  higher spin values, due to a transfer of angular momentum from dark
  matter to gas. For comparison, we show (upper panel) the spin
  parameter distributions for dark matter and gas at $z = 8$ (solid
  line) and $z = 0$ (dashed line), including the gas distribution derived 
  with a stricter mass cut (cyan) to exclude bias due to poorly resolved haloes. 
  \label{FigNRZ}}
\end{figure}

\subsection{Dependence of spin on halo mass} \label{Sec_DMmassdep}

In Fig.~\ref{FigDMMass} we show the dependence of the Peebles spin
parameter of FOF-haloes from Illustris-1-Dark on the halo mass
$M_{\rm FOF}$. This figure shows a two-dimensional histogram where
FOF-haloes have been binned according to their mass and spin parameter
in 50 equidistant logarithmic bins over the given range. As the
absolute number of haloes increases rapidly with decreasing mass, we
have normalised every mass bin to unity. The normalised number of
FOF-haloes in every mass bin is indicated by the grey shading ranging
from a fraction of $0$ to $0.2$ of the haloes in each mass
bin. Contours are drawn at constant fractions of $0$, $0.05$, $0.1$,
$0.15$, and $0.2$, respectively. The median spin parameter in every
mass bin is shown as a red line, the overall median spin parameter as
a blue line.

Fig.~\ref{FigDMMass} shows a small, but clearly present, systematic
increase of the median Peebles spin parameter with halo mass. This
trend is presumably caused by two related effects. More massive haloes
originating from larger initial density perturbations are not fully
collapsed at $z=0$ yet. Also, they are still actively accreting matter
that perturbs their outskirts and slows down the relaxation
process. The orbital angular momentum of the involved minor mergers
adds to the intrinsic angular momentum of the main halo and leads to
an enhanced spin.  As we have shown in Fig.~\ref{Figfofso}, FOF-haloes
are in general much more extended than SO-haloes, such that this
effect plays a role in FOF-haloes but not in SO-haloes. This is in
agreement with the results from \cite{Maccio07}, who find the trend of
spin with halo mass to be less distinctive and consistent with zero
for SO-haloes.

\section{Non-radiative simulation results} \label{Sec_NR}

\subsection{Intrinsic differences between dark matter and gas} \label{Sec_NRzdep} 

At high redshift, baryons experience the same gravitational torques
from the surrounding density field as dark matter, and thus are
assumed to have identical `initial' spins \citep{Fall80}. In the
absence of any additional physical processes such as star formation
and feedback one then naively expects the gas to sustain its initial
spin and the correspondence with the dark matter spin.

To test the assumption of gas and dark matter having identical initial
spin distributions, \cite{vdBosch02} performed non-radiative
cosmological simulations that they evolved until a redshift of
$z=3$. Fitting the standard lognormal function to the measured Bullock
spin parameter distributions of the dark matter and gas components of
their 378 SO-haloes, \cite{vdBosch02} derive mean dark matter and gas
spin values of $\lambda_{0,\rm dm} = 0.040$ and
$\lambda_{0,\rm gas} = 0.039$, respectively, confirming that gas and
dark matter have identical `initial' spins\footnote{Note that there is
  no clear definition of `initial' spin to be measured at a distinct
  cosmic time, as structures acquire their angular momentum
  continuously by tidal torques from the surrounding large scale
  gravitational field. However, `initial' spin usually refers to the
  spin structures have before strong non-linear interactions and
  galaxy formation physics start playing a significant role.}.

In the upper panel of Fig.~\ref{FigNRZ}, we show as solid lines the
`initial' dark matter and gas Peebles spin parameter distributions
from the non-radiative Illustris-2-NR simulation at $z=8$ and confirm
this result also for the Peebles spin parameter distribution. The
derived distributions have median Peebles spin parameters of
$\lambda_{{\rm dm},z=8} = 0.0335$ and
$\lambda_{{\rm gas},z=8} = 0.0324$, respectively, which are equal
within statistical fluctuations. The dashed lines in Fig.~\ref{FigNRZ}
show the same spin parameter distributions at $z=0$ derived from
$\sim 65,000$ FOF-haloes from Illustris-2-NR. Whereas the spin
parameter distribution of the dark matter component exhibits only
slightly higher values than the `initial' spin distribution with a
median of $\lambda_{{\rm dm},z=0} = 0.0377$ at $z=0$, consistent with
the trend in the dark matter only Illustris-1-Dark simulation, the gas
component has evolved to substantially higher values with a median
spin parameter of $\lambda_{{\rm gas},z=0} = 0.0493$. At $z=0$ this
yields a ratio of the median gas to dark matter spin parameter of
$\lambda_{\rm gas} / \lambda_{\rm dm} = 1.308$, i.e.~a $\sim 30\%$
higher spin of the gas than the dark matter.

Similar results have been obtained by \citet{Chen03} and
\citet{Sharma05} who performed non-radiative $\Lambda$CDM simulations
and estimated the mean gas and dark matter spin parameter values from
fitting the lognormal to the derived Bullock spin parameter
distributions. They arrive at a ratio of mean gas to dark matter spin
equal to $\lambda_{0,\rm gas} / \lambda_{0,\rm dm} = 1.19$ and
$\lambda_{0,\rm gas} / \lambda_{0,\rm dm} = 1.44$, respectively, using
however much smaller halo sample sizes of 48 and 41 SO-haloes,
respectively.  The small sample sizes are likely responsible for the
variations in the reported size of the effect, but the general trend
of having a higher specific angular momentum at $z=0$ in the gas
compared to the dark matter is consistent. \citet{Gottloeber07} quote
similar results for more than $10,000$ cluster-sized FOF-haloes with
masses larger than $ 5\times 10^{13}h^{-1} {\rm M}_\odot$.  Applying
the same method as above, modulo some uncertainty by applying the
Bullock spin parameter definition to FOF-haloes (see
Section~\ref{Sec_SpinDefs}), they find a spin parameter enhancement of
$\lambda_{0,\rm gas} / \lambda_{0,\rm dm} = 1.32$.

\cite{Sharma12} suggested that the different mechanisms by which dark
matter and gas achieve equilibrium lead to an inside-out transport of
angular momentum in the dark matter component and an outside-in
transport in the gas component. The inside-out transport of angular 
momentum in the dark matter by dynamical friction of mergers entering 
the inner halo was also already observed by \cite{Zavala08}.
As the dark matter is insensitive to
hydrodynamic interactions, and the total gas mass available for
gravitational interactions is small compared to the dark matter mass,
the processes taking place in the two components are largely decoupled
from each other. As SO-haloes exclude the outer regions of
gravitationally bound haloes (compare Fig.~\ref{Figfofso}), the
different transport mechanisms described by \cite{Sharma12} might
explain why the gas to dark matter spin ratio becomes larger than
unity for SO-haloes.

When the outer regions are fully included, such as in FOF-haloes, the
gas to dark matter spin ratio would be expected to approach unity
again in this picture.  However, we find an enhanced gas to dark
matter spin ratio of the same order of magnitude for FOF-haloes as
well, suggesting that different radial redistribution mechanisms of
angular momentum in the two components provide an insufficient
explanation. Instead, there must be an additional mechanism by which
gas acquires more specific angular momentum than dark matter.

In Fig.~\ref{FigNRZ} we show how the Peebles spin parameter
distributions of the dark matter (middle panel) and the gas (lower
panel) components of FOF-haloes from the Illustris-2-NR simulation
evolve with redshift. Black dots mark the median spin parameters at
the different redshifts. The dark matter spin parameter distribution
exhibits the same self-similarity in time as in the dark matter only
Illustris-1-Dark simulation. However, the spin parameter distribution
of the gas gradually shifts to higher spin values with decreasing
redshift, illustrating a continuous specific angular momentum
acquisition in the gas component throughout cosmic time.

This acquisition could be explained by mergers getting ram pressure
striped during infall, which leads to a decoupling of their gas and
dark matter components. The displacement of the centres of mass of the
two components produces a mutual torque of the gas and dark matter
components onto each other, allowing a net transfer of angular
momentum from the dark matter to the gas.
Hydrodynamic shocks and other instabilities occurring in the gas are then crucial for the redistribution of angular momentum inside the gas component, but cannot directly account for transfer of specific angular momentum between dark matter and the gas, as dark matter couples only gravitationally to the gas.
The gain of specific angular momentum in the gas is evident in the shift of the gas spin parameter distribution, whereas the corresponding loss of angular momentum in the dark matter distribution is barely visible, as its mass fraction is much larger than that of the gas.
Compared to the dark matter spin in the dark matter
only Illustris-1-Dark simulation, the gas component gains more
specific angular momentum than the dark matter by a factor of
$1.26$. To compensate for this gain in the gas component, given a
cosmic baryon fraction of $f_\text{b} = \Omega_b/\Omega_m = 0.1673$,
the dark matter component in the non-radiative run has to transfer
$\sim 5.2\ \%$ of its initial spin to the gas. The median dark matter
spin parameter in the dark matter only Illustris-1-Dark simulation is
$\lambda_{\rm DM} = 0.0391$, which leads to an expected reduced spin
parameter of the dark matter component in the non-radiative simulation
of $\lambda_{\rm NR,exp} =0.0371$ at $z=0$. In Illustris-2-NR we
measure a median spin of $\lambda_{\rm NR} =0.0377$ (also see
Fig.~\ref{Figdmcomp}), but as we observe a weak trend of spin with
halo mass (see Section~\ref{Sec_NRmassdep}), we expect such a
deviation caused by the different mass cuts in Illustris-1-Dark and
Illustris-2-NR.

\begin{figure}  
\centering
\includegraphics[width=0.49\textwidth,trim= 0 10 0 0,clip]{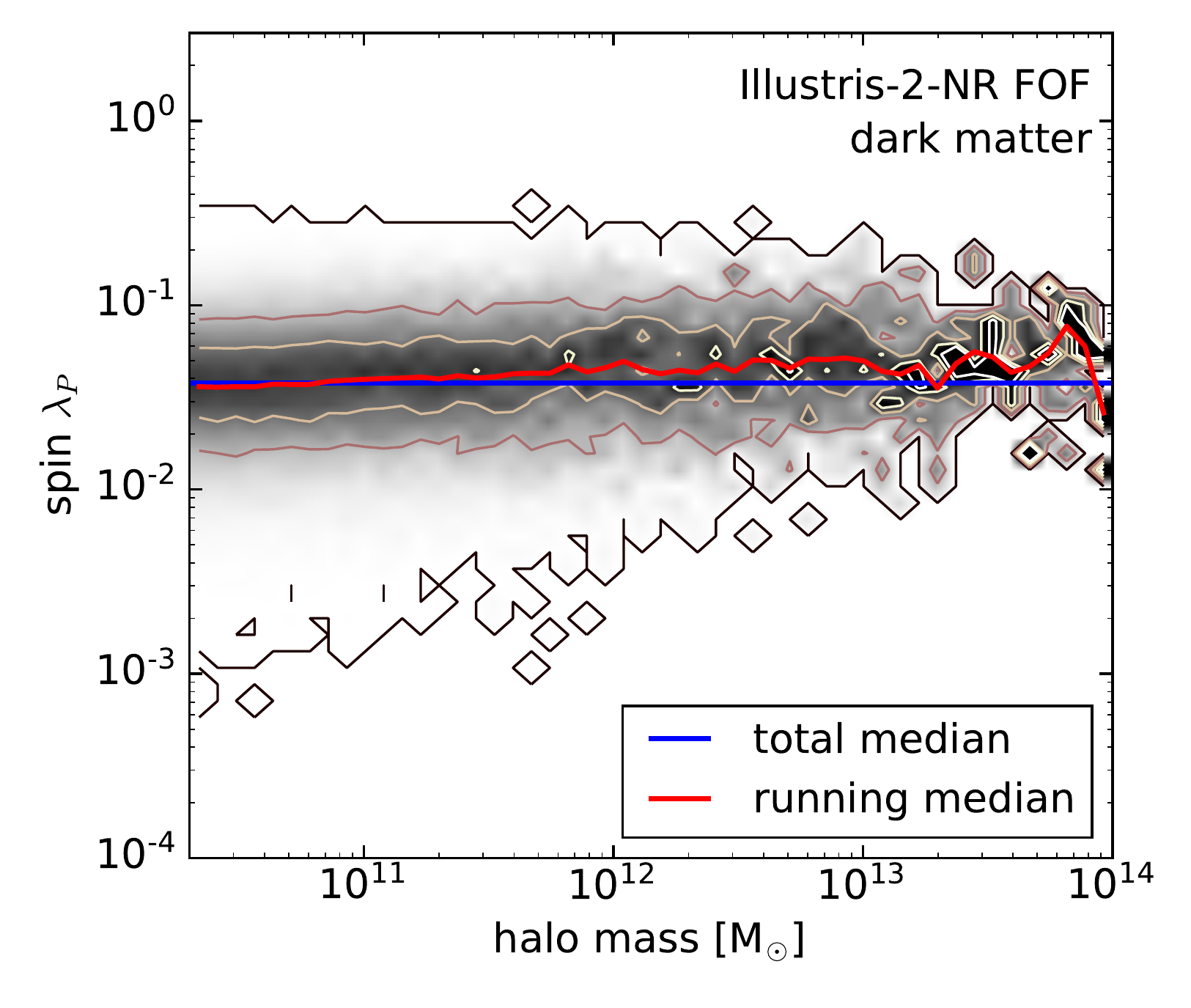}
\includegraphics[width=0.49\textwidth,trim= 0 10 0 0,clip]{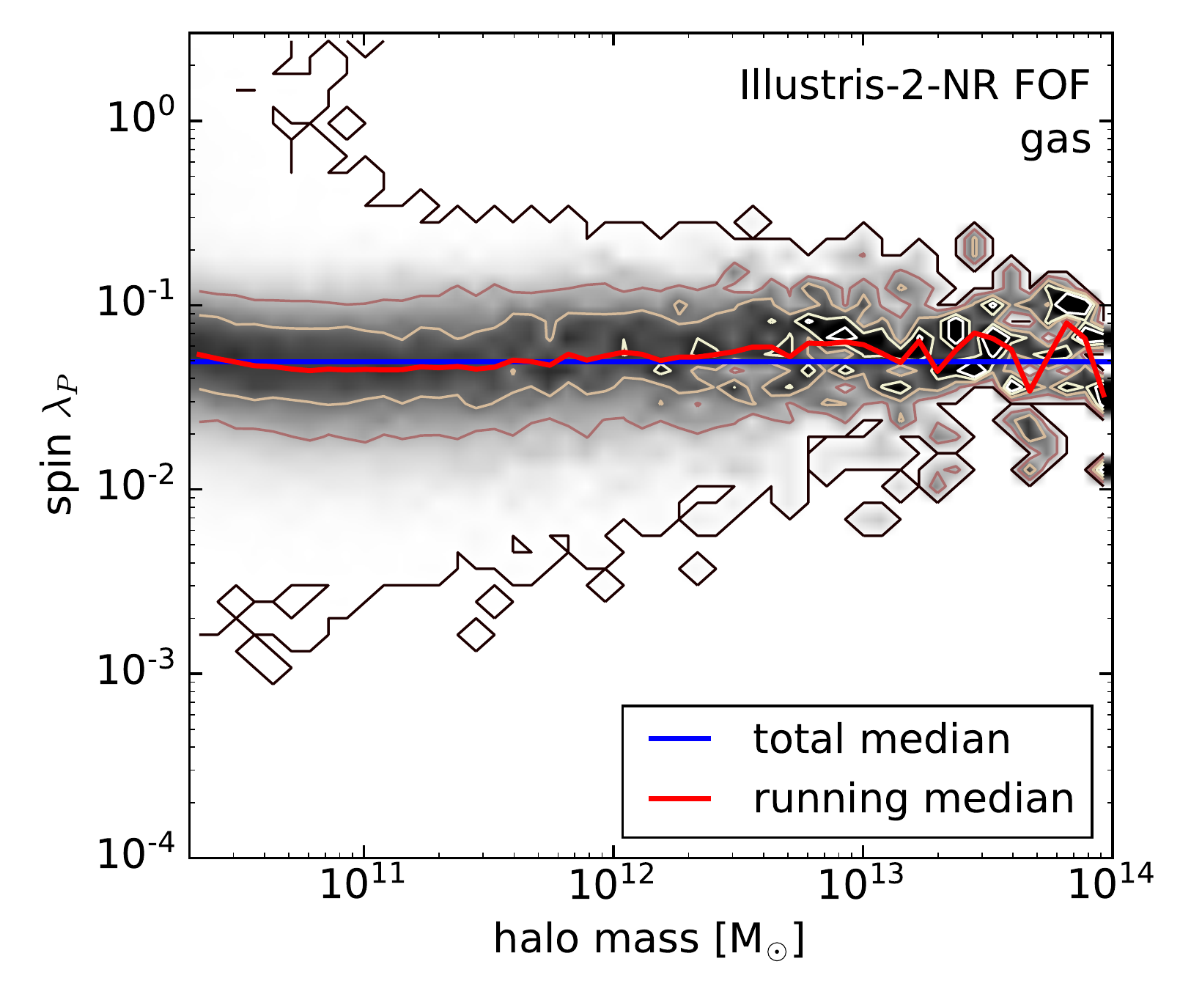}
\caption{Distribution of the dark matter (upper panel) and gas (lower
  panel) Peebles spin parameter as a function of FOF-halo mass
  $M_{\rm FOF}$ for Illustris-2-NR, normalised in every mass bin. The
  contours are drawn at constant fractions of $0$, $0.05$, $0.1$,
  $0.15$, and $0.2$, respectively. The total median spin parameter is
  shown as blue line, the median of every mass bin as red line. The
  dark matter spin shows the same trend as in Illustris-1-Dark; the
  gas spin follows this dark matter trend. \label{FigNRMass}}
\end{figure}

Furthermore, as we select our haloes based on a minimum number of dark matter particles only, some haloes in our sample have poorly resolved gas components. To quantify the impact from such objects, we show as cyan line in the upper panel of Fig.~\ref{FigNRZ} the gas spin parameter distribution obtained from haloes whose gas component is resolved by at least $300$ gas cells. 
This spin distribution is shifted to slightly smaller spin values compared with the gas distribution from the full sample, but otherwise exhibits the same behaviour. For this restricted sample we find a median value of $\lambda^*_{\rm gas,z=0} = 0.0458$, which corresponds to an enhancement factor of $1.17$ with respect to dark matter. The expected dark matter spin in the non-radiative simulation is then $\lambda^*_{\rm NR,exp} = 0.0376$, and thus almost identical to the measured value.  
However, as poor resolution affects small mass haloes, restricting the sample based on this criterium introduces an even larger discrepancy in the compared mass ranges, which is a source of bias due to the non-negligible trend of spin with halo mass.

Another possible mechanism that could contribute to the gain of
specific angular momentum by the gas component is if there is a
preferred orientation of the rotation axis of the gas component
perpendicular to the infall directions of merging matter. As we show
in Section~\ref{Sec_FPmis}, the gas and dark matter component of
FOF-haloes are on average misaligned by $\sim 35^\circ$, such that
given this preferred orientation the orbital angular momentum of
mergers could on average spin up the gas more than the dark
matter. Whether such a preferred orientation of the gas component with
infalling matter however exists and what may cause it is left for a future
investigation.

\subsection{Dependence of spin on halo mass} \label{Sec_NRmassdep}

In Fig.~\ref{FigNRMass} we show the dependence of the dark matter
(upper panel) and gas (lower panel) Peebles spin parameter of
FOF-haloes from Illustris-2-NR on halo mass $M_{\rm FOF}$. The
two-dimensional histogram was obtained in the same way as for
Fig.~\ref{FigDMMass}. The dark matter exhibits the same trend as
already observed in the dark matter only Illustris-1-Dark simulation,
with the median Peebles spin parameter increasing slightly with halo
mass. The spin parameter of the gas component, though being somewhat
higher, follows the same mass trend as observed for the dark
matter. The small upward trend in the least massive mass bins is
likely due to resolution effects, as the haloes are selected based on
being resolved by at least 300 dark matter particles, but there is no
limit on the minimum number of gas cells, such that the gas spin
parameter of the least massive haloes can be in principle based on
only a few dozen cells in gas poor haloes.

\section{Full physics simulation results} \label{Sec_FP}

\subsection{Dark matter spin statistics} \label{Sec_FPdm}

In this section we present the angular momentum properties of the dark
matter component of FOF-haloes from the full physics Illustris-1
simulation and compare them to the dark matter properties derived from
the dark matter only Illustris-1-Dark simulation. Illustris-1
comprises $\sim 320,000$ FOF-haloes fulfilling our selection criteria.

\begin{figure}
\centering
\includegraphics[width=0.49\textwidth,trim= 10 10 0 0,clip]{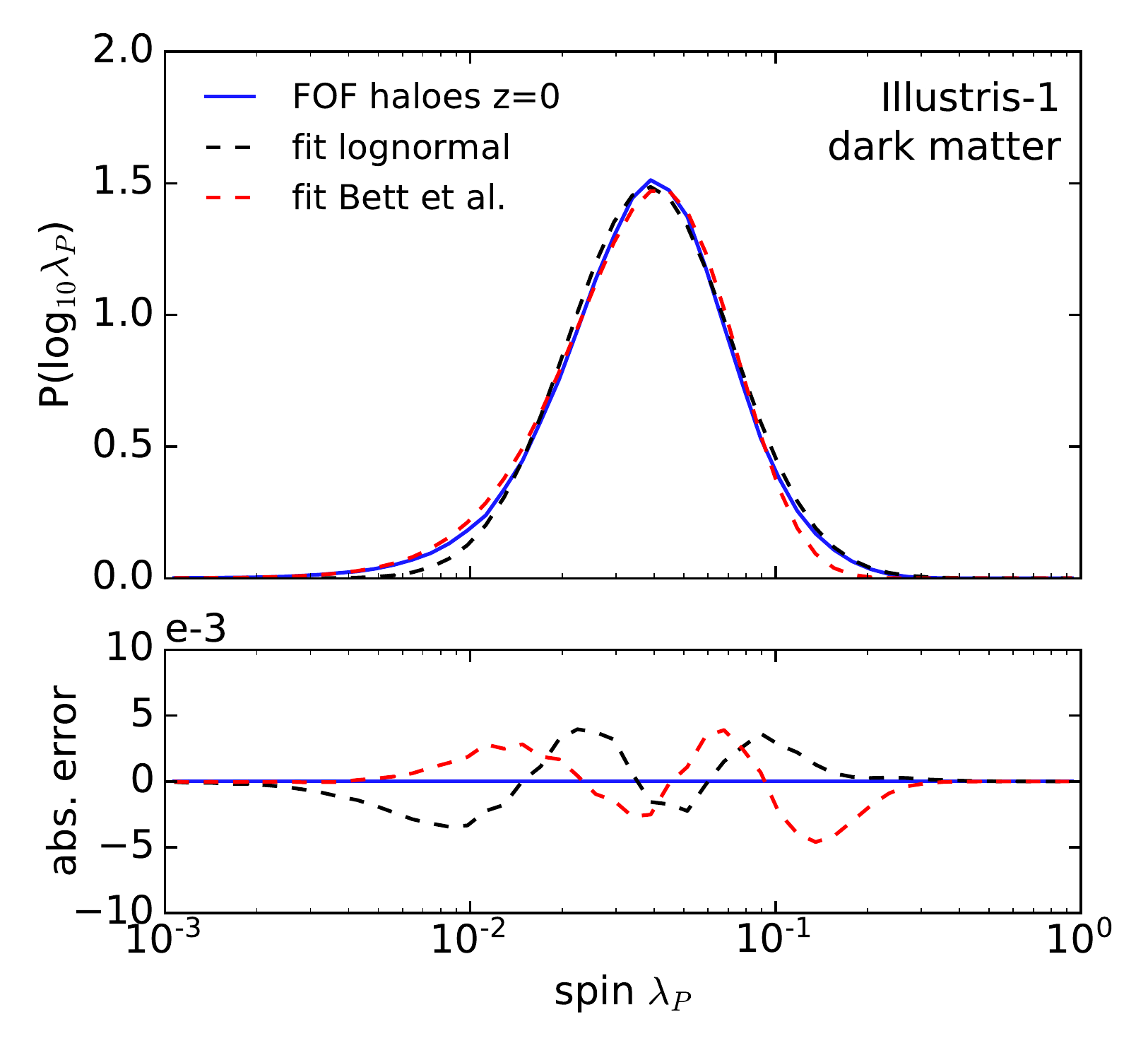}
\caption{Peebles spin parameter distributions of the dark matter
  component (blue) of FOF-haloes from Illustris-1 at $z=0$. The
  distribution is fitted with a lognormal function (dashed black) and
  a fitting function proposed by \citet{Bett07} (dashed red). For
  better comparison we show the absolute errors of the fitting
  functions with respect to the distribution obtained from
  Illustris-1.\label{FigFPdmz0fit}}
\end{figure}

\begin{figure} 
\centering
\includegraphics[width=0.5\textwidth,trim= 10 10 0 0,clip]{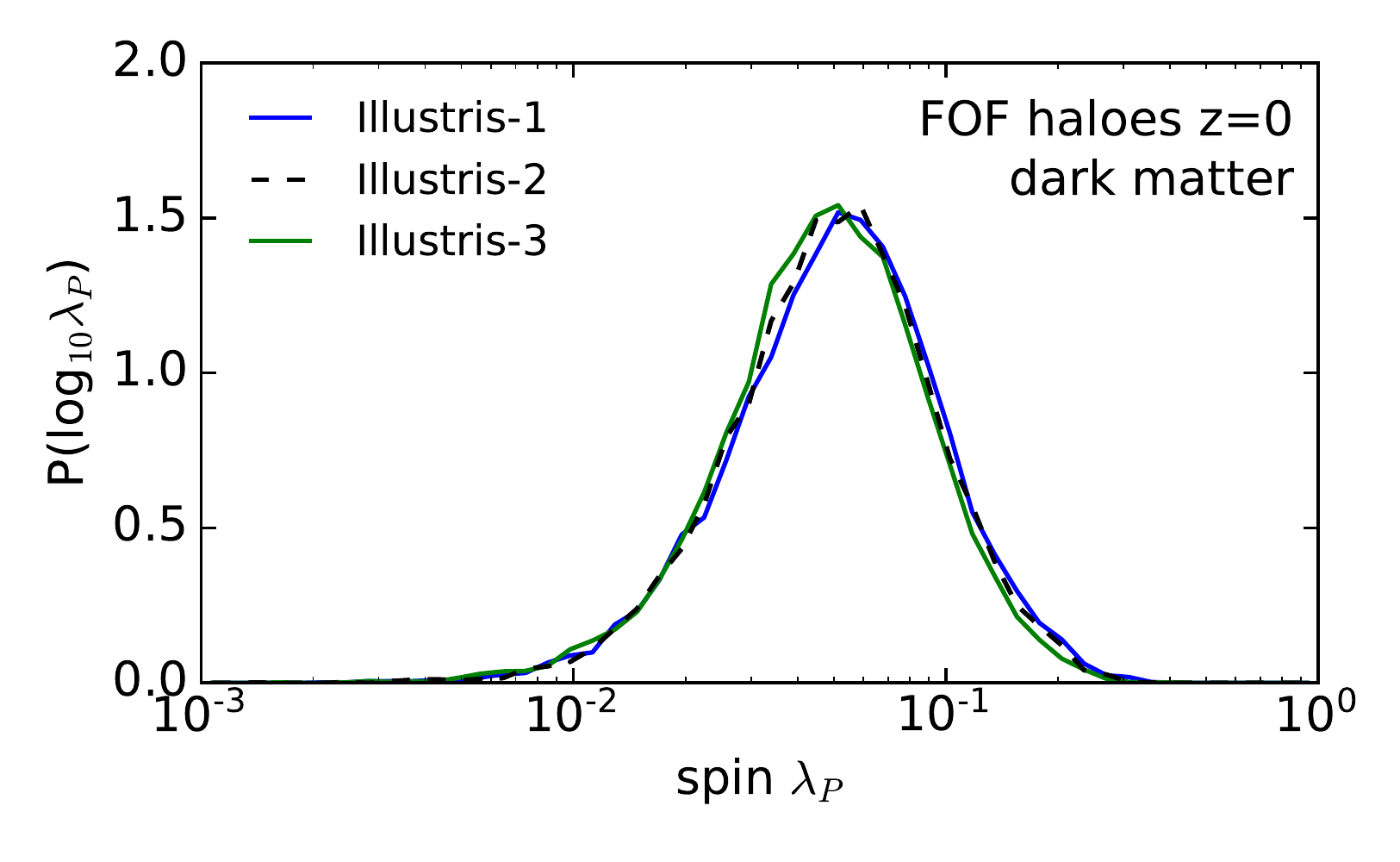}
\caption{Peebles spin parameter distributions of the dark matter
  component of FOF-haloes for the three different resolutions of the
  full physics {\em Illustris} simulations (see Tab.~\ref{TabSims}) at
  $z=0$. We find good convergence with the small deviations
  originating in the limited halo sample size. \label{FigFPdmconv}}
\end{figure}

In Fig.~\ref{FigFPdmz0fit} we show analytic fits of the lognormal and
the fitting function proposed by \citet{Bett07} to the Peebles spin
parameter distribution of the dark matter component at $z=0$. The best
fit parameters as well as the root mean square errors of the fits are
listed in Tab.~\ref{TabFPdmfit}. These parameters differ only
insignificantly from the best fit parameters derived for the Peebles
spin parameter distribution of dark matter only FOF-haloes from
Illustris-1-Dark that are listed in Tab.~\ref{TabDMfit}. Furthermore,
we show the convergence of the Peebles spin parameter distribution of
the dark matter component for the three different resolutions (see
Tab.~\ref{TabSims}) of the full physics {\em Illustris} simulations in
Fig.~\ref{FigFPdmconv}. We find very good convergence, of the same
quality as for Illustris-Dark, with the small deviations originating
in the limited halo sample size.

\begin{figure}
\centering
\includegraphics[width=0.5\textwidth,trim= 10 10 0 0,clip]{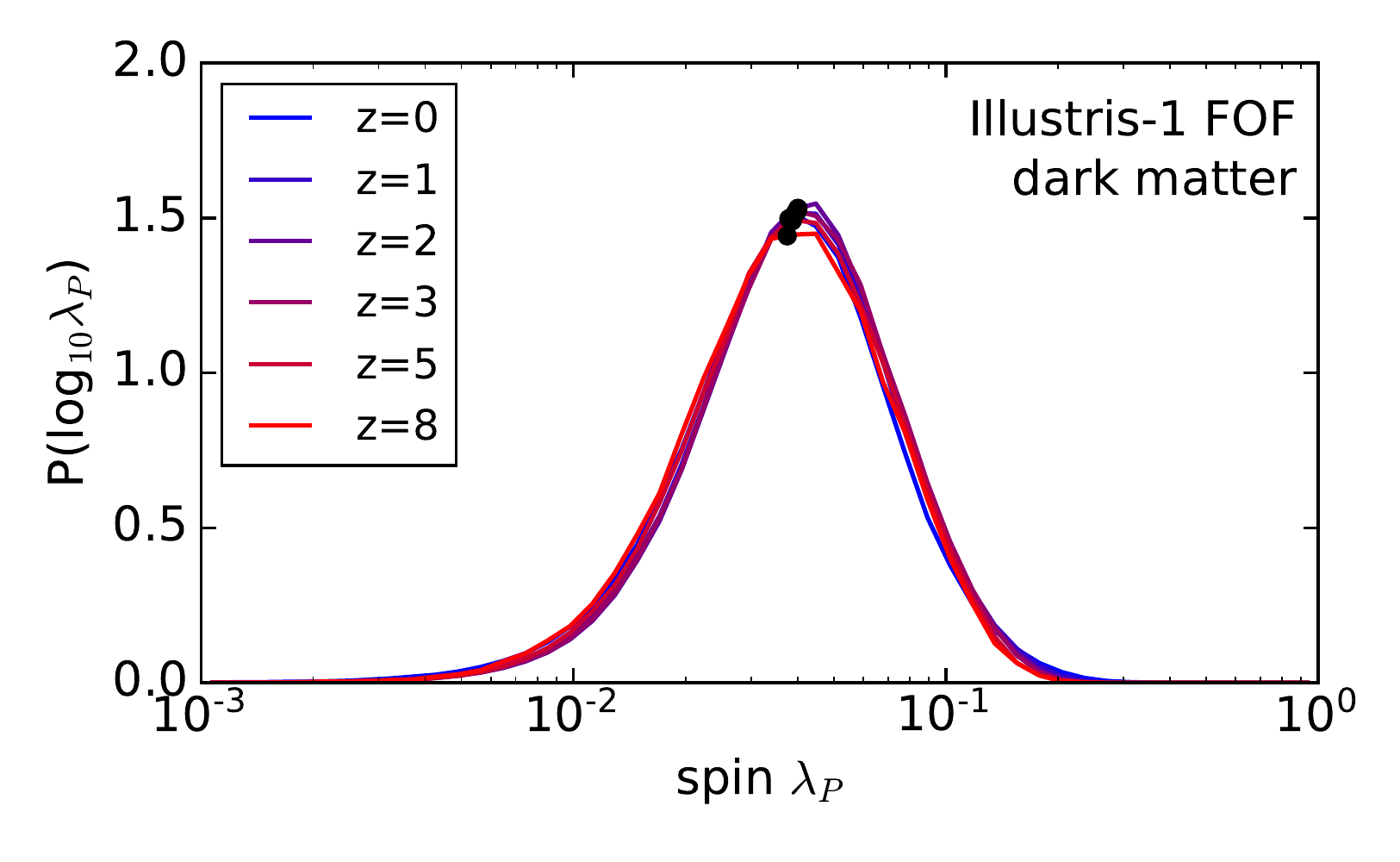}
\caption{Redshift evolution of the Peebles spin parameter distribution
  of the dark matter component of FOF-haloes from the full physics
  Illustris-1 simulation. Black dots mark the median spin parameter at
  every redshift. The spin parameter distribution is self-similar in
  time confirming the dark matter only results. \label{FigFPdmz}}
\end{figure}

The redshift dependence of the dark matter Peebles spin parameter
distribution from Illustris-1 is shown in Fig.~\ref{FigFPdmz}. Black
dots mark the median spin parameter at every redshift. We find the
same behaviour of the distribution as in the dark matter only
Illustris-1-Dark, in the form of an almost perfect self-similarity in
time. A small trend of the median spin towards higher values with
decreasing redshift is again present, as already observed in the dark
matter spin distributions from both Illustris-1-Dark and
Illustris-2-NR.

\begin{table}
\centering
\begin{tabular}{|l|l|l|l|}
\hline
fit lognormal & $\lambda_0 = 0.039$  & $\sigma = 0.27$ & $\epsilon_{\rm rms}= 0.0018$ \\ 
fit Bett et al. & $\lambda_0 = 0.042$ & $\alpha = 3.30$ & $\epsilon_{\rm rms}= 0.0018$ \\ \hline
\end{tabular}
\caption{Best fit parameters of the analytic fits to the Peebles 
  spin parameter distributions of the dark matter
  component derived from Illustris-1. 
  The fit parameters are remarkably similar to the parameters
  derived for Illustris-1-Dark 
  listed in Tab.~\ref{TabDMfit}, showing very good 
  convergence of the dark matter properties. \label{TabFPdmfit} }
\end{table}

\begin{figure} 
\centering
\includegraphics[width=0.5\textwidth,trim= 10 10 0 0,clip]{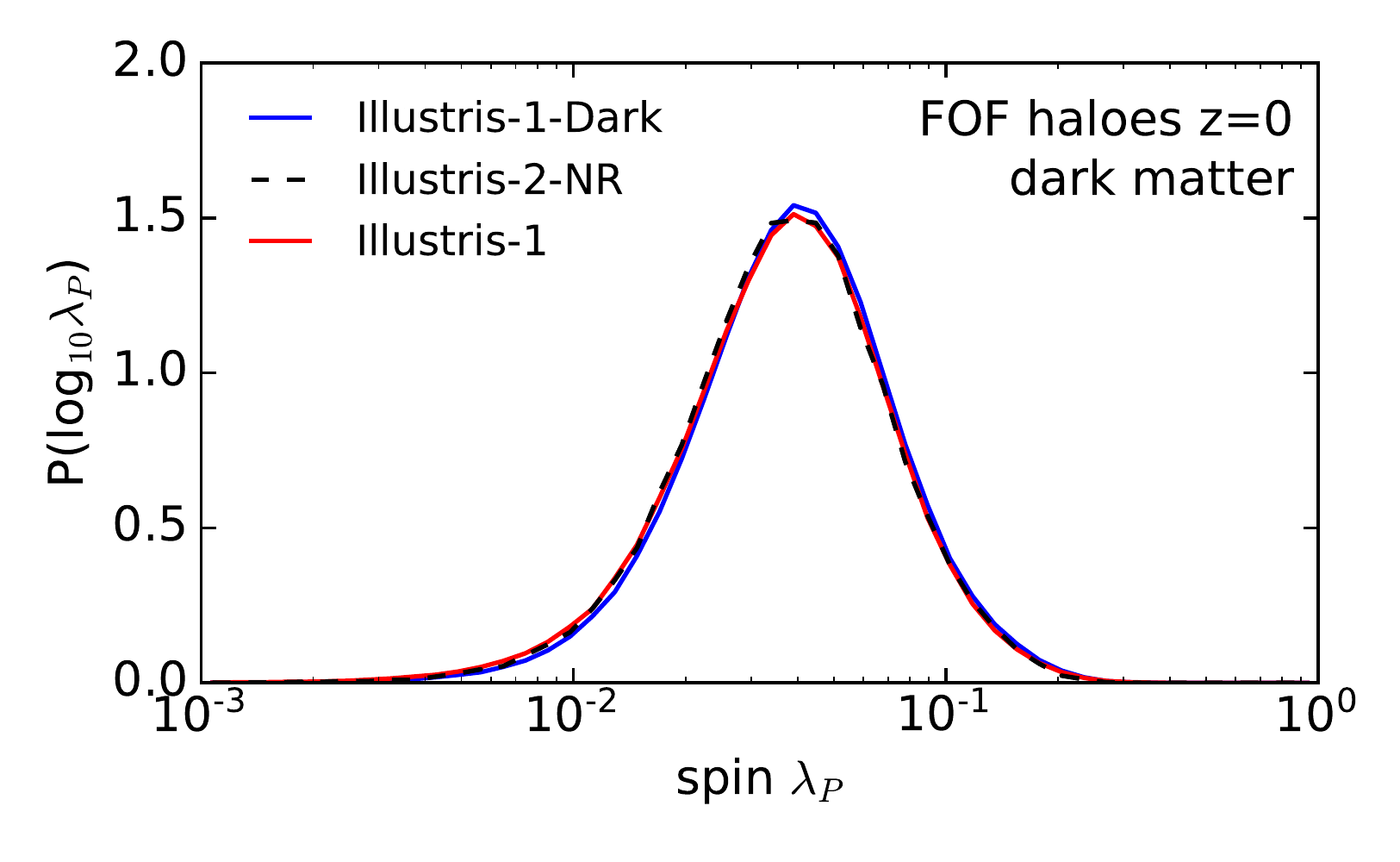}
\caption{Peebles spin parameter distribution of the dark matter
  components of FOF-haloes at $z=0$ from Illustris-1, Illustris-2-NR,
  and Illustris-1-Dark. The distributions derived from the three
  different types of simulations are remarkably similar. The median
  spin parameters are $\lambda_{\rm DM} =0.0391$ for Illustris-1-Dark,
  $\lambda_{\rm NR} =0.0377$ for Illustris-2-NR, and $\lambda =0.0379$
  for Illustris-1. Thus the simulations including baryons have a
  slightly smaller median dark matter spin than in the dark matter
  only simulation. The relative shift of the median values amounts to
  $3.6\%$ for Illustris-2-NR and $3.1\%$ for Illustris-1 compared to
  Illustris-1-Dark, consistent with a transfer of angular momentum
  from dark matter to gas during mergers. \label{Figdmcomp}}
\end{figure}

However, the presence of baryons does introduce subtle changes in the
dark matter component, which are barely visible at first sight. To
highlight this point, we show in Fig.~\ref{Figdmcomp} the dark matter
Peebles spin parameter distributions at $z=0$ derived from the three
different simulation types at the highest available resolution (see
Tab.~\ref{TabSims}), the dark matter only simulation Illustris-1-Dark,
the non-radiative Illustris-2-NR, and the full physics Illustris-1
simulation. The dark matter Peebles spin parameter distributions
derived from these three simulations are remarkably similar. The
median spin parameters at $z=0$ are $\lambda_{\rm DM} =0.0391$ for
Illustris-1-Dark, $\lambda_{\rm NR} =0.0377$ for Illustris-2-NR, and
$\lambda =0.0379$ for Illustris-1. The simulations including baryons
thus have slightly smaller median spin parameters than the dark matter
only one. The relative shift of the median values amounts to $3.6\%$
for the non-radiative Illustris-2-NR simulation and to $3.1\%$ for the
full physics Illustris-1 simulation compared to the dark matter only
Illustris-1-Dark. This shift of the dark matter spin distribution to
somewhat smaller values in the presence of a baryonic component can be
explained as a reflection of the transfer of angular momentum from the
dark matter to the gas during mergers, as discussed in
Section~\ref{Sec_NR}.

\subsection{Spin statistics of baryons in different halo regimes} \label{Sec_FPbary}

\begin{figure}
\centering
\includegraphics[width=0.5\textwidth,trim= 0 0 0 0,clip]{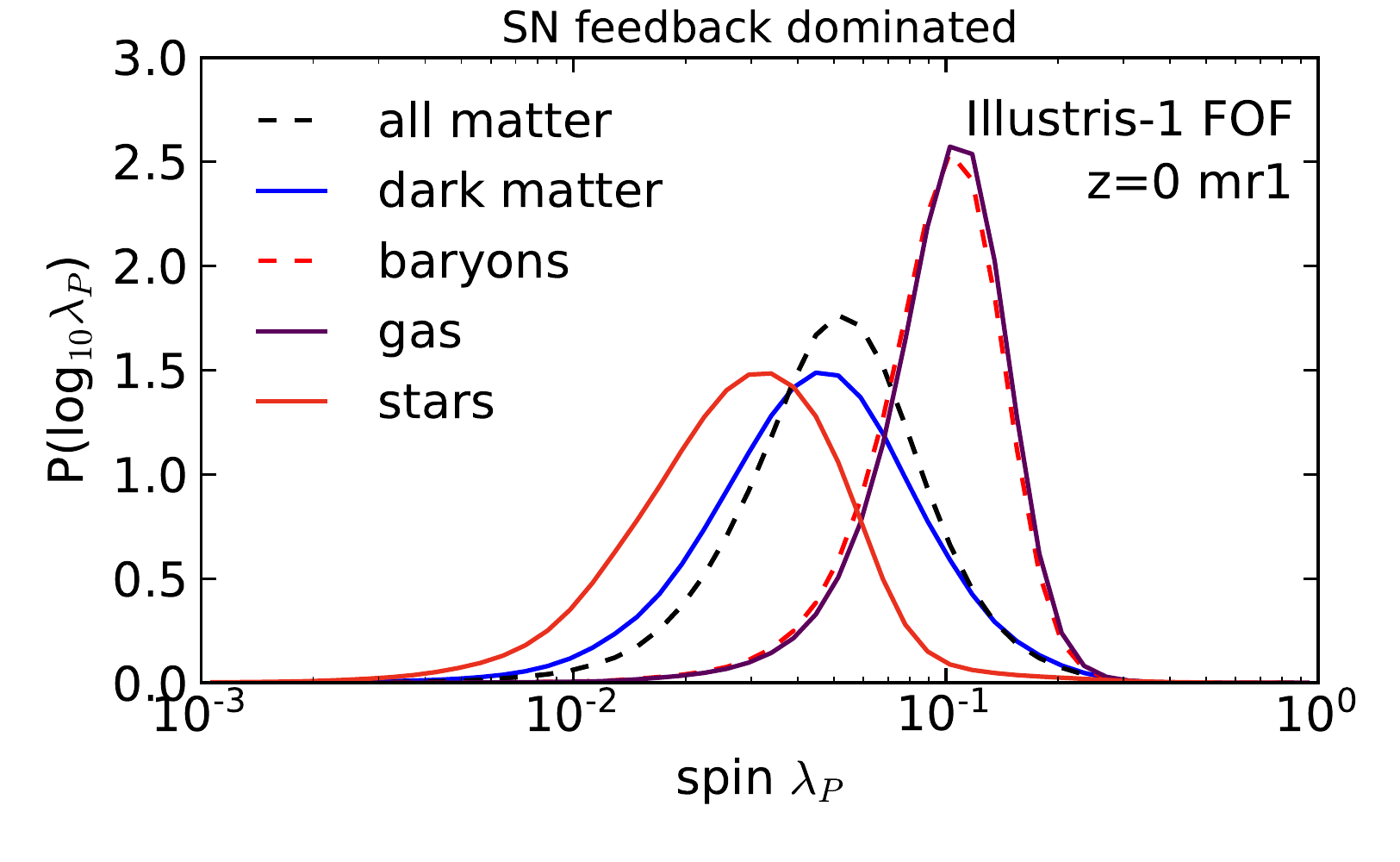}
\includegraphics[width=0.5\textwidth,trim= 0 0 0 0,clip]{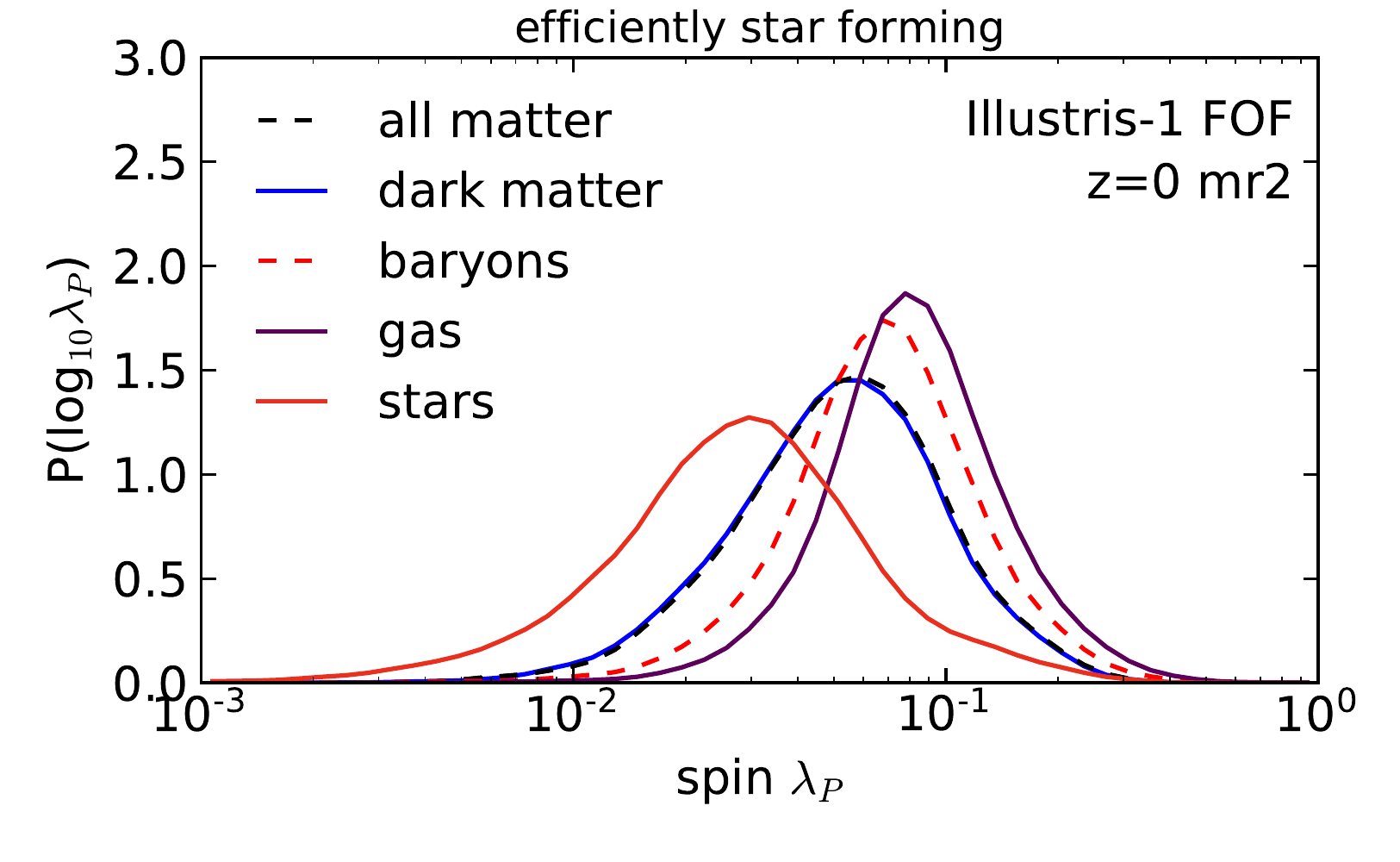}
\includegraphics[width=0.5\textwidth,trim= 0 0 0 0,clip]{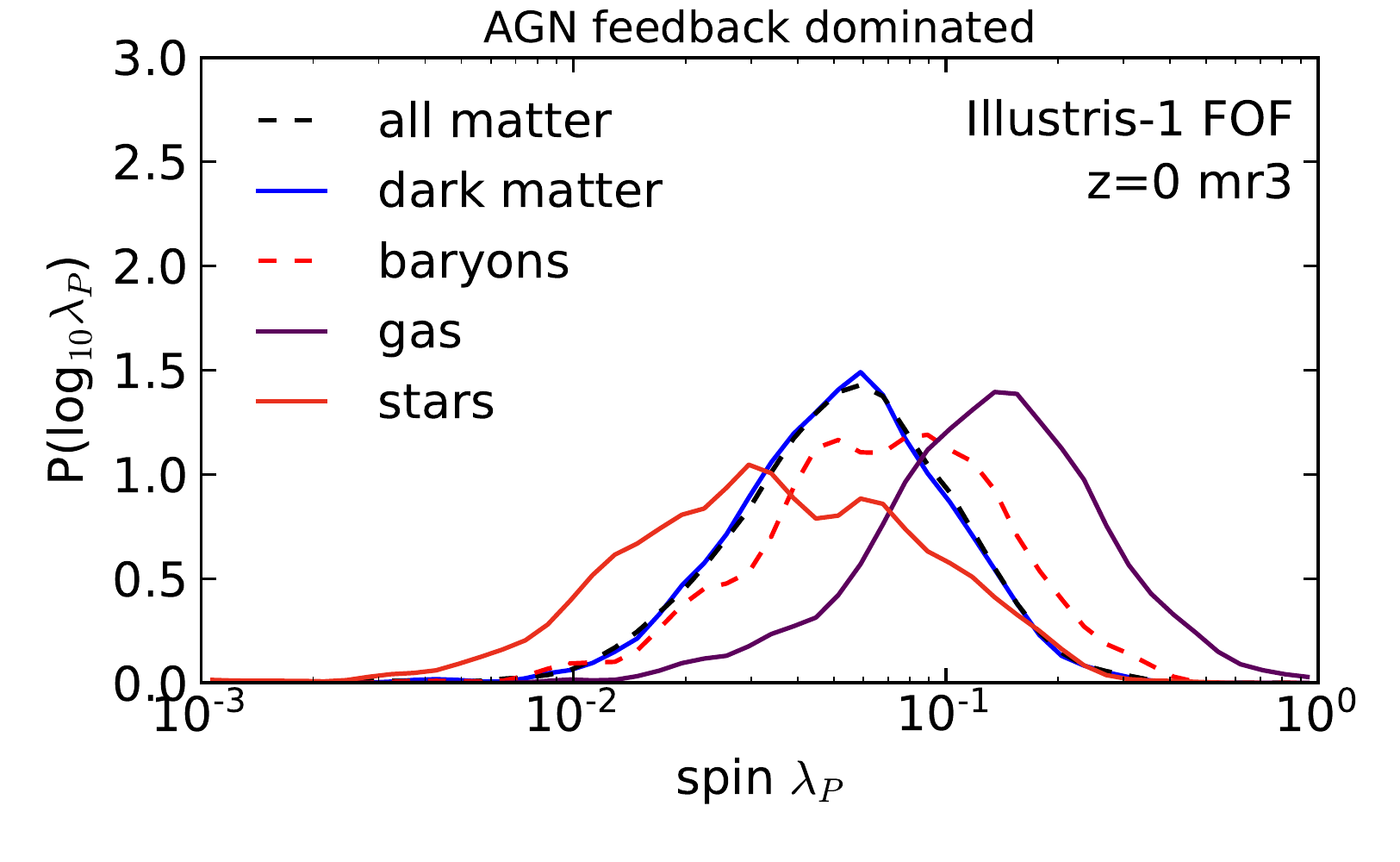}
\caption{Peebles spin parameter distributions of the dark matter, gas,
  and stellar component of FOF-haloes as well as the combined baryonic
  spin parameter distribution and the total spin parameter
  distribution in the three examined mass ranges 
  of the full physics simulation Illustris-1, as discussed in
  Section~\ref{Sec_FPbary}.  This figure demonstrates that galaxy
  formation physics breaks the self-similarity of the spin parameter
  distribution observed in dark matter only simulations.
  \label{FigFPbarySpinDist}}
\end{figure}

In this section, we investigate the spin statistics of gas and stars
within FOF-haloes from Illustris-1. The baryonic spin is strongly
affected by physical processes such as star formation and feedback,
which also indirectly impact the dark matter through the change in
baryonic density. These physical processes make the efficiency of
galaxy formation a strong function of halo mass, hence we expect the
baryonic spin to depend on halo mass as well. This motivates us to
define three mass ranges in which we are going to examine the spin
parameter distributions:
\begin{enumerate}
\item $[2\times 10^{10} {\rm M}_\odot,2\times 10^{11} $M$_\odot]$: SN feedback dominated
\item $[2\times 10^{11} {\rm M}_\odot,2\times 10^{12} $M$_\odot]$: efficiently star forming
\item $[2\times 10^{12} $M$_\odot,2\times 10^{13} $M$_\odot]$: AGN feedback dominated
\end{enumerate}
We discuss the association between these mass bins and the
individual feedback regimes further blow.

We show the Peebles spin parameter distributions of dark matter, gas,
and stars, as well as the combined baryonic and the total spin
parameter distributions for Illustris-1 in the above mass bins in
Fig.~\ref{FigFPbarySpinDist}. As the number of haloes increases
rapidly with decreasing halo mass (the number of haloes in the three
mass bins are 51383, 7286, and 834, respectively), the overall spin
statistic is dominated by haloes just above the lower limit of each
mass bin. To reduce noise in the highest mass bin, the spin parameter 
distributions in Fig.~\ref{FigFPbarySpinDist} have been smoothed with a 
Gaussian kernel with standard deviation of one bin in spin parameter space 
and truncated after four bins. The median Peebles spin parameter values in 
the different mass bins as well as the stellar to gas mass ratios are listed in
Tab.~\ref{TabFPmr}.

\begin{figure}  
\centering
\includegraphics[width=0.48\textwidth,trim= 0 10 0 0,clip]{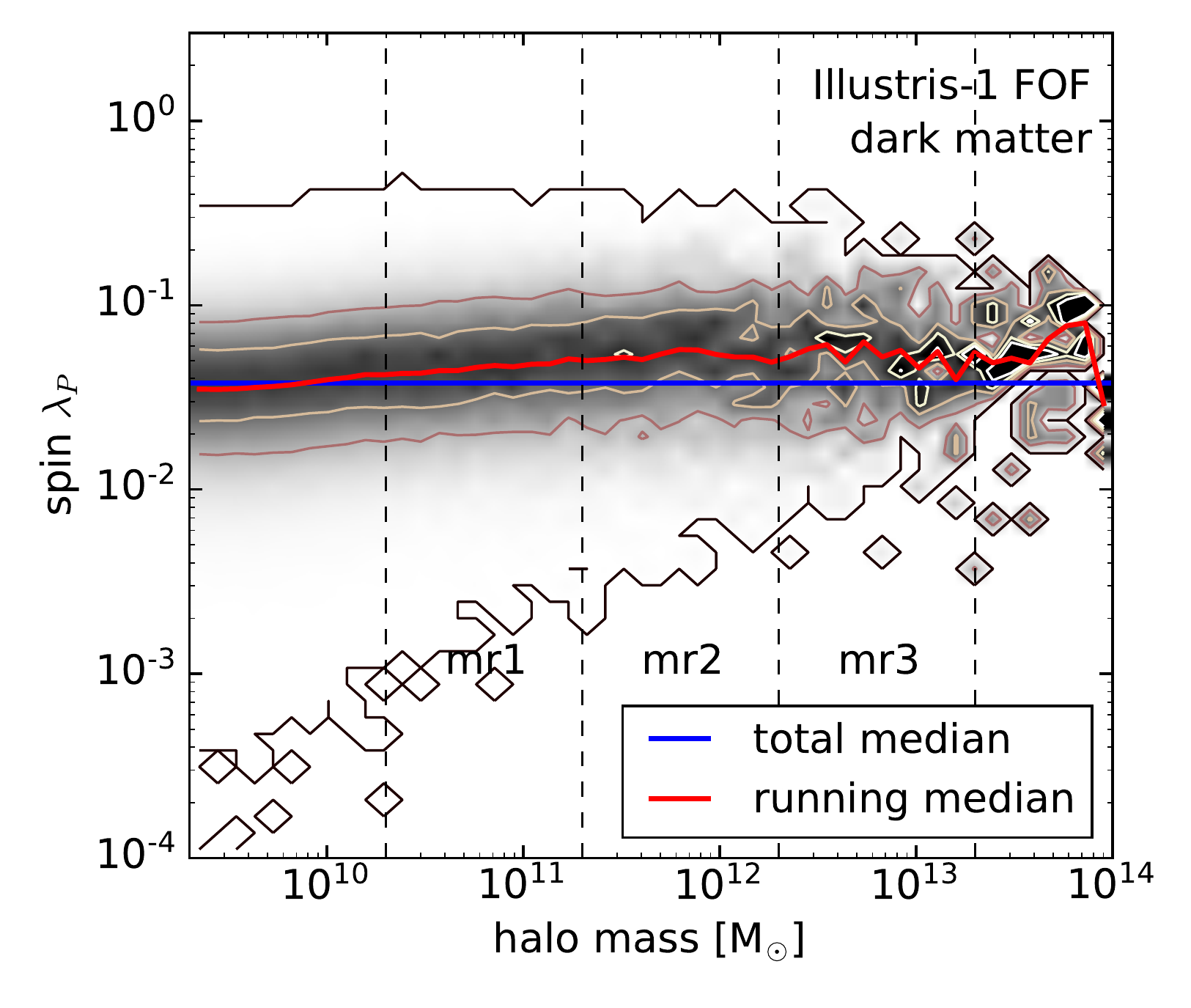}
\includegraphics[width=0.48\textwidth,trim= 0 10 0 0,clip]{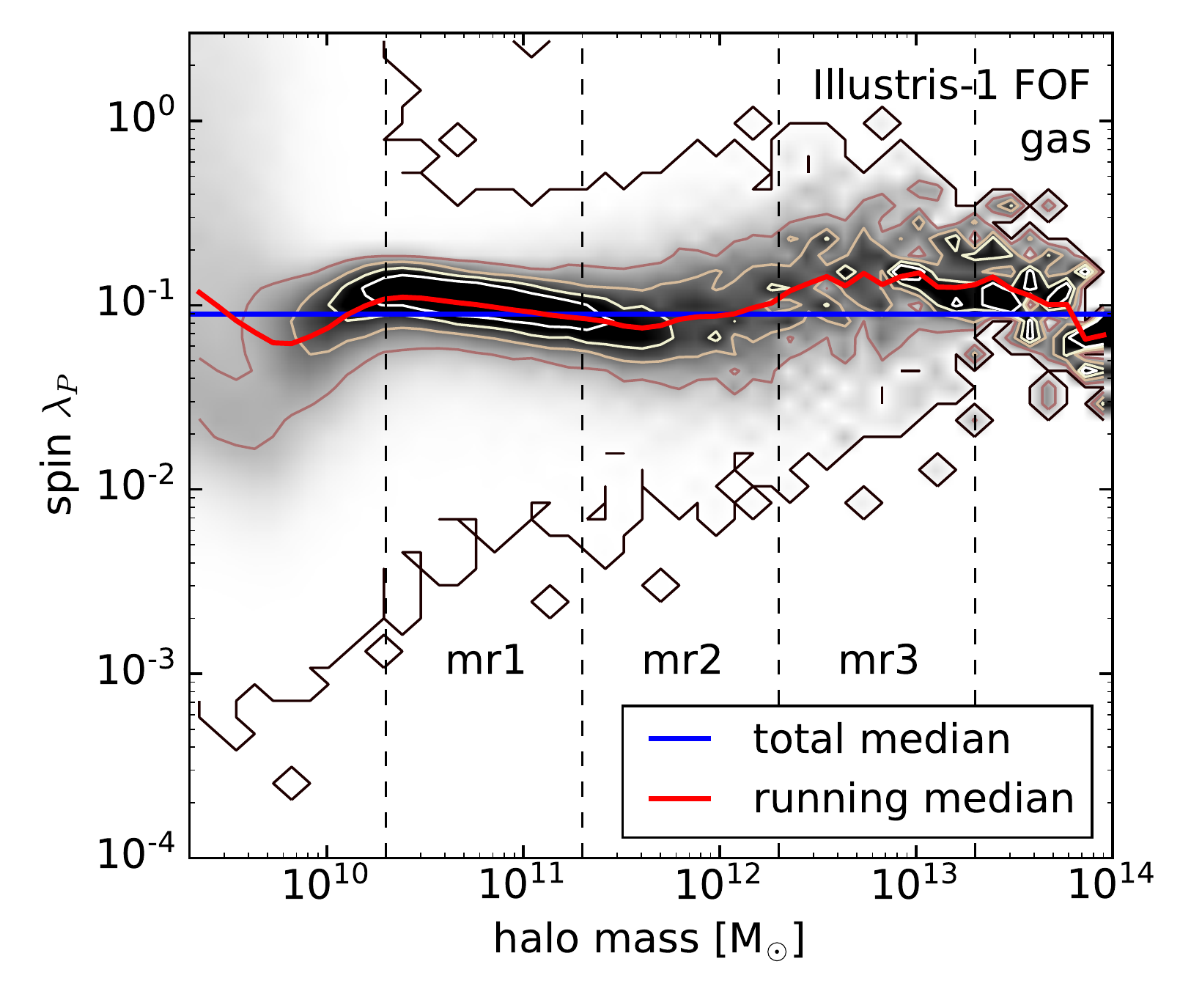}
\includegraphics[width=0.48\textwidth,trim= 0 10 0 0,clip]{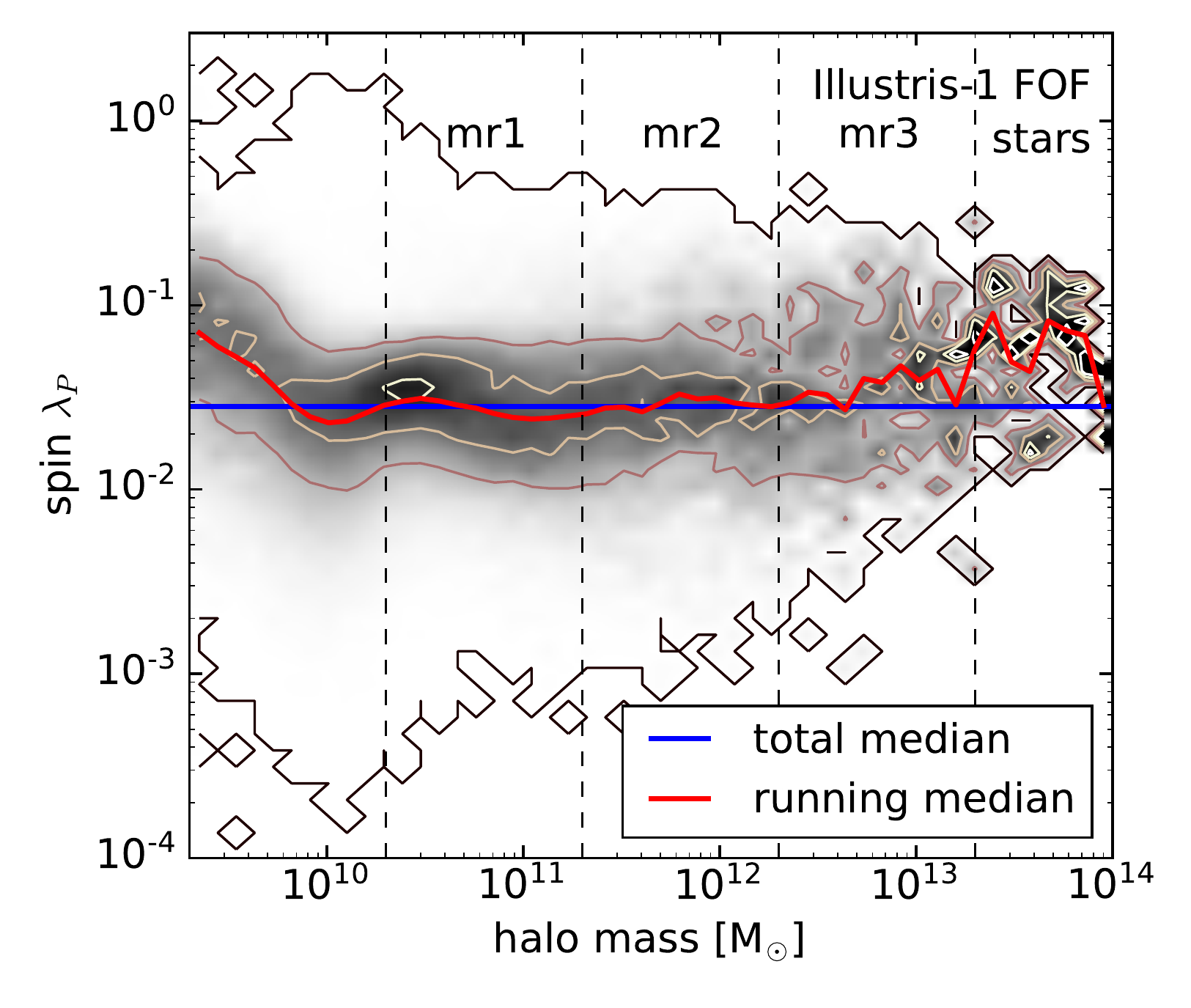}
\caption{Distribution of the dark matter (upper panel), gas (middle
  panel), and stellar (lower panel) Peebles spin parameter with halo
  mass $M_{\rm FOF}$ for Illustris-1 normalised in every mass bin. The
  grey shading indicates a fraction of $0$ to $0.2$ of all haloes in a
  given mass bin having a given spin. Contours are drawn at constant
  fractions of $0$, $0.05$, $0.1$, $0.15$, and $0.2$. The total median
  spin parameter is shown as a blue line, the median of every mass bin
  as a red line. The dark matter spin shows the same behaviour as in
  Illustris-1-Dark, the baryonic spin however exhibits a stronger
  trend with mass that is caused by the impact of feedback. We discuss
  this trend in more detail in
  Section~\ref{Sec_FPbary}. \label{FigFPMass}}
\end{figure}

Fig.~\ref{FigFPbarySpinDist} demonstrates that the scale-dependent
galaxy formation physics breaks the self-similarity of haloes with
respect to their spin distribution. In realistic simulations of galaxy
formation, the gas spin depends strongly on the halo mass (compare
also Fig.~\ref{FigFPMass}) and is on average about twice as high as
the dark matter spin, whereas the stellar spin is only about half as
large as the dark matter spin. The general trend of a higher spin of
the gas compared with the dark matter, and of a lower stellar spin, is
in agreement with findings from \cite{Teklu15} based on 622 haloes
with no restrictions on their dynamical state.

\begin{table}
\centering
\begin{tabular}{|l|c|c|c|c|c|c|}
\hline
range & all & dm & gas & baryons & stars & $\frac{M_{*}}{M_{\rm gas}}$  \\\hline
(i) \phantom{ii}mr1 & 0.0506 & 0.0449 & 0.102 \ & 0.0985 & 0.0288 & 0.07  \\ 
(ii) \phantom{i}mr2 & 0.0534 & 0.0524 & 0.0814 & 0.0681 & 0.0283 & 0.38  \\ 
(iii) mr3 & 0.0546 & 0.0541 & 0.133 \ & 0.0692 & 0.0346 & 1.61  \\ \hline
\end{tabular}
\caption{Median Peebles spin parameter values of the different
  components making up the halo for different halo mass ranges. \label{TabFPmr} }
\end{table}

\begin{figure}
\centering
\includegraphics[width=0.5\textwidth,trim= 0 0 0 0,clip]{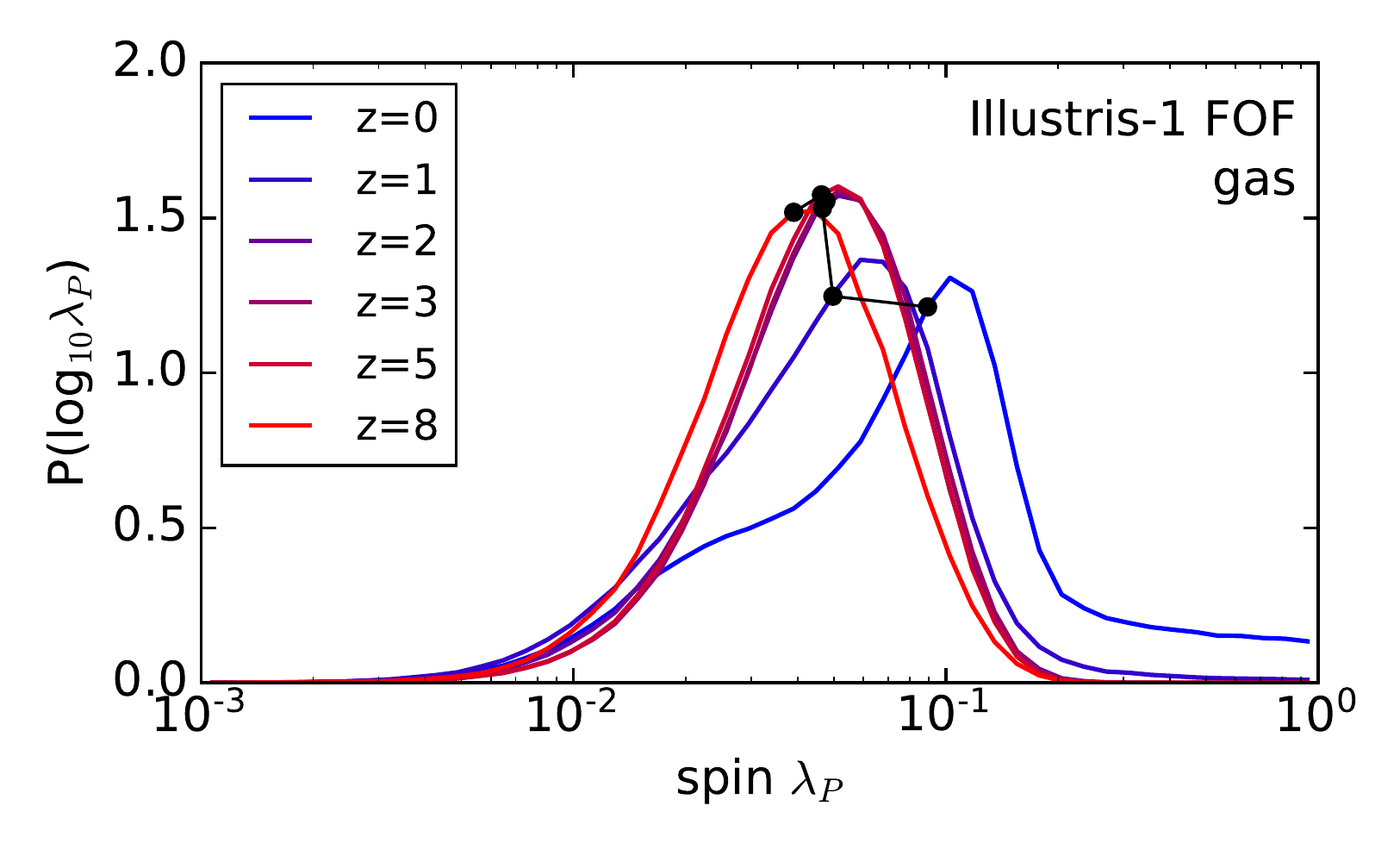}
\includegraphics[width=0.5\textwidth,trim= 0 0 0 0,clip]{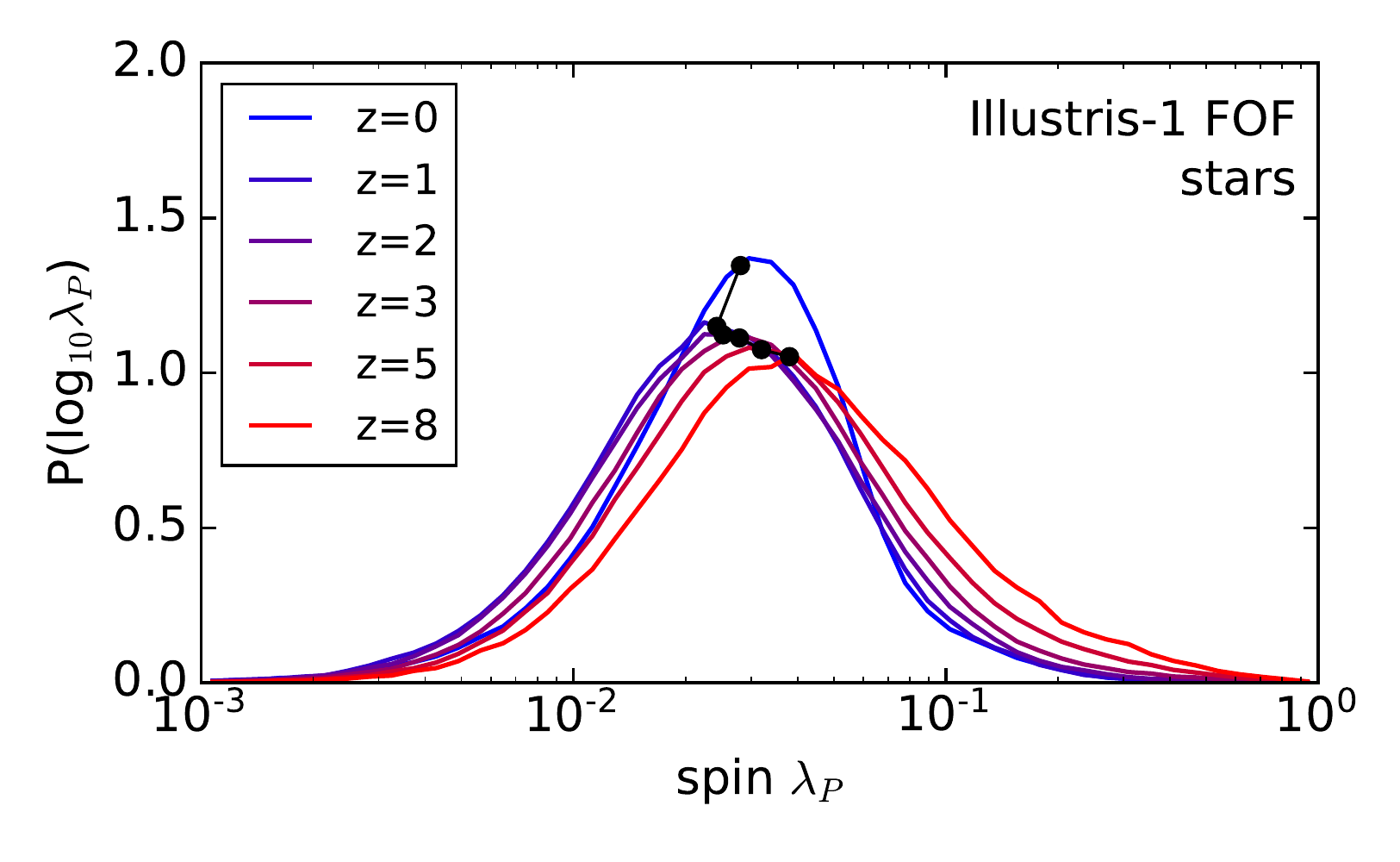}
\caption{Peebles spin parameter distribution of the gas and stellar
  component in the full physics Illustris-1 simulation as a function
  of redshift. Black dots mark the median spin parameter at every
  redshift. Gas increases its specific angular momentum content with
  cosmic time. The stellar component on the other hand evolves towards
  slightly smaller median spins. Both components exhibit changes in
  the shape of their spin distributions. \label{FigFPbaryz}}
\end{figure}

The baryonic spin distribution of low mass haloes (upper panel) is
almost completely determined by the gas spin, as those haloes contain
only few stars. With increasing halo mass, the stellar mass to gas
mass fraction increases rapidly, and the baryonic distribution shifts
to progressively lower spin values (middle and lower panels).  The
baryonic spin of all haloes from the three mass bins is on average a
factor of $\sim 1.8$ larger than the dark matter spin, which is
substantially more than the value of $\sim 1.3$ found in the
non-radiative Illustris-2-NR simulation. We investigate the origin of
this enhancement in detail in Section~\ref{Sec_FPgas}.  The total spin
of haloes is thus underestimated by dark matter only simulations by
$\sim 13\%$, $\sim 2\%$, and $\sim 1\%$ in the three bins,
respectively.

In Fig.~\ref{FigFPbaryz} we show how the Peebles spin parameter
distribution of the gas and stellar component of the full FOF-halo
sample from the full physics Illustris-1 simulation evolves as a
function of redshift. Black dots mark the median spin parameter at
every redshift. The specific angular momentum of the gas component
continuously grows with cosmic time. The shape of the gas distribution
is getting distorted from the classical lognormal after redshift $z=2$
due to a second bump emerging at the low tail of the
distribution. This low spin bump is caused purely by the smallest mass
haloes and vanishes completely if we enforce an additional mass cut 
on our halo sample at $2\times 10^{10} {\rm M}_\odot$, equal to the lower bound of the mr1 mass range. This bump could be explained by the gas component of low mass haloes not gaining
any angular momentum due to the lack of mergers and efficient feedback 
at dwarf halo masses, but also could be purely due to poor resolution 
of the gas component at these halo masses. The
stellar component exhibits a more subtle change in the shape of the
spin distribution but evolves to only slightly smaller average spin
with cosmic time.

In Fig.~\ref{FigFPMass} we show the detailed mass dependence of the
Peebles spin parameter of FOF-haloes from Illustris-1. This figure
shows a two-dimensional histogram where FOF-haloes have been binned
according to their mass and spin parameter and was obtained as in
Figs.~\ref{FigDMMass} and \ref{FigNRMass} with independently
normalised mass bins to account for the variation of halo number with
mass.  The spin parameter of the dark matter component (upper panel)
exhibits the same small increase of the median value with increasing
halo mass as observed in the dark matter only Illustris-1-Dark
simulation. However, the average spin parameters of the gas (middle
panel) and the stellar component (lower panel) show a more pronounced
trend with mass, which can be understood as a direct consequence of
the impact from feedback.

The two feedback processes perturbing the gas and altering the
distribution of baryons are galactic winds driven by supernova (SN)
explosions and AGN feedback. SNe occur in cold, dense, star-forming
gas regions in the inner halo, which have small specific angular
momentum. The wind velocity imparted on the gas by a SN event is taken
to be linearly proportional to the one-dimensional velocity dispersion
of the halo, as motivated by \cite{Okamoto10}. Details of the
prescription can be found in \cite{Vogelsberger13}. For low mass
haloes, the winds are more efficient in expelling a fraction of the
gas and depleting star formation in the inner regions of the halo. In
the {\em Illustris} simulation, SN feedback plays a major role below a
FOF-halo mass of $\sim 2\times 10^{11} {\rm M}_\odot$. Haloes above
this mass are able to retain most of their gas and to actively form
stars. Haloes with FOF-halo masses above
$\sim 2\times 10^{12} {\rm M}_\odot$ however grow massive enough black
holes such that AGN feedback becomes efficient in quenching their star
formation and expelling gas from the halo. As black holes are located
in the very centre of haloes, AGN feedback also tends to mostly expel
gas with low specific angular momentum from the inner halo.

Considering the halo masses at which the different feedback processes
are most efficient, it is easy to interpret the detailed trend of the
gas spin with halo mass. Below FOF-halo masses of
$\sim 2\times 10^{11} {\rm M}_\odot$, galactic winds become
increasingly effective in expelling low specific angular momentum gas,
which leads to an enhanced spin parameter of the remaining gas. Above
$\sim 2\times 10^{12} {\rm M}_\odot$, AGN feedback becomes strong and
efficient in expelling the same low specific angular momentum gas. In
between these two mass regimes the galaxies are able to hold on to
most of their gas and are efficiently star-forming. Correspondingly,
they have a lower than average gas spin parameter in this mass bin.

The trend of stellar spin with halo mass arises from the superposition
of two effects. Most stars form at around $z \approx 2$, where the gas
component is not yet depleted in its content of low specific angular
momentum material (compare to Fig.~\ref{FigFPbaryz}). Thus, the bulk
of stars forms from the cold, dense gas in the inner halo, which has
low specific angular momentum. 
Furthermore, \cite{Zavala16} find evidence for old stars to undergo a loss of specific angular momentum (up to $90\%$) to the outer dark matter halo through dynamical friction, similar to the loss that occurs in the inner dark matter halo to which old stars are attached. The overall low stellar spin determined by old stellar populations can thus be explained by a combination of both effects.
Later on, when low specific angular
momentum gas is either locked up in stars or expelled from the inner 
region due to feedback, star formation extends into the gas reservoir with 
higher specific angular momentum that now refills the inner halo. 
Thus young stars have
an enhanced spin on average, tracking the higher spin parameter of the
still available gas \citep[see also][]{Teklu15, Zavala16}. This imprints a trend
of spin with halo mass similar to the one exhibited by the gas spin on
top of the otherwise constant low stellar spin.

As pointed out by \cite{Zavala08}, and more recently by \cite{Zavala16} this behaviour is also closely related to the morphology of the galaxy forming at the halo centre. Those authors have shown that if most star formation takes place before turnaround, the stars are attached to dark matter clumps forming the inner halo and likewise loose a significant fraction of their specific angular momentum to the outer halo, leading to the formation of an elliptical galaxy. On the other hand, when star formation in the inner region is suppressed before turnaround, high specific angular momentum gas can penetrate the inner halo region at later times and form a stellar disc. Thus the precise strength and timing of the feedback events that determines the amount of expelled gas from the inner region and whether it overcomes the gravitational potential of the halo is crucial for the morphology of the central galaxy. In \emph{Illustris} we find a morphological mix of galaxies that is consistent with observations, which makes this simulation particularly suitable for the investigation of the impact of feedback onto the angular momentum properties. \cite{RodriguezGomez16} further investigate the connection between galaxy morphology, halo spin, and merger history. In future studies it will be crucial to disentangle the feedback induced mechanisms leading to the redistribution of matter and specific angular momentum inside of haloes and to quantify the impact on galaxy morphology.

We refrain from a detailed interpretation of the results at the very
high mass end of the distribution as it is affected by small number
statistics. Also, we disregard trends below a mass of
$10^{10} {\rm M}_\odot$ as our haloes were selected based on a minimum
dark matter particle number and thus the gas and stellar spin
parameters below this mass scale can be affected by resolution
effects.

\subsection{Specific angular momentum distributions of dark matter and gas} \label{Sec_Mj}

In order to understand whether galactic winds and AGN feedback can on
average expel a sufficient amount of low angular momentum gas from the
halo to account for the apparent gain of specific angular momentum in
the baryonic component we observe, we need to turn to the distribution
of specific angular momentum {\em inside} a halo. This has so far been
extensively studied for dark matter only haloes but not for the
baryonic component. \cite{Bullock01} suggested a universal angular
momentum profile for the distribution of specific angular momentum
inside a halo given by
\begin{equation} \label{eq:Mj}  
M(<j_{\rm sp}) = M_{\rm tot} \frac{\mu j_{\rm sp}}{j_0 + j_{\rm sp}},
\end{equation}
where $j_{\rm sp}$ is the specific angular momentum of dark matter
projected onto the rotation axis of the whole halo, $M(<j_{\rm sp})$
the cumulative dark matter mass of the dark matter with specific
angular momentum smaller than a given value $j_{\rm sp}$, and
$M_{\rm tot}$ denotes the total halo mass. $\mu$ and $j_0$ are two
fitting parameters that are not mutually independent. 
\cite{Bullock01} derived the universal angular momentum
profile from dark matter only SO-haloes, such that
$M_{\rm tot} = M_{200}$.

\begin{figure*}
\centering
\includegraphics[width=0.49\textwidth,trim= 0 0 0 0,clip]{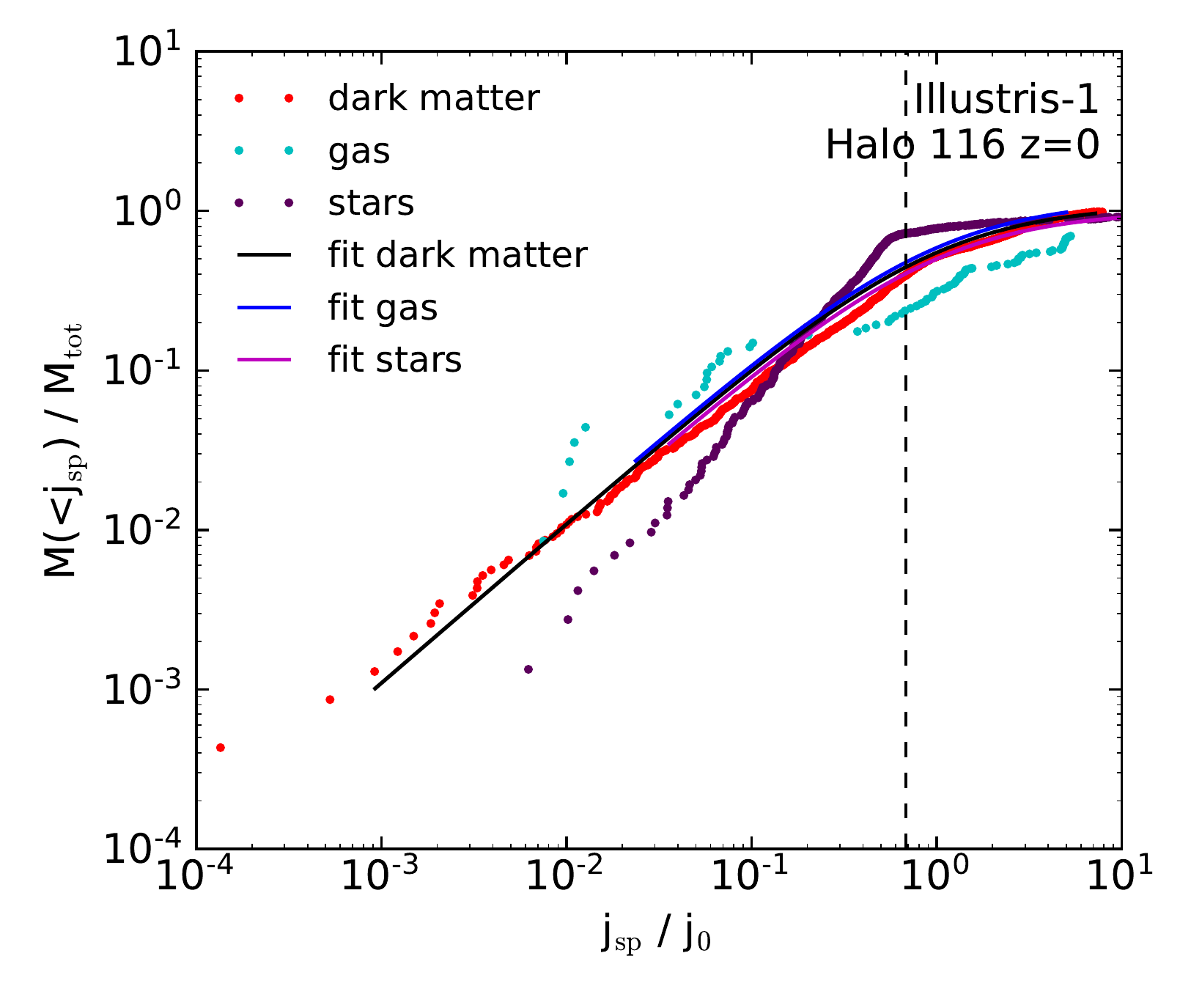}
\includegraphics[width=0.49\textwidth,trim= 0 0 0 0,clip]{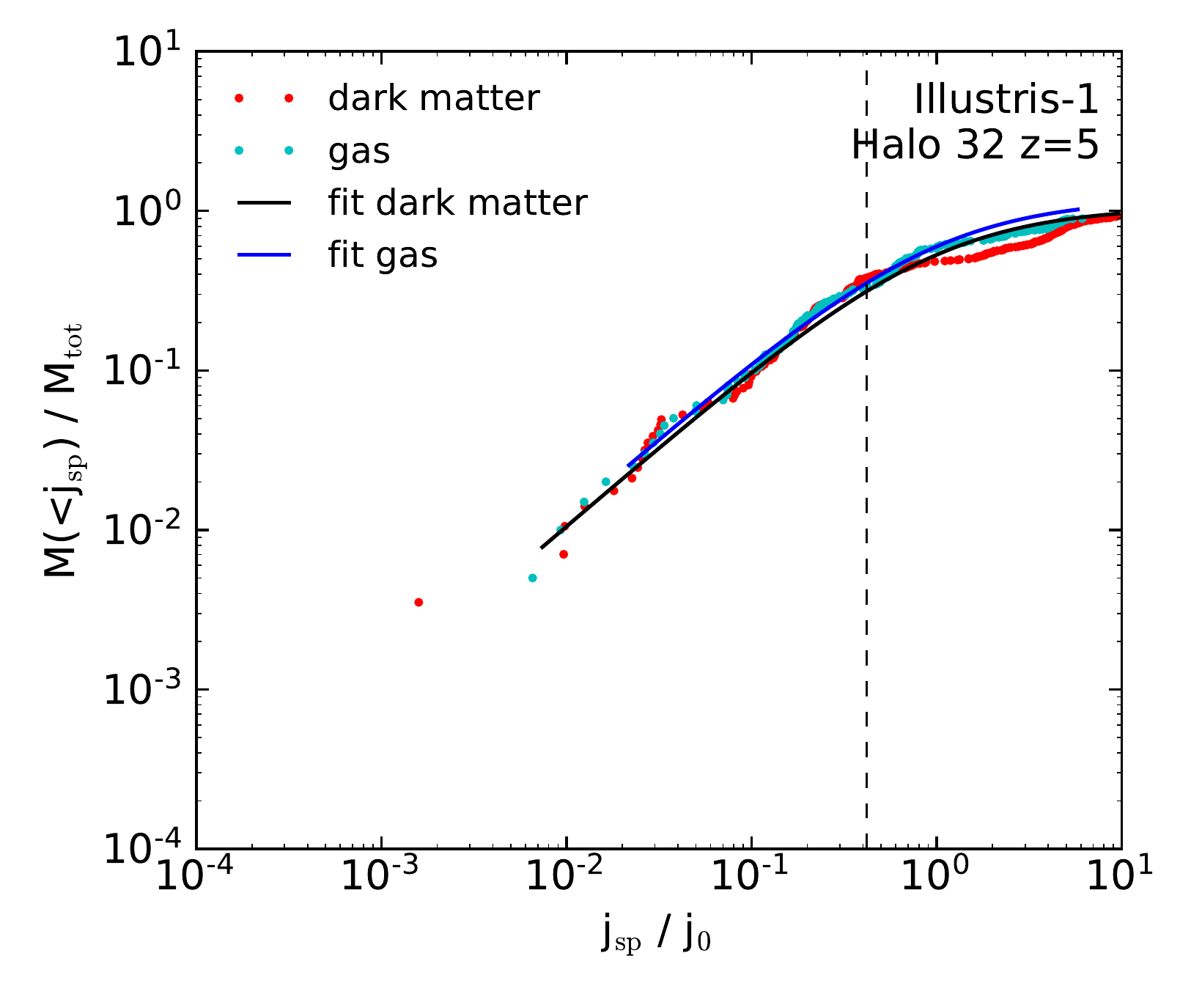}
\caption{Specific angular momentum distributions of a random halo from
  Illustris-1 at $z=0$ and $z=5$, obtained by binning the dark matter, gas,
  and stars in spherical bins of 1000 dark matter/star particles or gas cells
  each and calculating the specific angular momentum in each bin. The
  bins are then sorted by their specific angular momentum value
  projected onto the normalised total angular momentum of the whole
  halo. The distributions show the cumulative mass $M(< j_{\rm sp})$
  of all bins with $j_{\rm sp}$ smaller than a given value. Solid
  lines show least square error fits of the universal angular
  momentum profile to the derived distribution approximated by 50 mean
  values estimated in equidistant logarithmic bins. At high redshift,
  the universal profile is a good fit to both the dark matter and gas
  distributions. However, at $z=0$, this is true only for dark matter,
  indicating that the gas distribution gets highly perturbed by
  feedback processes during the subsequent evolution, causing 
  also the stellar distribution to be perturbed away from the universal profile.
  The black dashed line marks the average $j_{\rm sp}$ value below which 
  the gas is expelled from the halo due to feedback. See Section \ref{Sec_FPgas} for more details. \label{FigFPMj}}
\end{figure*} 

Above we deliberately used the term `dark matter' instead of `dark
matter particle', as the specific angular momentum distribution is
here meant to apply to the mean streaming velocity of the material,
not to individual particles. Due to the finite velocity dispersion of
the dark matter, a substantial fraction of the dark matter particles
can actually be counter-rotating with respect to the net rotation
direction. To account for this effect one needs to average over a
sufficiently large number of dark matter particles to obtain a fair
estimate of the mean streaming velocity and the specific angular
momentum. To derive specific angular momentum distributions we thus
bin the dark matter particles in spherical shells around the halo
centre of 1000 particles each, and use the specific angular momenta
and masses of the spherical bins as data points for the distribution.

Because searching the whole simulation volume for particles and cells
belonging to every SO-halo is computationally expensive in
post-processing, we simplify the present analysis by applying the
$R_{200}$ cut only to particles and cells belonging to the
corresponding FOF-halo. As we here analyse only very massive and
extended haloes, the number of particles/cells being part of the
SO-halo but not part of the corresponding FOF-halo is
negligible. However, we caution that this is not generally true,
especially for low mass haloes.  We also want to remark that deriving
specific angular momentum distributions for FOF-haloes can be more
problematic, as in the outskirts a significant fraction of the angular
momentum is carried by the orbital angular momentum of minor mergers,
which can have infall trajectories counter-rotating with respect to
the main halo. Simply removing the resulting bins with negative
specific angular momentum introduces a discrepancy in the total halo
mass, such that we refrain from adopting this approach.

In Fig.~\ref{FigFPMj} we show the specific angular momentum
distributions derived for the different halo components of a
randomly selected massive halo from Illustris-1 at $z=0$ and $z=5$. We
fit the universal angular momentum profiles given by Eq.~(\ref{eq:Mj})
to the dark matter, gas, and stellar specific angular momentum distributions. To
this end we bin the data points in 60 equidistant bins in the full
$j_{\rm sp}$-range and provide least square error fits to the average
values in those bins. We ignore the 10 lowest bins, as those contain
only few or no data points at all, leaving 50 values for determining
the best fit. Note that $\mu$ and $j_0$ are not independent fitting
parameters. Defining $m=M(<j_{\rm sp})/M_{\rm tot}$ one can rewrite
Eq.~(\ref{eq:Mj}) as
\begin{equation}
j_{\rm sp}(m) = \frac{m j_0}{\mu - m},
\end{equation}
such that the universal angular momentum profile can be reduced to a
one parameter function where the two fitting parameters $\mu$ and $j_0$
are related by
\begin{equation} \label{Jsp}
j_{\rm sp}^{\rm tot} = j_0 \int_0^1 \frac{m}{\mu - m} \text{d}m= j_0 \ [-\mu \ \text{ln}(1 - 1/\mu) - 1].
\end{equation}
Here $j_{\rm sp}^{\rm tot}$ is the absolute value of the total specific angular
momentum of the halo subset considered, and $\mu > 1$.

Fig.~\ref{FigFPMj} shows that at high redshifts the dark matter and
gas components of a halo have identical `initial' specific angular
momentum distributions and as such also spin parameters. This is a
direct consequence of the `initial' spin of dark matter and gas being
caused by large scale tidal torques from the surrounding gravitational
field, which acts the same way on all matter species. The universal
angular momentum profile, derived for haloes at $z=0$, turns out be a
reasonably good approximation of the specific angular momentum
distribution at high redshift, in agreement with dark matter mostly
sustaining its `initial' spin.

Consistently, the universal angular momentum profile is a good fit to
the dark matter specific angular momentum distribution at $z=0$. The
gas component at $z=0$ is however significantly perturbed by the
action of feedback which sets in at later times such that the gas
distribution does not follow the universal profile any
more. This also causes the angular momentum distribution of stars
to deviate from the universal profile.
Furthermore, the gas distribution lacks the low specific angular
momentum part of the gas that is still present in the dark matter, as
it was either locked up in stars or expelled by feedback. Though the
best fit universal angular momentum profiles are identical for all
halo components, the true distributions of the specific angular momentum of
gas and stars inside a halo exhibit very different features than prescribed by
the universal profile. We have investigated the gas specific angular
momentum distributions of many haloes and find a large variety in
their distributions with little commonality, thus not lending itself
to a description through a simple universal function. Instead, the
specific angular momentum distribution of gas inside a halo depends
strongly on the particular history of the individual halo.

\subsection{The origin of the baryonic spin enhancement} \label{Sec_FPgas}

In Section~\ref{Sec_FPbary} we have shown that the baryonic component
of haloes shows an enhanced spin compared to the dark matter.  Taking
the average of all haloes contained in the three examined mass ranges
we find a baryon to dark matter spin parameter ratio of $\sim 1.8$.
We can use the universal angular momentum profile from
\cite{Bullock01} to estimate the expected enhancement of the baryonic
spin parameter when a fraction of the low specific angular momentum
gas is expelled from the halo due to feedback, allowing us to
investigate whether this already explains the observed enhancement. To
this end, we however need to know how much baryonic mass is on average
still present in a halo at $z=0$.

In a quiet environment with no feedback at work the baryon to total
mass ratio of haloes should equal the universal cosmic baryon
fraction, $f_\text{b} = \Omega_b/\Omega_m = 0.1673$. With feedback
physics at work we find an average baryon to total mass fraction of
SO-haloes from Illustris-1 at $z=0$ of
$f_\text{halo} = M_\text{baryons} / (M_\text{dm} + M_\text{baryons}) =
0.0963$,
which is significantly below the cosmic baryon fraction. Compared to
the cosmic average, SO-haloes thus lose on average
$f = (f_\text{b}-f_\text{halo})/f_\text{b} \approx 42\%$ of their
initial gas mass.\footnote{For FOF-haloes we obtain $f \approx 48\%$
  which in the end yields very similar results. However, we carry out
  our calculation for SO-haloes, because the universal angular
  momentum profile was shown to hold for this type of halo.}
This is in very good agreement with \cite{Sharma12} who find an excess of low angular momentum gas in their non-radiative simulations that yields an angular momentum distribution incompatible with the exponential density profile of spiral galaxies. Those authors show that this discrepancy cannot be alleviated by different merger histories and the redistribution of angular momentum associated with those, and estimate that haloes have to loose $\sim 40\%$ of their low angular momentum gas for most of their haloes to host disc galaxies. 

As feedback occurs in star-forming regions that are comprised of cold,
dense, and slowly rotating gas, it expels gas with low specific
angular momentum. 
The high specific angular momentum is then redistributed within the remaining gas creating a new tail of low specific angular momentum gas. This new tail is due to gas that is now entering the inner halo, and will typically be less prominent than before the onset of feedback.
In our calculation we therefore assume that it is
always the gas with the lowest specific angular momentum that is
expelled from the halo. With this ansatz and making use of the fact
that gas (and thus baryons, as there are only few stars at high
redshift) and dark matter have identical `initial' universal angular
momentum profiles, we can calculate the specific angular momentum of
the remaining baryons $j_{\rm sp,f}$ by integrating the universal
angular momentum profile over the remaining mass, which simply
corresponds to integration limits from $f=0.42$ to 1,
\begin{equation} \label{Jspf}
\begin{split}
j_{\rm sp,f} &= \frac{j_0}{(1-f)} \int_f^1 \frac{m}{\mu - m} \text{d}m \\
&= \frac{j_0}{(1-f)} \ [ -\mu \ \text{ln}(\frac{\mu - 1}{\mu - f}) - (1-f) ].
\end{split}
\end{equation}
The factor $1/(1-f)$ is introduced to account for the fact that the
specific angular momentum $j_{\rm sp,f} = J_{\rm f}/(1-f)M_{\rm tot}$
is now obtained from the remaining baryons which have a reduced mass
of $(1-f)M_{\rm tot}$ compared to the initial mass of the baryonic
component. In Fig.~\ref{FigFPMj} we indicate the lower integration
limit $f$ as black dashed line.

\cite{Bullock01} found that $90\%$ of their haloes lie in a parameter
range of $\mu \in [1.06, 2]$. Inserting these two limiting values\footnote{We adopt the values given by \cite{Bullock01}, as those were derived from a statistical sample of $\sim 200$ haloes with $>6000$ resolution elements, and verified on $\sim 400$ haloes with $>1000$ resolution elements. We have analysed the angular momentum distributions only for $\sim 20$ individual haloes, but find shape parameters consistent with the given range.}
in Eq.~(\ref{Jsp}) and Eq.~(\ref{Jspf}), and taking the ratio of these
two equations, we arrive at an enhancement of the specific angular
momentum of the baryonic component by
\begin{equation}
 \frac{j_{\rm sp, f}}{j_{\rm sp}}\ [\mu = 1.06 - 2] = 1.63 - 1.49. 
\end{equation}
Thus we find a relative boost in the spin parameter of the baryonic
component of $\sim 1.55$ due to expulsion of low specific angular
momentum gas from the halo by feedback. Combined with the `inherent'
increase in gas spin discussed in Section~\ref{Sec_NR} by a factor of
$\sim 1.3$ due to a transfer of specific angular momentum from the
dark matter to gas, this yields an overall enhancement by a factor of
$\sim 2$ relative to the dark matter spin. This value is somewhat 
overestimated as it does not take into account that some
of the low specific angular momentum gas is locked up in stars or
otherwise remains in the halo and contributes to the average baryonic
spin at $z=0$. However, our estimate also assumes instantaneous
  removal of low angular momentum gas, whereas in reality feedback is
  a continuous process taking place at the inner halo which is constantly 
  refilled with relatively low angular momentum gas.   An analysis
  invoking tracer particles following the gas flow and recording its
  history in detail will be crucial to accurately pin down the fraction of low
  angular momentum gas still remaining in the halo.  In this study, we
  merely provide an estimate of an upper limit for the enhancement of
  baryonic spin based on the assumption that feedback instantaneously
  expels gas with the lowest specific angular momentum. In that
  sense,
the observed enhancement of the spin of the
baryonic component of $\sim 1.8$ can be solely explained by the combined effects of
the removal of low specific angular momentum gas from haloes by
feedback and the `inherent' spin enhancement of the gas component due
to angular momentum transfer from dark matter to the gas during halo
assembly, which is already captured in non-radiative simulations. 
As our analysis is based on more than $320\ 000$ haloes, this shows 
that (in a statistical sense) it is not necessary to invoke
cold filamentary gas accretion as an additional source for the enhanced baryonic spin.
The enhancement of baryonic spin due to cold flows was  discussed in detail by \citet{Danovich15} using 29 individually selected Milky Way sized galaxies at redshifts $z = 4-1.5$ from a cosmological simulation carried out with {\small ART}. Further support for this mechanism was recently provided by \citet{Stewart2016} based on simulations of a single Milky Way sized galaxy with multiple hydrodynamic codes, strengthening cold filamentary accretion as a code-independent viable mechanism for baryonic angular momentum gain in individual objects.
However, \cite{Sales12} showed that filamentary gas accretion from misaligned filaments can also lead to the opposite behaviour, a reduced spin of the baryonic component and the formation of an elliptical galaxy in the halo centre. The statistical relevance of cold flows as a source for baryonic spin enhancement thus still has to be established in future studies. Our results suggest that this mode may be relevant only in selected objects.

\begin{figure}
\centering
\includegraphics[width=0.5\textwidth,trim= 0 0 0 0,clip]{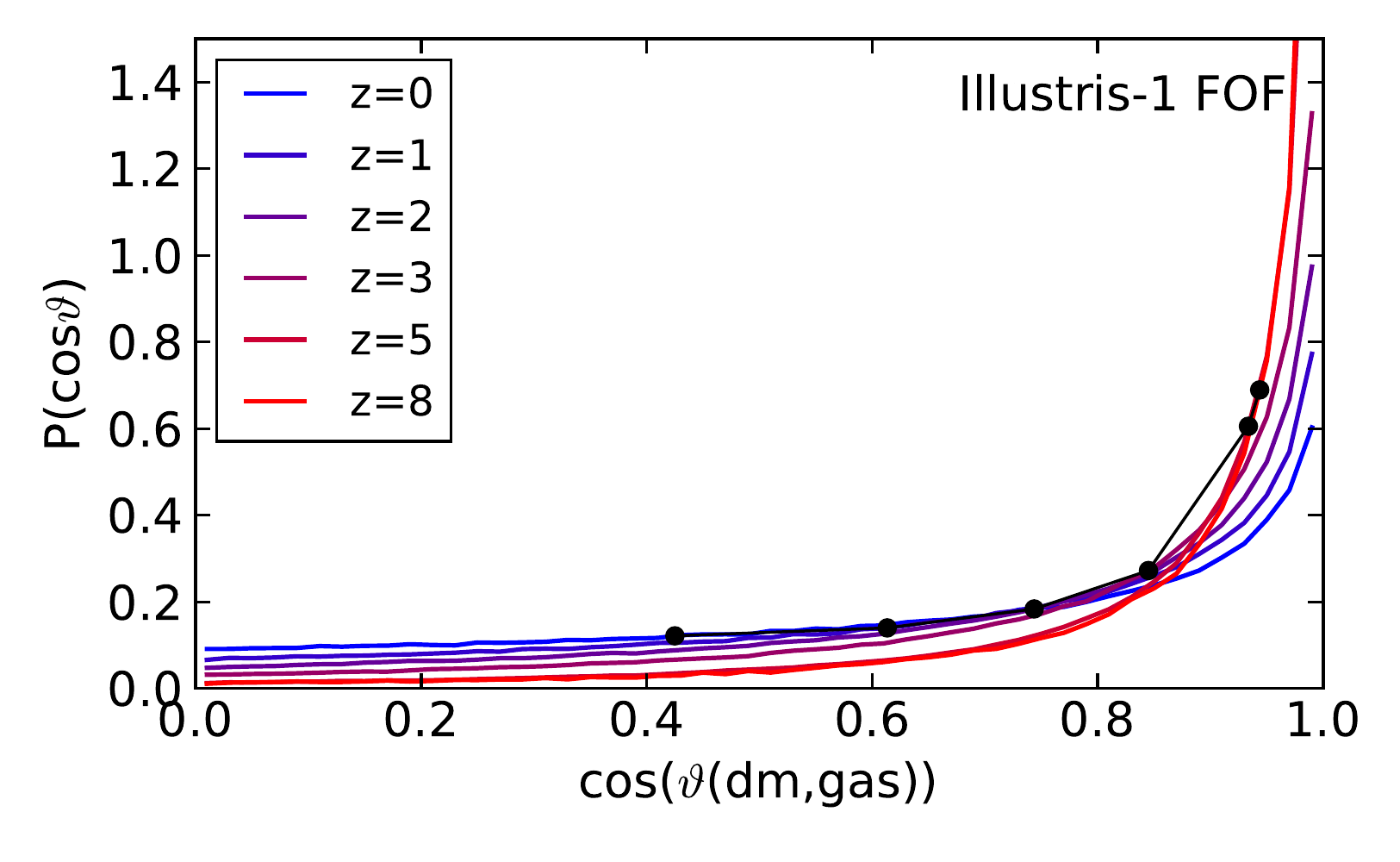}
\includegraphics[width=0.5\textwidth,trim= 0 0 0 0,clip]{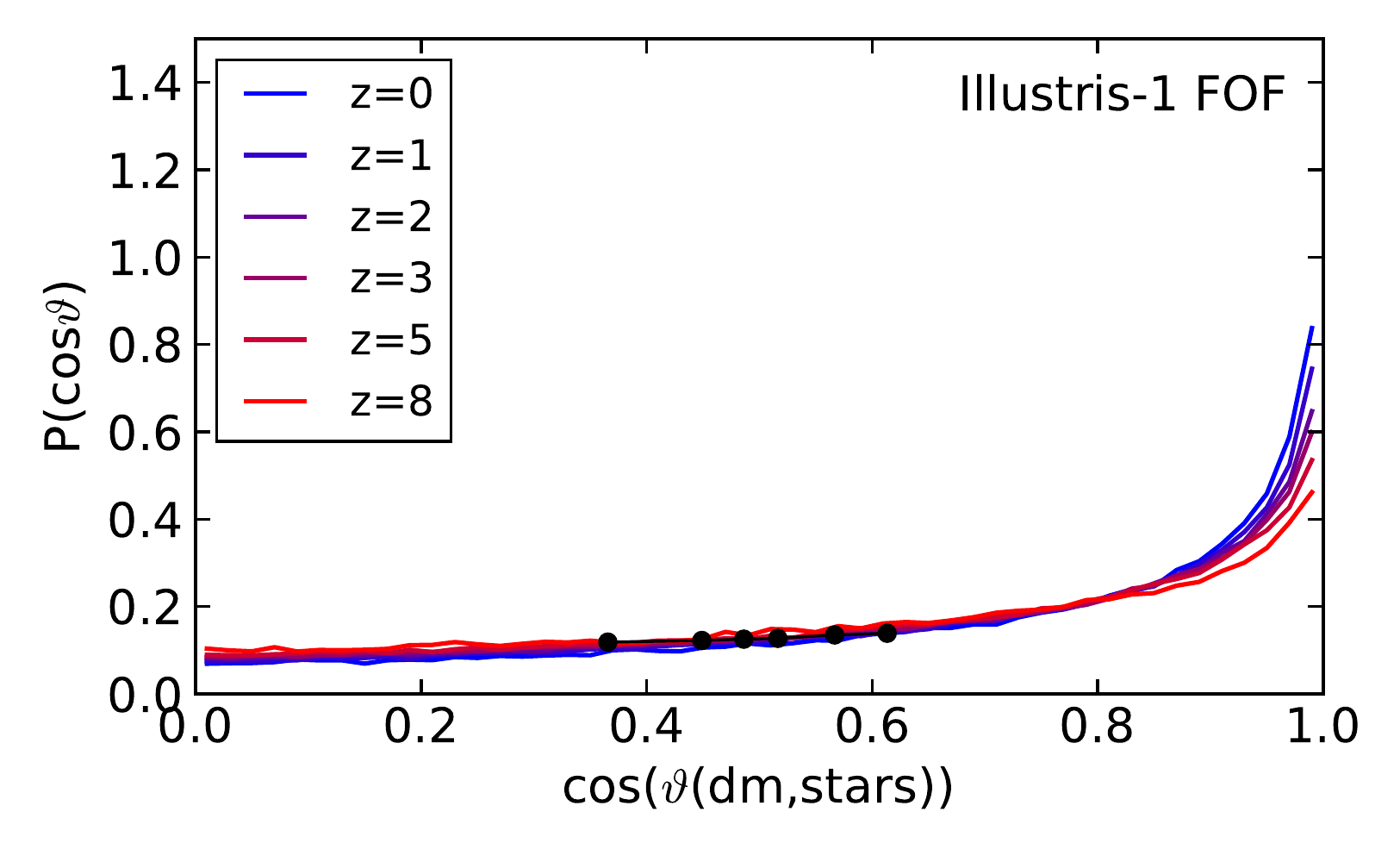}
\includegraphics[width=0.5\textwidth,trim= 0 0 0 0,clip]{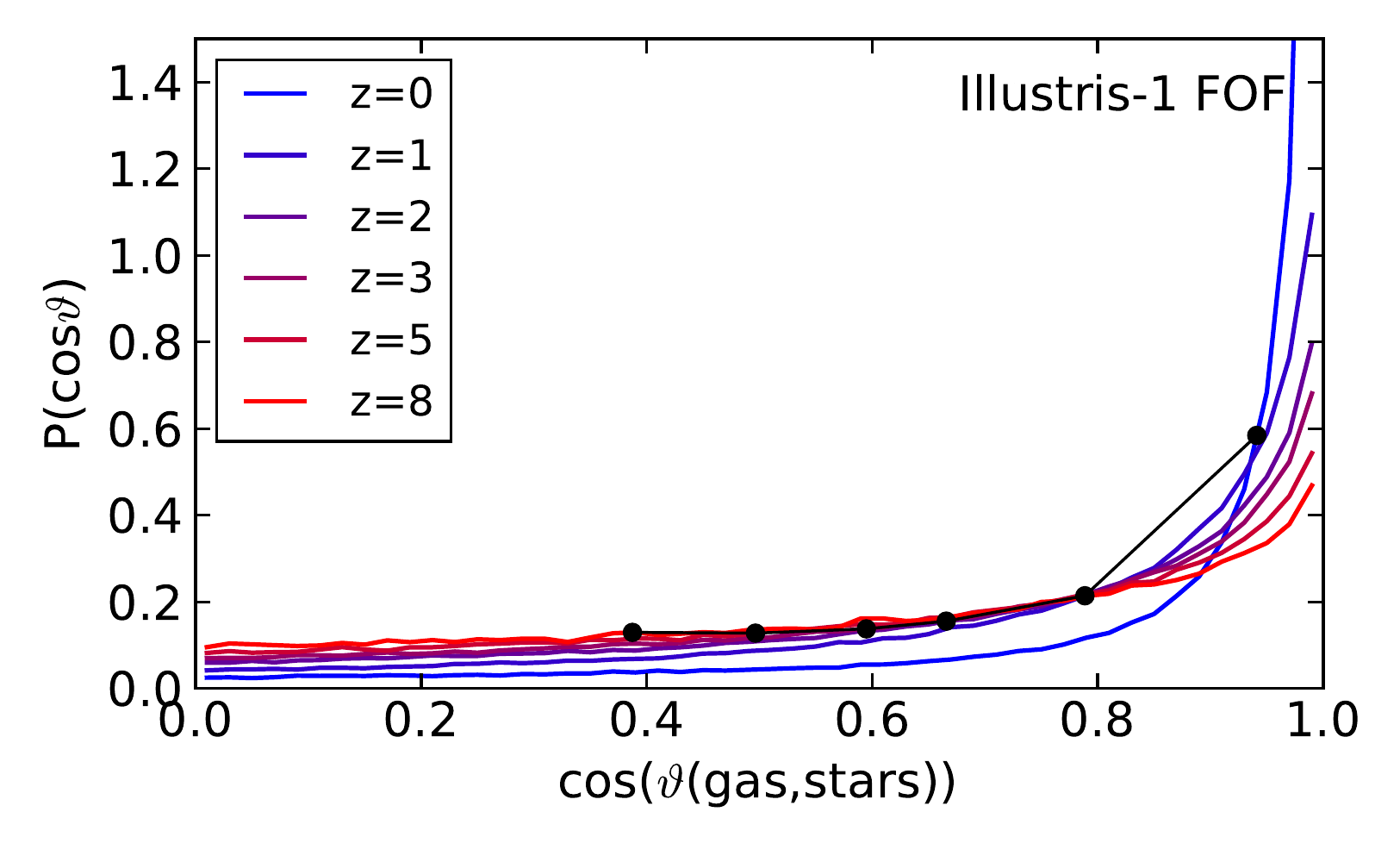}
\caption{Redshift evolution of the distribution of misalignment angles
  between the different halo components for FOF-haloes from
  Illustris-1. During cosmic evolution the dark matter and the
  stellar component (middle panel) as well as the gas and the stellar
  component (lower panel) become more aligned, whereas the
  misalignment between the dark matter and gas (upper panel) grows
  with cosmic time. 
  \label{FigFPmis}}
\end{figure}

\begin{figure}
\centering
\includegraphics[width=0.5\textwidth,trim= 10 0 0 5,clip]{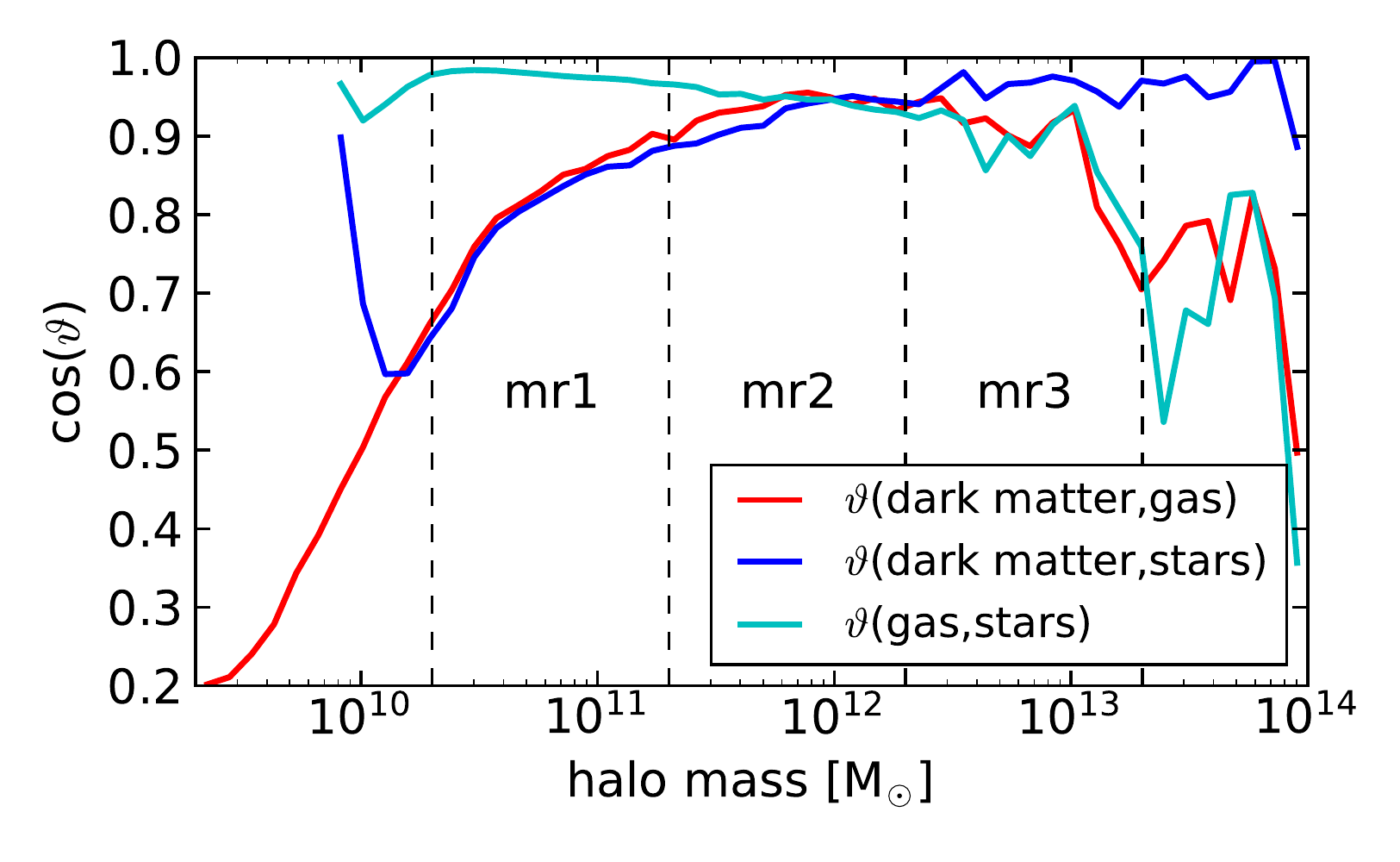}
\caption{Median misalignment between the different FOF-halo components from Illustris-1 as a function of halo mass. Best alignment between all halo components can be found in Milky Way sized haloes where the impact from feedback is weak. In general, the alignment of halo components is a strong function of halo mass, for a detailed discussion see Section~\ref{Sec_FPmis}. \label{FigFPctmass}}
\end{figure}

\subsection{Misalignment between the halo components} \label{Sec_FPmis}

Finally, we want to briefly analyse the distributions of misalignment
angles between the angular momentum vectors of the different halo
components and how they evolve with redshift. 
We quantify the misalignment by the cosine of the misalignment angle between two halo components,
\begin{equation}
\text{cos} (\vartheta) =  \frac{ \vec J_{1} \cdot \vec J_{2} }{ J_{1} \cdot J_{2} },
\end{equation} 
where $\vec J_{1/2}$ are the angular momentum vectors of two different
halo components, such as dark matter, gas, or stars and
$J_{1/2} = |\vec J_{1/2}|$ their absolute values. In Fig.~\ref{FigFPmis}
we show the distribution $P($cos$(\vartheta)$ of misalignments as a function of redshift. Black dots mark the median misalignment at every redshift. The distributions were derived by binning all FOF-haloes from Illustris-1 in 50 equidistant bins, covering the range of misalignments between cos$(0^\circ) = 1$ (perfectly aligned) and cos$(90^\circ) = 0$ (perpendicular), and normalising to the total number of haloes as well as the bin size. We refrain from showing the distributions up to cos$(180^\circ) = -1$ (anti-aligned), as in this range the distributions are a continuous extrapolation of the trend visible in the presented range.
We caution that the halo samples are always dominated by the lowest mass haloes whose gas and stellar components can be affected by poor resolution. However, we refrain from imposing an additional mass cut on our halo samples, because this would remove most of the haloes at high redshift.

We find that the dark matter and the stellar component (middle panel),
as well as the gas and the stellar component (lower panel) become
progressively more aligned towards lower redshift, whereas the misalignment between the dark matter and gas (upper panel) grows with cosmic time. The growing misalignment
between the gas and dark matter component is a natural consequence of
the feedback mechanisms continuously perturbing the gas but having
only indirect and weak effects on the dark matter through the change
of the baryonic density distribution. 

In Fig.~\ref{FigFPctmass} we show the median misalignment as a function of halo mass and thus feedback regime at $z=0$, derived by binning the FOF-haloes from Illustris-1 in 50 equidistant logarithmic bins in the given mass range and calculating the median misalignment cos$(\vartheta)$ in every bin. The vertical dashed lines single out the three mass ranges introduced in Section \ref{Sec_FPbary}. Below a FOF-halo mass of $\sim 10^{10} {\rm M}_\odot$ the median misalignment between stars and both the dark matter and gas exhibits a steep upward trend that is due to poor resolution of the stellar component. Above $\sim 2 \times 10^{13}\,{\rm  M}_\odot$ our results are affected by small number statistics.
Within the singled out intermediate mass ranges we find best alignments of all three components in Milky Way sized haloes where the impact from feedback is weak (see Section \ref{Sec_FPbary}). The alignment between dark matter and gas becomes worse in less and more massive haloes, as at those halo masses the gas component is perturbed by galactic winds and AGN feedback, respectively. The median misalignment angle across all three mass ranges is $\vartheta(\rm dm, gas) = 34.2^\circ$. Stars are almost perfectly aligned with the gas at small halo masses and exhibit the same misalignment with dark matter as the gas component. With increasing halo mass the stellar component becomes progressively better aligned with dark matter and correspondingly less well aligned with the gas. Within the three mass ranges we find median misalignment angles of $\vartheta(\rm dm, stars) = 35.7^\circ$ and $\vartheta(\rm gas, stars) = 12.1^\circ$. This trend can be possibly explained by massive haloes hosting elliptical galaxies comprised of mostly old stars, which are subject to gravitational interaction with dark matter but are not affected by the hydrodynamical interaction that the gas undergoes during the phase of late-time halo assembly. The stellar populations in low mass haloes will be typically younger and thus are expected to exhibit an angular momentum vector oriented along the rotational direction of the gas out of which these stars were formed.

Our results seem generally
consistent with previous studies, however a direct comparison is often
difficult due to the variety of adopted approaches in the literature,
such as measuring the misalignment for specific galaxy types, or only
in the inner region of haloes. We leave a detailed investigation of feedback induced misalignments between different halo components, as well as a study of radial trends in the spin alignment to future studies.

\section{Discussion and conclusions} \label{Sec_Dis}

In this work, we have analysed the distribution of halo spins in the
{\em Illustris} simulation suite, one of the first simulations of
galaxy formation with full hydrodynamics that produces a realistic
galaxy population in a sizeable volume, thus also yielding good
statistics, comparable to the best dark matter only simulations that
have been used for the study of these properties in the past. Our goal
has been a characterisation of the global angular momentum content of
haloes as a function of mass and time in this new generation of
hydrodynamical simulations, and to highlight the differences with
respect to dark matter only results.

To shed some light on the different approaches adopted in the
literature for measuring halo spin statistics, we have analysed the
systematic differences between the commonly used Peebles and Bullock
spin parameter definitions, and between the friends-of-friends (FOF)
and spherical overdensity (SO) halo definitions. Also, we have checked
the impact of sample selection criteria designed to single out
structures in quasi-equilibrium. For this investigation we have
employed a large sample of $\sim 400,000$ dark matter only FOF- and
$\sim 360,000$ SO-haloes from Illustris-1-Dark. The Peebles definition
yields robust spin parameter values for both halo definitions, an
advantage over the simpler Bullock parameter which has problems to
cope with the extended geometry of FOF-haloes. We find that for
SO-haloes the Bullock spin parameter needs to be rescaled by a
constant factor of 1.1 to yield the same mean value as the Peebles
parameter, when the concentration dependence is ignored. When
comparing spin parameters derived with the different definitions it is
thus necessary to bear this offset in mind. Finally, we find that
sample selection criteria have a small effect on the resulting spin
parameter distributions. The differences are of the same order of
magnitude as the variations between different results presented in the
literature, and thus can explain those discrepancies.

Interestingly, we find that only the Peebles spin parameter definition
applied to FOF-haloes yields a spin parameter distribution that is
self-similar in time to high accuracy. It is thus worthwhile to make
the extra effort of accurately calculating the gravitational binding
energy of haloes and use the Peebles definition to characterise the
angular momentum content of haloes. For this purpose, we use the
tree-gravity solver of {\small AREPO} for all haloes (as well as all
subhaloes, but these are not studied here) of the {\em Illustris}
simulation suite and include the corresponding results in a group
catalogue extension. Our augmented group catalogue furthermore
contains the kinetic energies of FOF- and SO-haloes (and subhaloes) as
well as the angular momentum vectors of the dark matter, gas, and
steller component of these. A full list of all properties available in
the group catalogue extension can be found in the Appendix, and these
data will be added to the public data release of \emph{Illustris}
\citep{Nelson15}.

With respect to the dark matter, we reproduce the well known result of
finding essentially no mass- and redshift dependence of the spin
parameter distribution. However, the subtle trend of spin with halo
mass and the fact that the number of haloes steeply increases with
decreasing halo mass causes the average spin of a halo sample to be
always dominated by the smaller haloes with lower spin and thus the 
resolution limit of the simulation. This has to be borne in mind when 
comparing literature results.

When baryons are added, the dark matter component retains its
properties. The baryons, however, exhibit a substantial gain in
specific angular momentum that increases towards low redshift. Already
in the non-radiative case, where one may naively expect both dark
matter and gas to retain their identical initial spins, we find an
enhanced gas spin by a factor of
$\lambda_{\rm gas} / \lambda_{\rm dm} \approx 1.3$ compared to the
dark matter.  This gain appears to arise from a transfer of specific
angular momentum from dark matter to the gas during late-time halo
assembly. This could be explained by infalling substructures getting their 
gas component ram pressure displaced, leading to a mutual torque between
the dark matter and gas components allowing for a net transfer of specific
angular momentum from dark matter onto gas. Such a transfer is also 
reflected in a small deficit of specific angular momentum at $z=0$ in the dark
matter component of the non-radiative simulation compared to the dark
matter only simulation. The amount of specific angular momentum
missing in the dark matter is exactly what is needed to balance the
gain observed in the gas component.

In simulations with active galaxy formation physics the enhancement of
the baryonic spin is even larger and leads to an average ratio of
$\lambda_{\rm gas} / \lambda_{\rm dm} \approx 1.8$. We derive this
value from a large sample of $\sim 320,000$ FOF-haloes from the
Illustris-1 simulation. If we assume that galactic winds and AGN
feedback expel preferentially low specific angular momentum gas from a
halo, we can estimate the expected apparent gain of specific angular
momentum in the baryonic component based on the total baryonic mass
lost from haloes. On average we find an expected enhancement by a
factor of $\sim 1.55$ from this effect. Combing this with the relative
enhancement of $\sim 1.3$ expected from the transfer of specific
angular momentum from dark matter to the gas as seen in the
non-radiative case, we arrive at a total enhancement by a factor of
roughly $\sim 2$ relative to the dark matter spin. Note that this
value represents a slight overestimate as our calculation does not
take into account some of the low specific angular momentum being
locked up in stars and remaining in the halo. However, the good
agreement with the actually measured enhancement factor of $\sim 1.8$
shows that these two effects combined are sufficient to explain the
higher specific angular momentum of the baryonic content in the full
physics simulations.  This also casts doubts onto recent suggestions
\citep{Stewart2016} that cold filamentary gas accretion is responsible
for the enhanced baryonic spin of haloes.

We also find that the different feedback mechanisms induce a strong
dependence of the gas spin on halo mass. In low and high mass haloes,
where galactic winds and AGN feedback are most efficient in expelling
low specific angular momentum gas, the gas spin is highest. The
stellar spin is far less affected, as star-formation takes place in
cold, dense gas in the inner region of haloes, where the material is
slowly rotating. The stellar component thus has small spin largely
independent of halo mass. The baryonic spin is ultimately determined
by the gas to stellar mass ratio of haloes which decreases with halo
mass. Another consequence from feedback processes perturbing the gas
component is a growing misalignment between the dark matter and gas
component with cosmic time, which again is largest at halo masses that
allow for most efficient feedback. Furthermore, we find that the alignment 
between the different halo components is a strong function of halo mass.

Our results thus clearly show that the baryonic spin sensitively
depends on the galaxy formation physics employed. Highly schematic
schemes for the evolution of the baryonic spin component, such as
invoked in simple inside-out scenarios for disk formation that assume
equal specific spin in dark matter and gas, need therefore be treated
with caution. It will be interesting to examine with future
simulations how strongly our results for the full physics simulations
depend on the details of the feedback modelling invoked to regulate
the galaxy formation process. Given the substantial impact of feedback
one may be inclined to anticipate a very large range of possible
outcomes. However, it appears also possible that the constraint to
reproduce basic observational facts such as the galaxy stellar mass
function effectively ties down the simulation predictions for the spin
properties, independent of the specific realisation of the feedback
physics.

\section*{Acknowledgments}

We thank the referee for insightful comments that helped to improve the paper.
JZ and VS acknowledge financial support from the Deutsche
Forschungsgemeinschaft through Transregio 33, ``The Dark Universe'',
and through the Klaus Tschira Foundation.  VS also acknowledges
support through the European Research Council under ERC-StG grant
EXAGAL-308037.  Some of the computations were performed on the
HazelHen supercomputer at the High-Performance Computing Center
Stuttgart (HLRS) as part of project GCS-ILLU of the Gauss Centre for
Supercomputing (GCS). JZ acknowledges support from International
Max-Planck Research School for Astronomy and Cosmic Physics at the
University of Heidelberg (IMPRS-HD).

\bibliography{am_paper.bbl}
\bibliographystyle{mn2e}

\begin{appendix}

\onecolumn

\section{Fields of extended group catalogue}

For completeness and as a reference, we here provide a full list of
all newly available halo and subhalo properties in the group catalogue
extension we computed for \emph{Illustris}. These additional
properties will be made available as part of the public data release
\citep{Nelson15} of \emph{Illustris}.

\begin{table*}
\begin{tabular}{ll}
\hline \hline
Name & description\\ \hline \hline
GroupEkin & kinetic energy of FOF-haloes \\ 
GroupEthr & thermal energy of the gas component of FOF-haloes\\
GroupEpot & potential energy of FOF-haloes\\ \hline
Group\_J&  total angular momentum of FOF-haloes\\
Group\_Jdm & angular momentum of the dark matter component of FOF-haloes\\
Group\_Jgas & angular momentum of the gas component of FOF-haloes\\
Group\_Jstars & angular momentum of the stellar component of FOF-haloes\\ 
Group\_CMFrac & total counter-rotating mass fraction of FOF-haloes\\
Group\_CMFracType & counter-rotating mass fractions per type: dark matter, gas, stars\\ \hline
Group\_Ekin\_* & kinetic energy of SO-haloes\\ 
Group\_Ethr\_* & thermal energy of the gas component of SO-haloes\\
Group\_Epot\_* & potential energy of SO-haloes\\ \hline
Group\_J\_* &  total angular momentum of SO-haloes\\
Group\_Jdm\_* &  angular momentum of the dark matter component of SO-haloes \\
Group\_Jgas\_* &  angular momentum of the gas component of SO-haloes \\
Group\_Jstars\_* & angular momentum of the stellar component of SO-haloes \\
Group\_CMFrac\_* &  total counter-rotating mass fraction of SO-haloes\\
Group\_LenType\_* & number of particles/cells of each matter type in SO-haloes\\
Group\_MassType\_* & mass per matter type in SO-haloes\\
Group\_CMFracType\_* & counter-rotating mass fractions per matter type in SO-haloes\\ \hline \hline
\end{tabular}
\caption{Full list of all newly available halo properties in the group
  catalogue extension. A halo can be either defined as a
  friends-of-friends (FOF) group or a spherical overdensity (SO). The
  latter can be either with respect to 200 times the critical density
  (Crit200), 500 times the critical density (Crit500), 200 times the
  mean density (Mean200), or with the redshift dependent overdensity
  expected for the generalised top-hat collapse model in a $\Lambda$-cosmology
  (TopHat200), see \citet{Bryan1998}. The corresponding properties can
  be accessed by replacing * with the terms in brackets. The different matter types in question are dark matter, gas, stars, and black holes.}
\end{table*}

\begin{table*}
\begin{tabular}{ll}
\hline \hline
Name & description \\ \hline \hline
SubhaloEkin & kinetic energy of subhaloes \\
SubhaloEthr & thermal energy of the gas component of subhaloes \\ 
SubhaloEpot & potential energy of subhaloes \\ \hline
Subhalo\_J(*) & total angular momentum of subhaloes\\
Subhalo\_Jdm(*) & angular momentum of the dark matter component of subhaloes\\
Subhalo\_Jgas(*) & angular momentum of the gas component of subhaloes \\
Subhalo\_Jstars(*) & angular momentum of the stellar component of subhaloes \\ 
Subhalo\_CMFrac(*) & total counter-rotating mass fraction of subhaloes\\
Subhalo\_CMFracType(*) & counter-rotating mass fractions per type: dark matter, gas, stars \\  \hline \hline
\end{tabular}
\caption{Full list of all newly available galaxy properties in the
  group catalogue extension. Subhaloes corresponding to galaxies are
  defined as all particles/cells that are gravitationally bound the
  same potential minimum. Furthermore, we include two more definitions
  of a galaxy that are closer to the observational approach and
  include only subhalo particles/cells that are in the stellar half
  mass radius (InHalfRad) or twice the  stellar half mass radius
  (InRad). The corresponding properties can be accessed by replacing * 
  with the terms in brackets. The different matter types in question 
  are dark matter, gas, stars, and black holes.}
\end{table*}

\end{appendix}

\label{lastpage}

\end{document}